\documentclass[12pt]{iopart}

\usepackage{graphics}
\usepackage{color}
\usepackage{epsfig}
\expandafter\let\csname equation*\endcsname\relax
\expandafter\let\csname endequation*\endcsname\relax%
\usepackage{amsmath}
\usepackage{mathrsfs}
\usepackage{amssymb}
\usepackage{bbm}
\usepackage{graphicx}
\usepackage{subfigure}
\usepackage{color}
\usepackage{hyperref}
\usepackage{slashed}
\usepackage{footmisc}
\usepackage{multirow}

\newcommand{\be}{\begin{equation}} \newcommand{\ee}{\end{equation}}
\newcommand{\ba}{\begin{array}{c}} \newcommand{\ea}{\end{array}}
\newcommand{\bea}{\begin{eqnarray}} \newcommand{\eea}{\end{eqnarray}}

\newcommand{\al}{&\!\!\!\!}

\begin{document}

\title[A Review on Partial-wave Dynamics]{A Review on Partial-wave Dynamics with Chiral Effective Field Theory and Dispersion Relation}

\author{De-Liang Yao$^{1,2,\dagger}$, Ling-Yun Dai$^{1,2,\dagger\dagger}$, Han-Qing Zheng$^{3,4,\ast}$, {\rm and} Zhi-Yong Zhou$^{5,\ast\ast}$}
\address{$^1$School of Physics and Electronics, Hunan University, 410082 Changsha, China}
\address{$^2$Hunan Provincial Key Laboratory of High-Energy Scale Physics and Applications, Hunan University, Changsha 410082, China}
\address{$^3$Department of Physics and State Key Laboratory of Nuclear Physics and Technology, Peking University, Beijing 100871, China}
\address{$^4$Collaborative Innovation Center of Quantum Matter, Beijing 100871, China}
\address{$^5$School of Physics, Southeast University, Nanjing 211189, China}
\ead{$^\dagger$yaodeliang@hnu.edu.cn}
\ead{$^{\dagger\dagger}$dailingyun@hnu.edu.cn~(corresponding author)}
\ead{$^{\ast}$zhenghq@pku.edu.cn}
\ead{$^{\ast\ast}$zhouzhy@seu.edu.cn}

\vspace{10pt}



%
%
%

\begin{abstract}
The description of strong interaction physics of low-lying resonances is out of the valid range of perturbative QCD. Chiral effective field theories have been developed to tackle the issue. Partial wave dynamics is the systematic tool to decode the underlying physics and reveal the properties of those resonances. It is extremely powerful and helpful for our understanding of the non-perturbative regime, especially when dispersion techniques are utilized simultaneously. Recently, plenty of exotic/ordinary hadrons have been reported by experiment collaborations, e.g. LHCb, Belle, and BESIII, etc..
In this review, we summarize the recent progress on the applications of partial wave dynamics combined with chiral effective field theories and dispersion relations, on related topics, with emphasis on $\pi\pi$, $\pi K$, $\pi N$ and $\bar{K}N$ scatterings.
\end{abstract}

\tableofcontents

\section{Introduction}
Hadron physics, the study of strong interaction of matter, is one of the most challenging and fascinating fields of modern science.  One of its primary goals is to scrutinize hadron structure and hadron spectrum, and particularly to explain the newly observed states in experiments.  As modern experiments are stepping into the era of precision measurements,  precise descriptions of the relevant phenomenologies become more and more essential, which strongly requires rigorous theoretical analyses of the underlying dynamics in a proper manner.

The underlying dynamics of strong interaction is governed by Quantum chromodynamics (QCD), a non-Abelian gauge field theory with quarks and gluons as degrees of freedom. Though at high energies perturbative treatments of the short-distance interactions are feasible and achievable based on the QCD Lagrangian,  it becomes extremely difficult in the low-energy region due to the fact that the low-energy strong-interaction phenomena is non-perturbative and has not yet been analytically solved from first principles.  As a result, instead of a direct use of QCD, various effective field theories (EFTs) or phenomenological models, motivated by the symmetries of QCD, or numerical approaches like lattice quantum chromodynamics (LQCD) have to be introduced to investigate hadron spectroscopy and hadron structures at low energies. Below we will follow the path of an EFT built based on chiral symmetry and, in particular, focus on its optimization by implementing axiomatic principles of deductive $S$-matrix theory.

Chiral perturbation theory ($\chi$PT), is one of the  EFTs of QCD in the low-energy domain, which continuously plays an important role in modern hadron physics and nuclear physics. It was originally set up for the study of the interactions and properties of Goldstone bosons, following the suggestion by Weinberg that the soft-pion results of current algebra can be recasted to the formalism of an EFT~\cite{Weinberg:1978kz}. The formalism was systematically developed by Gasser and Leutwyler in the meson sector for the two-flavor~\cite{Gasser:1983yg} and three-flavor~\cite{Gasser:1984gg} cases.\footnote{For a pioneering work on chiral logarithms and loops, see Ref.~\cite{Li:1971vr}.} Though $\chi$PT is an EFT realization of its underlying theory with the same symmetries, it has features of its own. Firstly, it is a perturbation theory feasible at low energies. Secondly, $\chi$PT is a non-renormalizable theory in the traditional sense, even though its fundamental theory is a renormalizable one. Thirdly, The degrees of freedom of $\chi$PT are no longer quarks and gluons in QCD but their hadronic asymptotic states that can be observed experimentally, making the comparison to experimental measurements convenient.

Over the years,  $\chi$PT gained a lot of achievements, which can be found in the reviews~\cite{Meissner:1993ah,Ecker:1996yy,Bijnens:2006zp,Scherer:2012xha}. Calculations of scattering processes and observables at one-loop (next-to-leading order, NLO) and two-loop (next-to-next-to-leading order, NNLO) orders became standard. For instance,  the NLO calculation of $\pi\pi$ scattering was performed in Ref.~\cite{Gasser:1983yg} and the NNLO calculation was performed by the authors of Refs~\cite{Bijnens:1995yn,Bijnens:1997ni}. For the SU(3) case, the one-loop calculation of $\pi K$ scattering can be found in Refs.~\cite{Bernard:1990kw,Bernard:1990kx}, and the two-loop one was done in Ref.~\cite{Bijnens:2004bu}. On the other hand, various versions of $\chi$PT have been developed to deal with specific degrees of freedom under consideration. Thereinto, one of the most important is baryon $\chi$PT (B$\chi$PT), which additionally takes baryons into account and becomes a powerful and systematical tool in nuclear physics. We refer the readers to e.g. Refs.~\cite{Bernard:2007zu,Scherer:2012xha} for reviews on one-baryon/nucleon sector and to e.g. Refs.~\cite{Bernard:1995dp,Epelbaum:2008ga,Hammer:2019poc} for reviews on few and many body nucleon systems, respectively. It is worth noting that, unlike Goldstone bosons in the purely mesonic $\chi$PT, the baryons have non-zero masses in the chiral limit, a consequence of which is that the power counting becomes subtle as pointed out in Ref.~\cite{Gasser:1987rb}. To tackle the power counting breaking (PCB) problem, various methods have been proposed during the past thirty years~\cite{Jenkins:1990jv, Bernard:1992qa,Ellis:1997kc,Becher:1999he,Fuchs:2003qc}.  Though there exist different versions of $\chi$PT, their common key ingredients are chiral symmetry of QCD and the concept of EFT. Therefore, throughout this review, they are all referred to as chiral effective field theory ($\chi$EFT) for brevity.

By construction, $\chi$EFT can be used to perturbatively calculate the transition amplitudes of a given process under the guidance of the so-called power counting rule, which corresponds to a perturbation expansion of the $S$-matrix. However, the obtained $\chi$EFT amplitude usually violates unitarity as required by the full $S$-matrix, since it is impossible, in reality, to perform a full calculation up to the infinite order. Furthermore, from the viewpoint of $S$-matrix theory, the perturbative terms do not yield the correct singularities on the unphysical Riemann sheets where dynamical properties, e.g. the formation of a resonance, is involved.\footnote{In $S$-matrix theory for a single channel interaction, a bound/virtual state is a singularity below threshold on the real axis on the physical/second sheet of the complex s-plane of the S-matrix, while resonances are poles on the unphysical sheets of the complex $s$-plane.} In practice, the restoration of unitarity can be established by performing a resummation of the perturbative scattering amplitude. The procedure of restoring unitarity is called unitarization. The most widely used unitarization approaches are e.g. $K$-matrix method, Pad\'e approximation, Lippmann-Schwinger and Bethe-Salpeter equations, etc..

Nevertheless, the abovementioned unitarization methods still have extra shortcomings. As the unitarization procedure is normally equivalent to a resummation of the scattering amplitudes in the s-channel, it breaks crossing symmetry. Besides, they might also cause violation of causality, resulting in the emergence of hazardous spurious poles on the physical sheet of the complex $s$-plane. Thus, more rigorous model-independent approaches have been developed by using dispersion relations in literature. For instances, Roy-type equations are derived  based on fixed-t dispersion relations and crossing symmetry can be regained, for an early attempt, see ref.~\cite{Hannah:2001ee}.  Another approach is the production representation method proposed in Refs~\cite{Zheng:2003rw,Xiao:2000kx,Zhou:2004ms}. It unitarizes the left-hand (and inelastic) cuts of the perturbative amplitude taken from $\chi$EFT so as to avoid the violation of causality. The consistency of this representation with crossing symmetry was later examined in Refs.~\cite{Guo:2007hm,Guo:2007ff}. Both of the Roy-equation approach and the production representation method have been applied to investigate $\pi\pi$ scattering, leading to a robust establishment of the existence of the $\sigma$ state~\cite{Ananthanarayan:2000ht,Xiao:2000kx,Pelaez:2015qba}.

To conclude, a rigorous manner to construct a trustworthy amplitude is to make a joint use of chiral amplitudes and general principles of $S$-matrix theory such as unitarity, analyticity and crossing symmetry. Usually, results from $\chi$EFT are utilized either as inputs like in the various unitarization approaches, or low-energy benchmarks for e.g. Roy-equation analyses. It should be emphasized that the proper singularities of the partial-wave amplitudes under consideration should be respected faithfully, especially when their contribution is not negligible. In this review,  we will go through the various schemes towards faithful scattering amplitudes of two-body interactions by imposing the $S$-matrix theory requirements, and discuss their great achievements in hadron spectroscopy by showing some selected important examples.

The outline of this review is as follows. In section~\ref{sec:2}, we will first give a short review on the theoretical approaches of constructing amplitudes with the constraints of unitarity and analyticity  and with the help of chiral results from $\chi$EFT. Specifically, the concept of final state interactions is discussed in section~\ref{sec:FSI}. Resummation methods relying on unitarity are presented in section~\ref{sec:2.2}, while dispersive approaches developed in accordance with analyticity are given in section~\ref{sec:2.3}. Section~\ref{sec:3} is devoted to the discussion of hunting hadron states  by applying unitarization methods and dispersion relations. Examples of light scalar mesons, light baryon resonances and exotic states are briefly reviewed in section~\ref{sec:lsm}, \ref{sec:3.2} and \ref{sec:3.3}, in order. We end with a brief summary and outlook in section~\ref{sec:sum}.

\section{Towards faithful scattering amplitudes}\label{sec:2}

\subsection{Final state interactions \label{sec:FSI}}
\begin{figure}[hpt]
\begin{center}
\includegraphics[width=0.8\textwidth,height=0.3\textheight]{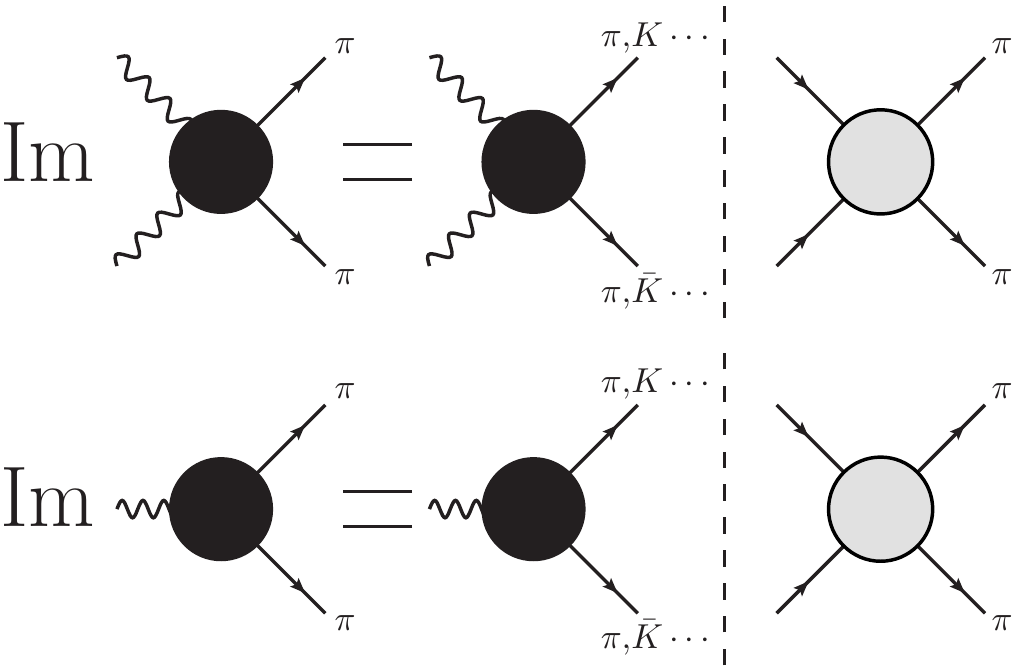}
\end{center}
\caption{\label{Fig:FSI} The illustrative diagrams to indicate the theorem of FSI in two-body scattering and decay processes, with two pions as the final states. The virtual dashed lines are the cuts. On the right side of the cut they are the hadronic scattering.   }
\end{figure}

Final-state interactions (FSI) are important to study hadron physics, which is a good topic to start with, showing how various non-perturbative methods are necessary in surpassing perturbation calculations. It describes the rescatterings of hadrons \lq weakly' produced. For simplicity, we only focus on two body scatterings\footnote{For FSI of three body re-scattering, it can be studied based on two body re-scattering amplitudes, such as the Faddeev equation \cite{Faddeev:1960su} and the Khuri-Treiman equation \cite{Khuri:1960zz}. There are some approaches trying to solve the Faddeev equation, for instance, the Alt-Grassberger-Sandras approach \cite{Alt:1967fx} and the so called fixed-center approximation \cite{Xie:2010ig,Ren:2018qhr}. For three body decay processes studied by Khuri-Treiman equation, we refer to e.g. Refs.~\cite{Kambor:1995yc,Guo:2015zqa} for $\eta\to\pi\pi\pi$  and Ref.~\cite{Niecknig:2012sj} for $\omega/\phi\to\pi\pi\pi$ and references therein.}.  For instance, the processes of two photons scattering into two pions or a virtual photon transiting into two pions can be considered as typical examples of FSI of two body re-scattering, as shown in Fig.\ref{Fig:FSI}.
$\chi$EFT works well in the low energy region, and it can be used to calculate the lowest order diagrams of the whole amplitude.
Dispersion relation is a strict method to obtain amplitudes through the discontinuity across the branch cut, which keeps analyticity, unitarity, and crossing symmetry.
Thus a combined use of $\chi$EFT and dispersion relation is very helpful to study the processes including FSI.

The theorem of FSI can be expressed as
\begin{equation}
{\rm Im} {\mathcal{F}_k}^I_J(s)=\sum_{l}{{\mathcal{F}_l}^I_J}^*(s) \rho_{l}(s) {T_{lk}}^I_J(s)\, . \label{eq:FSI}
\end{equation}
Here $\mathcal{F}^I_J(s)$ is the partial wave amplitude where the initial state(s) decays (scatter) into the channels related to the hadronic re-scattering amplitude ${T_{lk}}^I_J(s)$ with isospin $I$, spin $J$ and channel labels $l$, $k$.  The two-body phase space factor $\rho_{l}(s)$ is given by
\begin{equation}
\rho_l(s)=\frac{1}{s}\sqrt{\left(s-(M_l+m_l)^2\right)\left(s-(M_l-m_l)^2\right)} \,,\label{eq;rho}
\end{equation}
and  $M_l$ and $m_l$ are the masses of the two particles in the $l$-th channel, respectively. The theorem of FSI can be graphically elaborated with Fig.\ref{Fig:FSI}.
On the left-hand side of the equality, they are the imaginary parts of the full amplitude $\mathcal{F}^I_J(s)$ in the physical region.
This discontinuity is related to the product of the diagrams on the right-hand side of the equality. The two parts are obtained by cutting off the two intermediate propagators, guided by the Cutkosky rule from quantum field theory (QFT), where one of them is changed into the conjugate form and also a phase space factor should be included.
Notice that $\mathcal{F}^I_J(s)$ should be amplitudes of electromagnetic or weak interactions, e.g. $\gamma\gamma\to\pi\pi/\bar{K}K$ or $D\to\pi\pi e\nu$. The only requirement is that the FSI are of strong interaction nature.
For the hadronic scattering, the partial wave amplitudes $T^I_J(s)$ should satisfy the coupled channel unitarity
\begin{equation}
{\rm Im} {T_{nk}}^I_J(s)=\sum_{l}{{T_{nl}}^I_J}^*(s) \rho_{l}(s) {T_{lk}}^I_J(s)\, . \label{eq:T;unit}
\end{equation}

In the single channel case, it is easy to find that Eqs.~(\ref{eq:FSI},\ref{eq:T;unit}) ensure that $\mathcal{F}^I_J(s)$ and  $T^I_J(s)$ have the same phase, which is the well-known Watson's theorem \cite{Watson:1952ji}.  For coupled channel case, the phases of $\mathcal{F}^I_J(s)$ and $T^I_J(s)$ are the same only in the elastic region. In practice, $\mathcal{F}^I_J(s)$ can be represented by the Au-Morgan-Pennington (AMP) method \cite{Au:1986vs,Morgan:1993td}:
\begin{equation}
{\mathcal{F}_k}^I_{J}(s)\;=\;\sum_{l}{\alpha_l}^I_{J}(s)\,{T_{lk}}^I_J(s)\,, \label{eq:FSI;AMP}
\end{equation}
with the coupling functions $\alpha_l(s)$ real, in terms of polynomials of $s$, which simulates the contribution from the left hand cut (l.h.c.) and distant right hand cut (r.h.c.).

Even in the low energy region, the FSI could change the amplitudes obviously from the prediction of $\chi$EFT, see the difference between the Born amplitude of the $\gamma\gamma\to\pi\pi$ and the one given by dispersion relations \cite{Dai:2014zta,Dai:2016ytz}, there is a distinct difference between them not faraway from $\pi\pi$ threshold. In the high energy region, perturbation calculation with $\chi$EFT is impossible, thus many models are built with implementation of the principles of QFT. In these models, the $T^I_J(s)$ and $\mathcal{F}^I_J(s)$ amplitudes are related via the theorem of FSI, see e.g. Refs.~\cite{Au:1986vs,Dai:2014zta,Meissner:1990kz,Guo:2011pa,Kang:2013jaa,Chen:2015jgl,
Guo:2015zqa,Colangelo:2016jmc,Hanhart:2016pcd,Liu:2017vsf,Guo:2019twa,Danilkin:2018qfn,Dai:2019lmj}.

\subsection{Unitarization and resummation methods}\label{sec:2.2}
\subsubsection{$K$-matrix}
$K$-matrix is proposed by Wigner \cite{Wigner:1946fyr,Wigner:1947zz}, firstly applied to study the resonances appearing in nuclear reaction\cite{Wigner:1946fyr,Wigner:1947zz}. The single channel scattering partial wave should keep unitarity
\begin{equation}
\mathrm{Im}T^I_J(s)=\rho(s)|T^I_J(s)|^2\ , \label{eq:unit;inv;T}
\end{equation}
where $\sqrt{s}$ is the energy in the center of mass frame and $\rho(s)$ is the phase space factor defined in Eq.(\ref{eq;rho}). It is not hard to derive the so called inverse amplitude relation
\begin{equation}
\mathrm{Im}{T^I_J}^{-1}(s)=-\rho(s)\ .
\end{equation}
This also holds true for the multi-channel case where $T^I_J$ and $\rho$ become matrix forms. One could parameterize the single channel scattering amplitude as
\begin{eqnarray}
 T^I_J(s)=\frac{1}{M(s)-i\rho(s)}=\frac{K(s)}{1-i\rho(s) K(s)}\ ,\nonumber\\
 S^I_J(s)=\frac{1+i\rho(s) K(s)}{1-i\rho(s) K(s)}\ , \label{eq:KM;T}
\end{eqnarray}
where $K(s)$ and $M(s)=K^{-1}(s)$ are real in the physical region and may be represented by polynomials of $s$ and poles in many applications.
It can be generalized to the cases including FSI, together with Eq.~(\ref{eq:KM;T}) one has
\begin{equation}
\mathcal{F}^I_J(s)=F(s)(1-i\rho(s) K(s))^{-1}\, .\label{eq:KM;F}
\end{equation}
Here the $F(s)$ is real too. As done in \cite{Dai:2012pb}, one can input the tree amplitude $A^I_J(s)$ (multiplied by a polynomial and the $K(s)$) instead of the $F(s)$, then the $\mathcal{F}^I_J(s)$ amplitude changes into
\begin{equation}
\mathcal{F}^I_J(s)={A}^I_J(s)\alpha(s) T^I_J(s) \, .\label{eq:KM;F;1}
\end{equation}
The $\alpha(s)$ is a polynomial absorbing the l.h.c. and distant r.h.c.. Eq.~(\ref{eq:KM;F;1}) is nothing else but the one given by the AMP method, as shown in Eq.~(\ref{eq:FSI;AMP}).
It is not hard to find that it satisfies the Watson's theorem, which states that $\mathcal{F}^I_J(s)$ and $T^I_J(s)$ should have the same phase, as implied by Eq.~(\ref{eq:FSI}). In the low energy region ($s\to0$), $\mathcal{F}^I_J(s)$ should return to the tree amplitude, keeping chiral symmetry \cite{Dai:2012pb,Chen:2019mgp}.
All the equations given above can be generalized to coupled channel cases, and then the $S$-matrix is parameterized as
\begin{equation}
S^I_J(s)=[1+i\rho^{1/2}(s)K(s)\rho^{1/2}(s)][1-i\rho^{1/2}(s)K(s)\rho^{1/2}(s)]^{-1}\, ,
\end{equation}
where $\rho(s)=diag(\rho_1(s),\rho_2(s),\ldots \rho_n(s))$.
Here $K(s)$ is a real symmetric matrix, and the same for the $\mathcal{F}^I_J(s)$ amplitudes referring to the FSI.

In another formalism, the $K$-Matrix is rewritten as
\begin{equation}
T^I_J(s)=K(s)[1-C(s) K(s)]^{-1}\,,\label{Eq:KM;T}
\end{equation}
where $C(s)$ is the diagonal matrix of the canonical definition of Chew-Mandelstam functions \cite{Chew:1960iv,Edwards:1980sa}, and they can be represented by once subtracted dispersion relations:
\begin{eqnarray}
C_i(s)=\frac{s}{\pi}\int_{s_{thi}}^{\infty} ds'\frac{\rho_i(s')}{s'(s'-s)}.\label{eq:C}
\end{eqnarray}
Notice that the two-body phase space factor has only diagonal elements. The Chew-Mandelstam functions can be expressed explicitly as
\begin{eqnarray}
C_i(s)&=&\frac{1}{\pi}+\frac{M_i^2-m_i^2}{\pi s}\ln\left(\frac{m_i}{M_i}\right)-\frac{M_i^2+m_i^2}{\pi (M_i^2-m_i^2)}\ln\left(\frac{m_i}{M_i}\right)   \nonumber\\
&+&\frac{\rho_i(s)}{\pi}\ln\left(\frac{\sqrt{(M_i+m_i)^2-s}-\sqrt{(M_i-m_i)^2-s}}{\sqrt{(M_i+m_i)^2-s}+\sqrt{(M_i-m_i)^2-s}}\right)\,. \label{Eq:C;ana}
\end{eqnarray}
Since the unitary cut is embodied in once subtracted dispersion relation, the amplitude can be used to extract poles and residues in the complex $s$-plane near the real axis. This method is used to extract the pole locations of e.g. $f_0(980)$, $a_0(980)$ and $P_c(4312)$, $P_c(4440)$ in sections \ref{sec:f980}, \ref{sec:a980}  and \ref{sec:Pc}, respectively. Notice that these resonances are faraway from the elastic region so that the perturbative calculation of $\chi$EFT does not work, and other dispersive approaches, such as the PKU factorization and Roy equation, are difficult to be applied in lack of coupled channel unitarity, too.

\subsubsection{Pad\'e~approximation}
In the past decades several techniques have been developed to unitarize the $\chi$PT amplitudes\cite{Truong:1988zp,Gasser:1990bv,Dobado:1996ps,Pelaez:2003rv,Sun:2005uk}, named as unitarized chiral perturbation theory (U$\chi$PT).
Among them Pad\'e~approach is an intuitive way to derive the unitarization formalism \cite{Gasser:1990bv,Willenbrock:1990dq}, which constructs the amplitude from different order $\chi$PT results.
The inverse amplitude method (IAM) \cite{Truong:1991gv,Dobado:1989qm,Dobado:1996ps,Pelaez:2003rv} is somehow similar to this one and we will discuss it at the end of this section.
The on-shell Bethe-Salpeter (BS) equation \cite{Oller:1997ti} also belongs to the chiral unitary approach, but we will discuss it in the next section.
These U$\chi$PT approaches provide a systematic way to study the meson-meson scattering of the lightest octet pseudoscalars. Although they are not accurate enough to determine the precise pole locations of the resonances ($\sigma$, $\kappa$, $f_0(980)$, $a_0(980)$, $\rho$, $K^*(890)$ etc.) appearing as the intermediate states, it is impressive to see that U$\chi$PT can find out the poles of these resonances which are not written explicitly in the lagrangian of $\chi$PT. For details see discussions in section \ref{sec:lsm}.

One can use a Pad\'e  approximation function
\begin{eqnarray}
T^{[M,N]}(x)=\frac{Q_N (x)}{P_M (x)}, \nonumber
\end{eqnarray}
to make sure that its Taylor expansion is close to the original function when $x$ is small.
Here $Q_N (x)$ and $P_M (x)$ represent functions with the powers of the variables $x$ being N and M, respectively. Usually, the first $N+M$ terms of the Taylor expansion of the Pad\'e [N,M] approximation will be the same as that of the original function. Hence the Pad\'e approximation will restore perturbative unitarity of $\chi$PT with respect to perturbation expansion.  As a simple example, we give the Pad\'e approximation  \cite{Gasser:1990bv,Willenbrock:1990dq} for partial waves of $\pi\pi$ scattering \cite{Sun:2005uk}:
\begin{eqnarray}
T^{I~[1,1]}_{J}(s)&=&\frac{T^{I}_{J,2}(s)}{1-\frac{T^{I}_{J,4}(s)}{T^{I}_{J,2}(s)}}, \nonumber\\
T^{I~[1,2]}_{J}(s)&=&\frac{T^{I}_{J,2}(s)}{1-\frac{T^{I}_{J,4}(s)}{T^{I}_{J,2}(s)}
-\frac{T^{I}_{J,6}(s)}{T^{I}_{J,2}(s)}+\left(\frac{T^{I}_{J,4}(s)}{T^{I}_{J,2}(s)}\right)^2}. \label{eq:pade}
\end{eqnarray}
Here the $T$ amplitudes are calculated by $\chi$PT. $I$, $J$ and the number \lq 1, 2' after the comma represent the isospin, spin and the highest order of chiral expansions, respectively. When expanding the $\chi$PT amplitudes, one finds that
\begin{eqnarray}
\mathrm{Im}T^{I}_{J,2}&=&0 ,\nonumber\\
\mathrm{Im}T^{I}_{J,4}&=&\rho(s)(T^{I}_{J,2})^2  ,\nonumber\\
\mathrm{Im}T^{I}_{J,6}&=&2\rho(s)T^{I}_{J,2}~\mathrm{Re} T^{I}_{J,4}  . \label{eq:unit;chpt}
\end{eqnarray}
It is then easy to find that the amplitudes constructed by Pad\'e approximation satisfy unitarity:
\begin{eqnarray}
\mathrm{Im}T^{I[1,1]}_{J}(s)&=&\rho(s)|T^{I[1,1]}_{J}(s)|^2\ ,\nonumber\\
\mathrm{Im}T^{I[1,2]}_{J}(s)&=&\rho(s)|T^{I[1,2]}_{J}(s)|^2\ .
\end{eqnarray}
Meanwhile, these constructed amplitudes can restore the NLO (or NNLO) $\chi$PT results by perturbation expansion of small $s$, once the higher order terms are ignored.
The [1,1] and [1,2] Pad\'e approximants are helpful to study the $\pi\pi$ scattering \cite{Sun:2005uk} and also the $\pi\pi-K\bar{K}$ coupled channels \cite{Dai:2011bs}. However, it has several caveats: it violates crossing symmetry and spurious poles and cuts will be taken into the physical amplitudes unexpectedly, hence it violates the causality, as pointed out in Ref.\cite{Qin:2002hk}.

The IAM holds similar formulae but comes from different considerations. It is built through the fact that the imaginary part of the inverse amplitude in the physical region is exactly the phase space factor with a minus sign \cite{Dobado:1996ps}, see Eq.~(\ref{eq:unit;inv;T}). As an example, one could write the dispersion relations on the $T_2^2(s)/T(s)$ for single channel scattering, that is,
\begin{eqnarray}
G(s)=\frac{T_2^2(s)}{T(s)}=P_3(s)+\frac{s^3}{\pi}\int_L \frac{G(s')}{s'-s}+\frac{s^3}{\pi}\int_R \frac{G(s')}{s'-s}   \, .\nonumber
\end{eqnarray}
Here $P_3(s)$ is a third order subtraction polynomial.
Together with Eq.~(\ref{eq:unit;chpt}), one can derive Eq.~(\ref{eq:pade}) by assuming that the left hand cut of the $G(s)$ is the same as that of $-{\rm Im}_L T_4(s)$. This gives an estimation of the l.h.c.. The method could also be extended to higher orders. In Ref.~\cite{Guo:2011pa} an $N/D$ method is used to unitarize the $\chi$PT amplitudes, and it is classified to the U$\chi$PT, too.

\subsubsection{Lippmann-Schwinger and Bethe-Salpeter equations}
The decomposed partial wave amplitude, can be solved by the Lippmann-Schwinger (LS) equation, which is expressed as
\begin{eqnarray}\label{eq:LS}
T(z)&=&V(z)+ V(z)G_0(z)T(z)  \, ,\nonumber
\end{eqnarray}
where the equation is an integral equation, $G_0(z)$ is the Green function and $z$ is the energy dependence of the operator $T$.
In the view-point of QFT, this equation describes the resummation of all Feynmann diagrams.
The $T$ is solved by inputting a potential $V$, which can be given by perturbation calculation. For example in nucleon anti-nucleon interaction, the reaction amplitudes are obtained from the solution of a relativistic
LS equation \cite{Kang:2013uia,Dai:2017ont}:
{ \setlength{\mathindent}{-0.7cm}
\begin{eqnarray}
\indent T_{L''L'}(p'',p';E_k)&=&V_{L''L'}(p'',p') +\sum_{L}\int_0^\infty \frac{dpp^2}{(2\pi)^3} \, V_{L''L}(p'',p)
\frac{1}{2E_k-2E_p+i0^+}T_{LL'}(p,p';E_k). \nonumber\\
\label{LS}
\end{eqnarray} }
Here, the subscript $L$, $L'$, and $L''$ are the orbital angular momenta that can be used to represent different channels of the coupled partial waves.
$E_k=\sqrt{m^2+k^2}$, where $k$ is the on-shell momentum and $V$ is the potential calculated by $\chi$EFT.
This method can be extended to the production process, with the distorted-wave Born approximation (DWBA) applied \cite{Dai:2018tlc}
\begin{eqnarray}
A(z) &=&  A^0(z) + A^0 G_0 (z) T(z)\,,  \label{eq:LSA}
\end{eqnarray}
where $A_0(z)$ is the production amplitude and $T(z)$ is the hadronic scattering amplitude of the final states.

The Bethe-Salpeter~(BS) equation is similar to LS equation, but in BS equation one uses the four-momentum to do the integration, while in LS equation one uses the three-momentum and energy to do the integral.
A typical BS integration equation reads
{ \setlength{\mathindent}{0.2cm}
\begin{eqnarray}
T_{jk}(p_1,p_2;q)&=&V_{jk}(p_1,p_2;q)+ \sum_{l}\int \frac{d^4q}{(2\pi)^4} \,
\frac{V_{jl}(p_1,p_2;q)}{q^2-m_{1l}^2+i0^+}\frac{T_{lk}(p'_1,p'_2;q)}{(p_1+p_2-q)^2-m_{2l}^2+i0^+}\,.\nonumber\\
 \label{eq:BS;1}
\end{eqnarray} }
Here \lq $j$, $k$, $l$' represent different channels and \lq 1, 2' are for the incoming two particles.
Examples of the application of LS and BS equations on the hadron physics  can be found in Refs.\cite{Dai:2017fwx,Wang:2017smo,Pavao:2018xub,Liu:2019rpm,Haidenbauer:2020wyp}, etc..

However, the LS and BS equations are difficult to be solved. For simplicity one may use \lq on-shell' approximation \cite{Oller:1997ti} to be relieved from the non-trivial numerical calculation of the integral equation. In that case the potential $V$ and scattering amplitude $T$ can be extracted out from the integral and one has
\begin{eqnarray}
T(s)&=&\frac{V(s)}{1-V(s)G(s)}\,, \label{eq:BS;OS}
\label{BS}
\end{eqnarray}
where the $G(s)$ function is diagonal:
\begin{eqnarray}
G_{ll}(s)&=&i \int \frac{d^4q}{(2\pi)^4} \,
\frac{1}{q^2-m_{1l}^2+i0^+}\frac{1}{(p_1+p_2-q)^2-m_{2l}^2+i0^+}\,. \label{eq:BS;G}
\label{BS}
\end{eqnarray}
Note that it is divergent and needs to be regularized.
To deal with the integral, the cut-off scheme would cause some severe problems:  the amplitude can not be extended above (or even close to) the cut-off momentum, { and there is a singularity rising from the possibility of the denominator in the integrand being zero.}
The singularity will disappear in the limit that the cut-off runs into infinity. To avoid such problems one could integrate it out in the dimensional regularization method. One has \cite{Guo:2005wp}
{ \setlength{\mathindent}{0.2cm}
\begin{eqnarray}
G_{ll}(s)&=&\frac{1}{16\pi^2}\left\{ \frac{i\rho(s)}{2}\left[\;\ln[s-\Delta+is\rho(s)]+\ln[s+\Delta+is\rho(s)]-\ln[-s+\Delta+is\rho(s)] \right.    \right. \nonumber\\
&&\left.\left.-\ln[-s-\Delta+is\rho(s)]  \;\right]
+a(\mu)+\ln\frac{m_{1l}^2}{\mu^2}+\frac{\Delta-s}{2s}\ln\frac{m_{1l}^2}{m_{2l}^2} \right\} \,.
 \label{eq:BS;G;ana}
\end{eqnarray} }
Here $\rho(s)$ is the phase factor as defined before, $a(\mu)$ is the subtraction constant with $\mu$ the renormalization scale, $\Delta=m_{1l}^2-m_{2l}^2$. For earlier discussions about study of $G(s)$ in cut-off and/or dimensional regularization, we refer to Refs.\cite{Oller:1998hw,Meissner:1999vr} and references therein.
The merit of such an equation is that they are based on QFT and thus are less model dependent. It keeps unitarity (also in coupled channels), contains less parameters and could restore the perturbative calculations in the low energy region.  However, in the point view of \lq resummation', it only sums over the bubble chains of the $s$-channel and thus the crossing symmetry is not ensured. The choice of the subtraction constant is model dependent, too.
The \lq on-shell' BS equations are firstly used to study the light meson scattering, and now it is widely applied to various kinds of hadrons, such as the $\Lambda(1405)$, the $D^*_{s0}(2317)$ and $D^*_0(2400)$. See sections \ref{sec:1405} and \ref{sec:D2317} for further discussions.

\subsubsection{Relativistic Friedrichs-Lee scheme}

The Friedrichs-Lee model is a solvable model developed to understand the resonance phenomena when the bare states are coupled to the continuum states~\cite{Friedrichs:1948,Lee:1954iq}. {The simplest form of the model includes a free Hamiltonian
$H_0$ which has a simple continuous spectrum with a range of $[0,\infty)$, plus
an discrete eigenvalue $\omega_0$ embedded in the continuous
spectrum~($\omega_0>0$).}  If an interaction $V$ between the continuous
and discrete parts is introduced,  the discrete state of $H_0$ is
dissolved in the continuous spectrum and resonances are generated.

One could denote  the discrete state of $H_0$ by $|1\rangle$ and the continuous state by the $|\omega\rangle$, so
the free Hamiltonian is written down as
\begin{eqnarray}
H_0=\omega_0|1\rangle\langle 1|+\int_0^\infty \omega|\omega\rangle\langle\omega |\mathrm{d}\omega,
\end{eqnarray}
and the interaction $V$ is expressed as
\begin{eqnarray}
V=\lambda\int_0^\infty [f(\omega)|\omega\rangle\langle 1 |+f^*(\omega)|1\rangle\langle \omega |]\mathrm{d}\omega,
\end{eqnarray}
where $f(\omega)$ denotes the non-trivial coupling form factor governing the interaction.
Provided $|1\rangle$ and $|\omega\rangle$ form an orthogonal complete set, the normalizations are
\begin{eqnarray}
\langle 1|1\rangle=1,\langle 1|\omega\rangle=\langle \omega|1\rangle=0, \langle\omega|\omega'\rangle=\langle\omega'|\omega\rangle=\delta(\omega-\omega').
\end{eqnarray}
The eigenfunction of $H=H_0+V$ for an arbitrary real eigenvalue $x$ can be written down as
\begin{eqnarray}\label{eigenf}
H\Psi(x)=x\Psi(x).
\end{eqnarray}
Since $|1\rangle$ and $|\omega\rangle$ form a complete set, the eigen-wavefunction can be expressed as
\begin{eqnarray}
\Psi(x)=\alpha(x)|1\rangle+\int_0^\infty\psi(x,\omega)|\omega\rangle \mathrm{d}\omega.
\label{eq:EigenV}\end{eqnarray}
This model is exactly soluble and the in-state and out-state solutions can be obtained by standard derivations and written down as
\begin{align}
|\Psi_\pm(x)\rangle=&|x\rangle+\lambda\frac{f(x)}{\eta^\pm(x)}\Big[|1\rangle+\lambda\int_0^\infty\mathrm{d}\omega\frac{f(\omega)}{x-\omega\pm
i \epsilon}|\omega\rangle\Big]
\,,
\end{align}
where the inverse of the resolvent function
\begin{eqnarray}
\eta^{\pm}(x)=x-\omega_0-\lambda^2\int_0^\infty\frac{f(\omega)f^*(\omega)}{x-\omega\pm
i \epsilon}\mathrm{d}\omega .
\end{eqnarray}
$\eta^{\pm}(x)$ can be analytically continued to the complex energy plane and
are the upper edge and lower edge of an analytic function $\eta(x)$ on
the cut from $0$ to $\infty$ on the real axis. This resolvent function is similar to a non-subtracted dispersion relation in the non-relativistic form.  The $S$-matrix of one continuum state can also be obtained by inner product of the in-states
and the out-states as
\begin{align}\label{smatrixI}
S(E;E')=\delta(E-E')\Big(1-2\pi
i\frac{f(E)f^*(E)}{\eta_+(E)}\Big)\,.
\end{align}
That means the inverse of the resolvent function $\eta(x)$ plays an important role in describing the scattering processes which will be addressed later.

Actually, the Hamiltonian eigenfunction can be generalized in the
rigged Hilbert space~(RHS) to have a complex eigenvalue of $H$ with
eigenvectors not belonging to the conventional Hilbert space~\cite{Gadella:2004}. That means, the state satisfying
\begin{eqnarray}\label{generaleigenf}
H\Psi(z)=z\Psi(z)
\end{eqnarray}
with $z$ being a complex value can be well defined in RHS and used
to describe the resonance, i.e. Gamow
states~\cite{Bohm:1989,Gadella:2004}. By analyzing the model carefully, it is found that the
complex eigenvalues of Eq.~(\ref{generaleigenf}) are just the poles of
the scattering amplitude, which corresponds to the zero point of
$\eta(z)$ function satisfying
\begin{eqnarray}
z-\omega_0-\lambda^2\int_0^\infty\frac{f(\omega)f^*(\omega)}{z-\omega}\mathrm{d}\omega=0.
\end{eqnarray}
From the cut starting from
threshold to the infinity, the $\eta(z)$ function can be
analytically continued to the second Riemann sheet (RS). The zeros of
$\eta(z)$~(poles of the resolvent function) could only be found on
RS-II or the real axis below the threshold, which
represent resonances, virtual states, or bound states of the
scattering processes.

Actually, besides the poles originating from the bare state at
$\omega_0$, other dynamical poles could appear. Such kinds of
dynamical poles could also be described by the wavefunctions similar to the resonance shifted from the bare state.
In Ref.~\cite{Xiao:2016dsx},
it is shown that some virtual state or resonance poles will be
generated in different kinds of interaction form factors, which implies
that the nontrivial interactions will determine the properties of the
extra states.

To study the light hadrons, the scheme should be extended to a relativistic form, because in principle the light
constituent quarks involved move fast comparable to the light speed~\cite{zhou:2020}.
An important
difference from the non-relativistic scenario is that the creation and
annihilation operators are involved.
The eigenvalue problem is equivalent to
finding the solution of
\begin{eqnarray}
[P_\mu,b^\dag(E,\mathbf{p})]=p_\mu b^\dag(E,\mathbf{p}),
\label{eq:eigeneq}
\end{eqnarray}
where $b^\dag(E,\mathbf{p})$ has both contributions from the operators of the discrete and continuum states.
By a direct calculation of the commutation relation, the continuum in-state and out-state creation operator can be written down explicitly~\cite{zhou:2020}. The resolvent function $\eta_\pm(E,\mathbf{p})$ is expressed as
\begin{eqnarray}
\eta_\pm(E,\mathbf{p})=E^2-\omega(\mathbf{p})^2-\int
\mathrm{d}E'^2[\frac{2\omega(\mathbf{p})\beta(E')\alpha(k(E',\mathbf
p))^2}{(E^2-E'^2\pm i0)}],
\end{eqnarray}
where $\beta(E^\prime)$ denotes an integration measure and $\alpha$ is the coupling vertex between the discrete state and the continuum state. It is obviously similar to $\eta_\pm(E)$ in the non-relativistic Friedrichs model~\cite{Xiao:2016mon,Zhou:2017txt}, and it has a right hand cut
starting from threshold energy squared of the two-particle continuum.
In the c.m. frame, $\mathbf{p}=\mathbf{0}$, the variable is changed to
the invariant mass $W$ of two-particle system, or more concisely to $s=W^2$, so that the  $\eta$ function can be written down as
\begin{eqnarray}\label{DR:s}
\eta_\pm(s)=s-\omega_0^2-\int_{s_{th}} ds' \frac{\rho(s')}{s-s'\pm i0},
\end{eqnarray}
where $s_{th}=(\mu_1+\mu_2)^2$ and the spectral function
$\rho(s)=2\omega_0\frac{k(s)\varepsilon_1(s)\varepsilon_2(s)}{\sqrt{s}}\alpha(k(s))^2$
in which the coupling form factor $\alpha(k)$ can be obtained by using
the relativistic quark pair creation~(RQPC) model~\cite{Fuda:2012xd,zhou:2020} or some other
relativistic models for consistency.  The $\eta$
function is Lorentz invariant and just similar to the relativistic
dispersion relation.
Its main difference from the non-relativistic case is that the relation is in terms of  the energy squared $s$ instead of the energy $E$.
The $S$-matrix of one continuum state can also be obtained by inner product of the in-states
and the out-states,
\begin{eqnarray}
S(E,\mathbf p;E',\mathbf p')=\delta^{(3)}(\mathbf p-\mathbf p')\delta(E-E')\Big(1-2\pi
i\frac{\rho(s)}{\eta_+(s)}\Big)\,.
\end{eqnarray}
In general, since there is only one continuum state here, i.e. one
unitarity cut,  every
bare discrete state will be shifted to become two conjugate poles on RS-II representing a resonance or remaining on the real axis
being virtual or bound state poles. When the coupling is tuned down,
these poles will move to the bare position of the discrete state. These two ``bare" poles contribute only one resonance or one state.

In addition, there could also be dynamically generated poles which do not
move to the bare states and normally run towards the
singularities of the form factor when the coupling is switched
off. These are dynamically generated by the interaction between the
discrete state and the continuum. The properties of the extra
``dynamical" poles depend on the concrete form of the non-trivial
interaction form factor. These ``dynamical" poles behave as
another state perhaps shown as another lineshape peak in the cross section for
example.

The Friedrichs-Lee model can be generalized to include more discrete bare
states and continuum bare states. With more discrete bare states, the $\eta$
function will become a matrix whose dimension is equal to the
number of the discrete bare states. With more continuum bare states, more
dispersion integrals will be added in the $\eta$ functions, with each
integral corresponding to a continuum threshold.
Although it is a dynamical model without complying with crossing symmetry, unitarity of the scattering amplitude is implemented from the beginning. Furthermore, analyticity is also respected and easily analyzed. {In its recent applications on the light scalar meson states as $\sigma, \kappa, a_0(980)$ and  $f_0(980)$~\cite{Zhou:2020pyo},  the scheme presents a general consistent picture, and its results can be crosschecked with other dispersive approaches.  See section \ref{sec:lsm} for details.}

\subsection{Dispersive approaches} \label{sec:2.3}
\subsubsection{$N/D$ method}
The $N/D$ method is derived by Ref.~\cite{Chew:1960iv} to restore the unitarity of the $\pi\pi$ partial wave amplitudes.
For a single channel scattering, one has
\begin{equation}
T(s)=N(s)/D(s)\, .
\end{equation}
Here the $D(s)$ function contains the r.h.c. in the physical region and the Castillejo-Dalitz-Dyson (CDD) poles \cite{Castillejo:1955ed} below the threshold, otherwise it is analytical in the whole complex $s$-plane. Excluding possible bound states, the $N(s)$ function is analytical in the whole complex-$s$ plane except on the l.h.c.. From unitarity and anlyticity, one finds
\begin{eqnarray}
N(s)&=&\frac{1}{\pi}\int_{-\infty}^{0}\frac{{\rm Im}_L T(s')D(s')}{s'-s}d s' ,\label{eq:ND;1}\\
D(s)&=&1-\frac{1}{\pi}\int_{s_{th}}^{\infty}\frac{\rho(s')N(s')}{s'-s}d s' . \label{eq:ND;2}
\end{eqnarray}
Inserting Eq.~(\ref{eq:ND;1}) into Eq.~(\ref{eq:ND;2}), one can rewrite it as
\begin{eqnarray}
D(s)&=&1-\frac{1}{\pi^2}\int_L d s' {\rm Im}_L T(s') D(s') G(s,s')   \,,\label{eq:ND}
\end{eqnarray}
where $G(s,s')$ is given as
{ \setlength{\mathindent}{0.5cm}
\begin{eqnarray}
G(s,s')&=& \frac{1}{(s'-s)s'}\left[ -(M^2-m^2)\ln\left(\frac{m}{M}\right)+\sqrt{[(M-m)^2-s'][(M+m)^2-s']} \right. \nonumber \\
&&~~~~~~~~~~\left.\ln\left(\frac{\sqrt{(M+m)^2-s'}-\sqrt{(M-m)^2-s'}}{\sqrt{(M+m)^2-s'}+\sqrt{(M-m)^2-s'}}\right) \right] +\left\{s\leftrightarrow s'\right\}   \,. \label{eq:ND;G;}
\end{eqnarray} }
Note that \lq m, M' represent the two unequal masses of the particles in the scatterings. The  $G(s,s')$ function is reduced to a simpler form for two equal mass (m) particles
\begin{eqnarray}
G(s,s')&=& \frac{\rho(s')}{s'-s}\ln\left(\frac{\rho(s')-1}{\rho(s')+1}\right)
+\left\{s\leftrightarrow s'\right\}   \,,\label{eq:ND;Gmm;}
\end{eqnarray}
with $\rho(s')=\sqrt{1-4m^2/s'}$.

\indent In some analyses, one can use the Omn$\grave{e}$s function instead of the inverse of the $D(s)$ function. Hence one can write the dispersion relation as $T(s)=P(s)\Omega(s)$ \cite{Dai:2014lza,Dai:2014zta}, where $P(s)$ contains l.h.c. too,
to be discussed in next sections. The $N/D$ method has also been used in the context of $\chi$PT already in, for instance, Refs.\cite{Meissner:1999vr,Oller:1998zr}. Indeed the l.h.c. has been ignored in these works and they are classified as U$\chi$PT in this review. See sections \ref{sec:f980} and \ref{sec:a980}. Nevertheless, they are quite helpful to study the resonances (such as $f_0(980)$ and $a_0(980)$) for lack of effective dispersive approaches to deal with the l.h.c..
Also one can use $\chi$EFT to constrain the l.h.c.. With the constraint on l.h.c., the $D$ function can be solved once there are enough data given in the physical region, following Eq.~(\ref{eq:ND}).
One may also need to include the CDD poles \cite{Castillejo:1955ed} in the $D$ function,
\begin{eqnarray}
D(s)&=&\sum_i\frac{g_i}{s-s_i}+1-\frac{1}{\pi}\int_{s_{th}}^{\infty}\frac{\rho(s')N(s')}{s'-s}d s' ,
\end{eqnarray}
These CDD poles correspond to zeros in the $T$ amplitude. It is difficult to describe the $\pi\pi$ scattering amplitude in the low energy region in lack of the CDD poles \cite{Castillejo:1955ed}. Indeed, the Adler zeros of the S-waves can be written in terms of the CDD poles in the $D(s)$ function.
For the vertex function $\mathcal{F}(s)$, one has the same formalism but the $N(s)$ function is different.
\begin{equation}
\mathcal{F}(s)=f(s)/D(s)\, .
\end{equation}
The $N/D$ formalism can be generalized to the coupled channel case. The coupled unitarity requires that the scattering amplitude $T_{ij}$ and the vertex function $F_j$ satisfy
\begin{eqnarray}
\frac{1}{2i}[T_{ij}(s+i\epsilon)-T_{ij}(s-i\epsilon)]&=&\sum_k \rho_k(s)T_{ik}(s-i\epsilon)T_{kj}(s+i\epsilon)\, ,\nonumber\\
\frac{1}{2i}[F_j(s+i\epsilon)-F_j(s-i\epsilon)]&=&\sum_k \rho_k(s)F_k(s-i\epsilon)T_{kj}(s+i\epsilon)\, .
\end{eqnarray}
One can define the $d_{ij}$ as the ($i$-th,$j$-th) component of the $D$ matrix and get
\begin{equation}
d_{ij}=\delta_{ij}-\frac{1}{\pi}\int_{s_{th_i}}^{\infty}ds'\frac{\rho_i(s')N_{ij}(s')}{s'-s}\, .
\end{equation}
The $d_{ij}$ function is analytical except for the cut starting from threshold $s_{th_i}$, and one has
\begin{equation}
T_{ij}=\sum_k\frac{N_{ik}(s)D_{(jk)}}{D}\, ,
\end{equation}
with $D=\det(d_{ij})$ and $\delta_{ij}D=\sum_k d_{ik}D_{(jk)}$.  $D_{(jk)}$ is the cofactor of the D matrix. The vertex function can be represented as
\begin{equation}
\mathcal{F}_j(s)=\sum_k\frac{f_k D_{(jk)}}{D}\, ,
\end{equation}
{with $f_k$ an analytical function \cite{Bjorken:1960zz}. }

\subsubsection{Roy and Roy-Steiner equations}
Roy equation proposed by \cite{Roy:1971tc} is built from fixed-$t$ dispersion relation on the scattering amplitude, including the crossing relation which is used to connect the amplitudes in $s-u$ crossing. It should be noted that a similar representation of $\pi N$ scattering based on fixed-$t$ dispersion relations were indeed formulated by Refs.\cite{Baacke:1970mi,Steiner:1970mh,Steiner:1971ms}, too. The merit of this method is that the unknown l.h.c. can be expressed by the r.h.c. in the physical region of partial waves. Single channel unitarity is also well imposed by keeping the real part of the partial wave amplitudes the same as what is calculated by the phase shift directly. Notice that in elastic region, the whole amplitude is known due to unitarity once the phase shift is given.
After the pioneering paper \cite{Roy:1971tc}, it has been soon applied to the $\pi\pi$ scattering \cite{Basdevant:1972uv,Pennington:1973hs,Epele:1977un,Pomponiu:1975bi}. In Ref. \cite{Pennington:1973hs} it is used to check the iso-scalar S-wave phase shift near the $\bar{K}K$ threshold and is helpful to solve the up-down ambiguity. In Ref.\cite{Ananthanarayan:1996gj}, a Roy equation analysis on $\pi\pi$ phase shift is performed and the LECs $\bar{l}_1, \bar{l}_2$ of $SU(2)\times SU(2)$ $\chi$PT are extracted.

In the original paper of Roy \cite{Roy:1971tc}, combining analyticity and crossing symmetry, it builds a set of non-linear integral equations for the partial wave amplitudes. The $\pi\pi$ scattering amplitude $T^I(s,t)$  can be expressed as
{ \setlength{\mathindent}{0.5cm}
\begin{eqnarray}\label{Roy}
\vec{T}(s,t)&=&g_1(s,t)\vec{a}+\int_{4m_\pi^2}^{\infty}ds'[g_2(s,t,s'){\rm Im}\vec{T}(s',0)+g_3(s,t,s'){\rm Im}\vec{T}(s',t)]   \, , \label{eq:Roy;F}
\end{eqnarray}
}
where the \lq vector' represents the matrix form of isospin amplitudes $T^I(s,t)$: $(T^0(s,t),T^1(s,t),T^2(s,t))^{\rm t}$. Similarly $\vec{a}$ is the matrix form of scattering lengths,  and the $g_i$ functions are written in the matrix form as  \cite{Roy:1971tc}:
{ \setlength{\mathindent}{0.cm}
\begin{eqnarray}
g_1(s,t)&=&s(1-C_{su})+t(C_{st}-C_{su})+4C_{su}\,,\nonumber\\
g_2(s,t,s')&=&C_{st}\left(\frac{1+C_{tu}}{2}+\frac{(2s+t-4m_\pi^2)(1-C_{tu})}{2(t-4m_\pi^2)}\right)\nonumber\\
&&\frac{1}{\pi s'^2}\left[\frac{t^2}{s'-t}+\frac{(4m_\pi^2-t)^2 C_{su}}{s'-4m_\pi^2+t}-\frac{4t m_\pi^2+4m_\pi^2(4m_\pi^2-t)C_{su} }{s'-4m_\pi^2}\right]\,,\nonumber\\
g_3(s,t,s')&=&\frac{1}{\pi s'^2}\left[\frac{s^2}{s'-s}+\frac{u^2 C_{su}}{s'-u}-\frac{(4m_\pi^2-t)^2 }{s'-4m_\pi^2+t}\left( \frac{1+C_{su}}{2}+\frac{(2s+t-4m_\pi^2)(C_{su}-1) }{2(t-4m_\pi^2)}\right)\right]\,.\nonumber\\  \label{eq:Roy;g}
\end{eqnarray} }
Here the crossing matrix are given as:
\begin{eqnarray}
&&C_{st}=C_{ts}= \left(
\begin{array}{c c c}
 \displaystyle 1/3 &  ~\displaystyle 1    &~ \displaystyle 5/3 \\
 \displaystyle 1/3 &  ~\displaystyle 1/2  &~ \displaystyle-5/6 \\
 \displaystyle 1/3 &  ~\displaystyle -1/2 &~ \displaystyle 1/6
\end{array}
\right)\, , \label{eq:Roy;Cst} \\
&&C_{su}=C_{us}= \left(
\begin{array}{c c c}
 \displaystyle 1/3  &~  \displaystyle -1  &~ \displaystyle 5/3 \\
 \displaystyle -1/3 &~  \displaystyle 1/2 &~ \displaystyle 5/6 \\
 \displaystyle 1/3  &~  \displaystyle 1/2 &~ \displaystyle 1/6
\end{array}
\right)\, . \label{eq:Roy;Csu}
\end{eqnarray}
They are involved in the crossing relations relating the $s$- and $t$-channel $\pi\pi$ scattering amplitudes
\begin{eqnarray}
T^{I_t}(s,t)=\sum_{I'_s=0}^{2} C_{st}^{I_t I'_s}T^{I'_s}(s,t)\,.  \label{eq:Roy;CR}
\end{eqnarray}
The absorptive part of the amplitudes can be expressed by the partial wave amplitudes
\begin{eqnarray}\label{Roy:F}
{\rm Im}\vec{T}(s,t)&=&64\pi \sum_{l=0}^{\infty}(2l+1){\rm Im}\vec{T}_l(s)P_l\left(1+\frac{2t}{s-4m_\pi^2} \right)  \, . \label{eq:Roy;t}
\end{eqnarray}
Note that from Eq.~(\ref{eq:Roy;t}) there should be an infinite set of partial waves, but in practice, it is convenient to keep only the S- and P-waves for the $\pi\pi$ scattering, and the higher partial waves are much smaller and can be ignored. In Ref.\cite{Kaminski:2011vj}, some higher partial waves, such as the D- and F-waves, are included in the Roy equations, but with a bit more complicated formalism.
It should also be noted that less subtractions (once-subtracted dispersion relation) have been used by \cite{GarciaMartin:2011cn}. It is also called GKPY equations sometimes in the literature.\footnote{Note that the limits of the number of subtractions have been discussed in Ref.\cite{Jin:1964zza}, where it is pointed out that the number of subtractions in fixed-t dispersion relation should not be larger than 2, with the requirement of Froissart bound.}.
For the subtraction constants, one could use $\chi$PT to constrain them. On the other hand, the low energy amplitudes of $\chi$PT can be represented by polynomials of Mandelstam variables $s,t,u$, exactly the same form as the subtraction constants in the dispersion relation. With this constraint, the accurate pole location of $\sigma$ is obtained by \cite{Caprini:2005zr}. If there is only one subtraction constant, there will be less constraint from the low energy $\chi$PT amplitudes but the experimental data will compensate for determining the pole locations. What is more, the NA48/2 data \cite{Batley:2010zza} are much more precise than the previous data, hence further study on pole locations of the $\sigma$ by Roy equations \cite{Moussallam:2011zg} and by Roy-like equations \cite{GarciaMartin:2011cn} are available.

For unequal mass case, things are more complicated as the $t$-channel physical cut is different from the $s$-channel one and the crossing relations would give less constraints due to the loss of some symmetries of identical particles.  The Roy-Steiner equation \cite{Roy:1971tc,Steiner:1971ms} is proposed and has been applied to many scattering processes with unequal mass particles, e.g. $\pi K\to\pi K$ \cite{Buettiker:2003pp,DescotesGenon:2006uk}, $\pi N\to\pi N$ \cite{Hoferichter:2015hva} and $\gamma\gamma\to\pi\pi$ \cite{Hoferichter:2011wk}, etc..  These equations work efficiently in elastic region.
There are different ways to write the Roy-Steiner equation.
As an example, in $\pi K$ scatterings, the $\kappa$ resonance pole is quite far away from the real axis of the $s$-plane. Thus it would be more convenient to choose the fixed-$us$ (or the so called hyperbolic) dispersion relation rather than the fixed-$t$ one, where the domain of the former is larger in the direction of \lq${\rm Im}s$' and thus it makes sure that the pole in the RS-II locates in the working domain \cite{DescotesGenon:2006uk}.
The $\pi K$ scattering amplitude can be redefined as \cite{Johannesson:1976qp}
\begin{eqnarray}
F^+(s,t)&=&\frac{1}{3}F^{I_s=1/2}(s,t)+\frac{2}{3}F^{I_s=3/2}(s,t)=\frac{1}{\sqrt{6}}F^{I_t=0}(s,t)\,, \nonumber\\
F^-(s,t)&=&\frac{1}{3}F^{I_s=1/2}(s,t)-\frac{1}{3}F^{I_s=3/2}(s,t)=\frac{1}{2}F^{I_t=1}(s,t)\,, \label{eq:RS;F}
\end{eqnarray}
and the partial wave expansion in the s-channel is given as
\begin{eqnarray}
F^{I_s}(s,t)=16\pi\sum_l (2l+1)P_l(z_s) f^{I_s}_l(s) \,. \label{eq:RS;iso}
\end{eqnarray}
With $s-t$ crossing, the partial waves of $\pi\pi\to \bar{K}K$ scattering amplitudes are
\begin{eqnarray}
F^{I_t}(s,t)=16\pi\sqrt{2}\sum_{l+I_t~even} (2l+1)[q_\pi(t) q_K(t)]^lP_l(z_t) g^{I_t}_l(t) \,. \label{eq:RS;iso;t}
\end{eqnarray}
To give an impression about the approach, we list some of the
Roy-Steiner equations of the $\pi K$ scattering in the hyperbolic form. They are given as \cite{Buettiker:2003pp}:
{ \setlength{\mathindent}{0.cm} \begin{eqnarray}
&&{\rm Re} f_l^{1\over2}(s)= k^{1\over2}_l(s) +{1\over\pi}\int_{4m^2_\pi}^{\infty} dt' \Big\{ K^0_{l0}(s,t') {\rm Im} g_0^0(t')+2 K^1_{l1}(s,t') {\rm Im}g_1^1(t')\Big\}+d^{1\over2}_l(s) \nonumber\\
&&~~ +{1\over\pi}~{\rm P.V.}\int_{m^2_+}^{\infty} ds'\,\sum_{l'=0,1}
\Bigg\{ \left(\delta_{ll'}{\lambda_{s}\over(s'-s)\lambda_{s'}}-{1\over3} K^\alpha_{ll'}(s,s')\right){\rm Im}\, f_{l'}^{1\over2}(s')
+{4\over3} K^\alpha_{ll'}(s,s') {\rm Im}\, f_{l'}^{3\over2}(s') \Bigg\} \,,\nonumber\\
&&{\rm Re} f_l^{3\over2}(s)= k^{3\over2}_l(s)+{1\over\pi}\int_{4m^2_\pi}^{\infty} dt' \Big\{ K^0_{l0}(s,t') {\rm Im}\, g_0^0(t')
- K^1_{l1}(s,t') {\rm Im}\, g_1^1(t')\Big\} + d^{3\over2}_l(s)\nonumber\\
&&~~ +{1\over\pi}~{\rm P.V.}\int_{m^2_+}^{\infty} ds'\,\sum_{l'=0,1}\Bigg\{ \left(\delta_{ll'}{\lambda_{s}\over(s'-s)\lambda_{s'}} +{1\over3} K^\alpha_{ll'}(s,s')\right) {\rm Im}\, f_{l'}^{3\over2}(s')+{2\over3} K^\alpha_{ll'}(s,s') {\rm Im}\, f_{l'}^{1\over2}(s') \Bigg\}\,,
\nonumber \label{eq:roy;f}
\end{eqnarray}}
and $g^I_l(t)$ are those of the $\pi \pi\to K\bar{K}$ scattering partial waves in the hyperbolic form.
Here only S- and P-waves are taken into account. The r.h.c. can be obtained by fitting to the experimental data of phase shift and inelasticity,
then by a fitting procedure to make sure the amplitude calculated in the left side of the equality reproduces the same r.h.c., and thus unitarity is implemented.

Roy and Roy-like equations give accurate pole locations of the lightest resonances, such as $\sigma$, $\rho$, $\kappa$ and so on, see e.g.~\cite{ Caprini:2005zr,DescotesGenon:2006uk,Buettiker:2003pp,Ananthanarayan:2000ht,GarciaMartin:2011jx}.
For details about how Roy and Roy-Steiner equations determined the pole locations of the $\sigma$ and $\kappa$, we refer to sections \ref{sec:sigma} and \ref{sec:kappa}, respectively.
Besides, the threshold parameters such as scattering length and slope parameter, which can be expressed as subtraction constants, can also be well extracted by Roy and Roy-like equations.

\subsubsection{PKU factorization }
A production representation, being called PKU factorization, is proposed by \cite{Zheng:2003rw,Xiao:2000kx,Zhou:2004ms}. It is based on dispersion relation and a factorization on the $S$ matrix in single channel scatterings.
The S matrix is represented as
\begin{equation}
S(s)=\tilde{F}(s)+i\rho(s)F(s) \,, \label{eq:pku,S}
\end{equation}
where
\begin{eqnarray}
F(s)=\frac{1}{2i \rho(s)}(S(s)-\frac{1}{S(s)}) \,, \, \tilde{F}(s)=\frac{1}{2}(S(s)+\frac{1}{S(s)}) \,, \label{eq:pku,F}
\end{eqnarray}
and
\begin{eqnarray}
\rho(s)F(s)=\sin 2\delta \, , \quad \tilde{F}(s)=\cos 2\delta \,.  \label{eq:pku,cos}
\end{eqnarray}
The $F(s)$ and $\tilde{F}(s)$ can be written in dispersion relations as
\begin{eqnarray}
F(s)&=&\alpha+\sum{\rm poles}+\frac{1}{\pi}\int_L \frac{{\rm Im}_L F(s')d s'}{s'-s} +\frac{1}{\pi}\int_R \frac{{\rm Im}_R F(s')d s'}{s'-s} \,,  \label{eq:pku,F}\nonumber\\
\tilde{F}(s)&=&\tilde{\alpha}+\sum_i\frac{\beta_i}{2(s-s_i)}+\sum_j\frac{1}{2S'(z_j^{II})(s-z_j^{II})} \nonumber\\
&&+\frac{1}{\pi}\int_L \frac{{\rm Im}_L \tilde{F}(s')d s'}{s'-s} +\frac{1}{\pi}\int_R \frac{{\rm Im}_R \tilde{F}(s')d s'}{s'-s} \,,  \label{eq:pku,Fh}
\end{eqnarray}
which clearly separate the contributions of poles and the cuts, including the l.h.c. and inelastic r.h.c.. It is worth pointing out that the inelastic r.h.c starts from the inelastic threshold since the elastic cut disappears by construction.
It is not difficult to find that the S-matrix can be written as
\begin{eqnarray}
S(s)&=&\cos[2\delta(s)]+i\sin[2\delta(s)]\,,\label{eq:pku,S;cos}
\end{eqnarray}
and it satisfies
\begin{eqnarray}
\cos^2[2\delta(s)]+\sin^2[2\delta(s)]\equiv 1\,,\label{eq:pku,unit}
\end{eqnarray}
which is called the generalized unitarity relation and holds on the entire complex $s$ plane.
This is a strong constraint for the partial wave amplitudes on the complex-$s$ plane {and it serves to find the simple solutions of $S$-matrix that contain either only one pole or zero on the real
axis, or a pair of conjugated poles on the second sheet of the complex $s$ plane.}
The poles can be classified as bound states below threshold on the RS-I, virtual states and resonances on the RS-II. They can be respectively parameterized as:
\begin{eqnarray}\label{eq:PKUbvr}
S^b(s)&=&\frac{1-i \rho(s)\mid a \mid}{1+i \rho(s)\mid a \mid}\,,\nonumber\\[2.5mm]
S^v(s)&=&\frac{1+i \rho(s)\mid a \mid}{1-i \rho(s)\mid a \mid}\,,\nonumber\\[2.5mm]
S^R(s)&=&\frac{M^2[z_0]-s+i\rho(s)s~ G[z_0]}{ M^2[z_0]-s-i\rho(s) s ~G[z_0]}\, ,
\end{eqnarray}
where $b$, $s$ and $R$ stand for bound states, virtual states and resonances, respectively. The parameter $a$ is the scattering length, $z_0$ denotes the pole position of the resonance, and
\begin{eqnarray}\label{z0depen}
M^2[z_0]&=&{\rm Re}[z_0] + {\rm Im}[z_0]\frac{{\rm Im}[z_0\rho(z_0)]}{{\rm Re}[z_0\rho(z_0)]}\, ,\nonumber\\
   G[z_0]&=&\frac{{\rm Im}[z_0]}{{\rm Re}[z_0\rho(z_0)]}\,.
\end{eqnarray}

{One can always express the physical $S$ matrix as a product of contributions of poles and cuts:}
\begin{eqnarray}
S(s)=\prod_i S_i^{pole}\cdot S^{cut}\, .\label{eq:pku,S;pro}
\end{eqnarray}
It is easy to prove that the cut part, $S^{cut}$, still satisfy the generalized unitarity, one could parametrize it as:
\begin{eqnarray}
S^{cut}&=&e^{2i\rho(s)f(s)}\,,\nonumber\\[3mm]
f(s)&=&\frac{s}{\pi}\int_L \frac{{\rm Im}_L f(s')d s'}{s'(s'-s)}+\frac{s}{\pi}\int_R \frac{{\rm Im}_R f(s')d s'}{s'(s'-s)}\, .
\end{eqnarray}
In this approach, unitarity and analyticity are ensured, and crossing symmetry is included by implementing the BNR relations \cite{Balachandran:1968zza,Roskies:1969pe}, which describe the crossing relations between the $\pi\pi$ scattering partial wave amplitudes. The $\chi$EFT can supply the l.h.c as well as the low energy amplitudes. They can be input into the  $S^{cut}$. Since $\chi$EFT only works in the low energy region, there could be some uncertainty due to the distant l.h.c., though one expects that it is strongly suppressed. This approach reveals that the $\sigma$ meson is essential to adjust $\chi$PT to experiment, see Ref.  \cite{Xiao:2000kx}, where it is shown clearly that the l.h.c. gives negative phase shift and thus one always needs a $\sigma$ pole to fit to the data of $K_{e4}$ decay \cite{Rosselet:1976pu} well.  PKU factorization presents the phase shift as additive contributions from different poles and cuts and successfully elaborates the existence of $\sigma$ and $\kappa$ in a model-independent way~\cite{Zheng:2003rw,Zhou:2004ms}, with generalized unitarity, analyticity and crossing relations. However, it is still not clear how to generalize this approach to the coupled channel case. Together with Roy and Roy-Steiner equations, the existence and the precise pole locations of $\sigma$ and $\kappa$ are now well determined, see sections \ref{sec:sigma} and \ref{sec:kappa} for further discussions.


\subsubsection{Representations with Omn$\grave{e}$s functions} \label{sec:Omnes}
Omn$\grave{e}$s function~\cite{Omnes:1958hv} is a general solution of single-channel integral equation. For instance, for $\pi\pi$ scattering, the partial wave amplitude can be expressed as
\begin{equation}\label{eq:TP}
T^I_J(s)=P^I_J(s)\Omega^I_J(s),
\end{equation}
where the Omn$\grave{e}$s function is given by
\begin{equation}\label{eq:Omnes}
\Omega^I_{J}(s)=\exp\left(\frac{s}{\pi} \int^\infty_{s_{th}} ds' \frac{\varphi^I_{J}(s')}{s'(s'-s)}\right) \,.
\end{equation}
In a simplified case, the l.h.c. and distant r.h.c. can be absorbed into the polynomials while the r.h.c. contribution is included in the Omn$\grave{e}$s function.
The single channel unitarity has been kept and analyticity is respected in the physical region.
For example, the $\gamma\gamma\to\pi\pi$ amplitude can be expressed as
\begin{equation}\label{eq:FP}
\mathcal{F}^I_{J\lambda}(s)=\tilde{P}^I_{J\lambda}(s)\Omega^I_J(s).
\end{equation}
Here $\lambda$ is the helicity.
The Watson theorem requires that the phases of the $\mathcal{F}^I_J(s)$ amplitude and the Omn$\grave{e}$s function should be the same below the inelastic threshold. By twice subtracted dispersion relations on $[\mathcal{F}^I_{J\lambda}(s)-B^I_{J\lambda}(s)]{\Omega^I_J}^{-1}(s)$, the S-waves of $\gamma\gamma\to\pi\pi$ amplitudes can be expressed as  \cite{Dai:2014zta}
\begin{eqnarray}
\mathcal{F}^{I}(s)\;&=&\;{\mathcal B}^I(s)+b^{I} s~\Omega^{I}(s)
 +\frac{s^2~\Omega^{I}(s)}{\pi}\int_L ds'\frac{{\rm Im}\left[ \mathcal{L}^{I}(s')\right]\Omega^{I}(s')^{-1} }{s'^2(s'-s)} \nonumber \\[2.5mm]
              &-&\frac{s^2\;\Omega^{I}(s)}{\pi}\int_R ds'\frac{{\mathcal B}^I(s')\;{\rm Im}\left[ \Omega^{I}(s')^{-1}\right] }{s'^2(s'-s)}\, . \label{eq:F;ampS}
\end{eqnarray}

For coupled channel case, one can use the Muskhelishivili-Omn$\grave{e}$s (MO) function, see Ref.~\cite{GarciaMartin:2010cw} and references therein. In the couple channel case it has
\begin{eqnarray}\label{MO:T}
{\rm Im}\bar{T}^I_J(s)&=& \bar{\Omega}^{I*}_J(s) \rho \bar{T}^I_J(s)  \, ,\nonumber
\end{eqnarray}
where the symbol \lq bar' represents the matrix form,
\begin{eqnarray}\label{eq:MO:matrix}
\bar{\Omega}^I_J(s)&=& \left(
          \begin{array}{cc}
   {\Omega^I_J}_{11}(s)      & {\Omega^I_J}_{12}(s)  \\
   {\Omega^I_J}_{21}(s)      & {\Omega^I_J}_{22}(s)   \\
          \end{array}
        \right)  \, ,\nonumber
\end{eqnarray}
in which the MO function $\bar{\Omega}(s)$ only contains the r.h.c..
The coupled channel unitarity can be expressed as
\begin{eqnarray}\label{MO}
{\rm Im}\bar{\Omega}^I_J(s)&=& \bar{\Omega}^{I*}_J(s) \rho(s) \bar{T}^I_J(s)  \, .\nonumber
\end{eqnarray}
The $\mathcal{F}^I_J(s)$ amplitudes of the processes, related to the FSI of these channels, for instance, the $\gamma\gamma\to\pi\pi,\bar{K}K$ amplitudes, can be expressed by once subtracted dispersion relations
\begin{eqnarray}\label{MO}
\vec{\mathcal{F}}(s)&=&\vec{\mathcal{F}}(0)\bar{\Omega}(s)+\frac{s}{\pi}\int_L ds' \frac{{\rm Im}_L [\vec{\mathcal{F}}(s)] \bar{\Omega}^{-1}(s)}{s'(s'-s)}    \bar{\Omega}^*(s)\rho(s) \bar{T}(s)  \, ,\label{eq:MO}
\end{eqnarray}
where the symbol \lq vector' denotes a matrix formalism.
The ${\rm Im}_L [\vec{\mathcal{F}}(s)]$ contains the l.h.c..
In practice, the l.h.c. in the above equation has to be input. Here we employ $\chi$EFT results which automatically respect crossing symmetry. It could be partly from loops of light pseudoscalars (such as pions and kaons) or exchanging resonances in the crossing channels. The former can be calculated by $\chi$PT and the latter by the phenomenological Lagrangian or by the Resonance chiral theory (R$\chi$T) within $1/N_C$ expansion.
For instance, in Refs.\cite{Dai:2014zta,GarciaMartin:2010cw,Mao:2009cc}, the l.h.c. of the $\gamma\gamma\to \pi\pi$ is simulated by the resonance exchange which appears in the t-channel, where the resonances with masses smaller than 1.3~GeV are all included.
These methods respect analyticity, unitarity,  and partly crossing symmetry that the l.h.c. can be somehow predicted by $\chi$EFT. However, the Omn$\grave{e}$s functions appearing in Eqs.(\ref{eq:F;ampS}, \ref{eq:MO}) need to be well determined. This usually demands that one has plenty of data, e.g. phase shifts/phases of the hadronic scattering processes.  Unfortunately, the relevant data are sometimes inadequate extremely, for instance, in the $\pi\eta-K\bar{K}$ coupled channels.

There are some other methods using Omn\`es function to develop the techniques of solving the scattering amplitude of single channel scatterings.
In Ref.~\cite{Dai:2019zao} a dispersion relation for $\ln T(s)$ is presented, and the partial wave amplitude is given as
\begin{eqnarray}
T^I_J(s)=T^I_J(s_0){\Omega^I_J}_L(s){\Omega^I_J}_R(s)\,,\label{eq:T;Omnes}
\end{eqnarray}
where  the phase (rather than the phase shift) is used to get the Omn\`es function.
Combining with unitarity, a representation for the $T^I_J(s)$ amplitude is found in the elastic region, \begin{eqnarray}
T^I_J(s)=-\frac{{\rm Im}[{\Omega^I_J}_R(s)^{-1}]{\Omega^I_J}_R(s)}{\rho(s)}\,.\label{eq:unitarity;ph}
\end{eqnarray}
Notice that Eq.~(\ref{eq:unitarity;ph}) also works for the coupled channel case in the physical region.
Here the phase for the l.h.c. is correlated with that of the r.h.c. in the elastic region by
\begin{eqnarray}
{\Omega^I_J}_L(s)=-\frac{{\rm Im}[{\Omega^I_J}_R(s)^{-1}]}{\rho(s) T^I_J(s_0)}\,.\label{eq:unitarity;lhc}
\end{eqnarray}
One then can calculate the two-body scattering partial wave amplitude in two steps:
Using Eq.~(\ref{eq:unitarity;ph}) to fit the Omn\`es function of the r.h.c. to  experimental data, and
then using Eq.~(\ref{eq:unitarity;lhc}) and other constraints below the threshold to fix the l.h.c. part.
This method is somehow similar to the N/D method, but it is much easier to perform numerical computation and the l.h.c. is clearly given in terms of the phase too. Nevertheless, it is necessary to take into account the obvious constraints of crossing relations between partial waves, which needs further studies.
These Omn$\grave{e}$s representations are used to extract the pole locations of the $f_0(980)$ and $a_0(980)$ in sections.\ref{sec:f980} and \ref{sec:a980}, respectively.
%

In the end of this section, it is worth stressing that dispersive approaches are efficient to extract information on resonances. Especially, PKU factorization is helpful to establish the existence of broad resonances faraway from the real axis, while Roy equation (or Roy-like equation) is one of the most powerful tools for precise determination of pole locations. However, the two methods are limited to single channel scatterings and hence, for the coupled channel cases, unitarized models such as K-matrix, Pade approximation and IAM still play a crucial role in the studies of resonances located above inelastic thresholds.

\section{Hunting for hadron states}\label{sec:3}
In the low energy region of strong interaction, the physics is rather fruitful and various kinds of mystery hadrons appear. To study them one needs the information of mass, width, and couplings, etc.. The FSI tools discussed in the above section are helpful to obtain the scattering amplitudes and extract the precise resonance information. Furthermore, recently experiment collaborations of LHCb, Belle, BESIII and so on have discovered some exotic hadrons, which started a new era of hadron physics as obviously they are not of normal $q\bar{q}$ or  $qqq$ structure.
All these demand a careful study about the scattering amplitudes and the intermediate hadron states.
In this section, researches on light scalar mesons, light baryon resonances as well as some exotic states are briefly discussed.

\subsection{Light scalar mesons\label{sec:lsm}}

The lightest scalars ($\sigma$, $\kappa$, $f_0(980)$ and $a_0(980)$) are rather interesting \cite{Pennington:2007yt,Pennington:2010dc,Jaffe:1976ig,Jaffe:2004ph,Pelaez:2015qba}. In particular, the isoscalar scalar mesons have the same quantum numbers as that of the QCD vacuum.
On the other hand, the heavier scalars ($f_0(1370)$, $f_0(1500)$, $K_0^*(1430)$, and $a_0(1450)$ ), though it is still controversial, are more likely to be $q\bar{q}$ states, while the light ones are not of such structure.
The lightest scalars referring to $\pi\pi$,  $\pi K$,  $\pi \eta$ scattering and there are lots of measurements about these processes. To extract the pole information and coupling, one also needs to do partial wave decomposition. The final-state interactions of the scattering partial wave amplitudes referring to these resonances are important to study the lightest scalars \cite{Au:1986vs,Morgan:1993td}.
$\chi$EFT could give information about the amplitudes in the low energy region with respect to Lorentz invariance, gauge symmetry, crossing symmetry, discrete symmetries etc.. However, as argued before, there are unknown couplings which need to be determined in the $\chi$EFT, hence one needs both experimental data and theoretical principles to constrain these LECs\footnote{It is worth to point out that the LECs nowadays can also be determined from lattice QCD, see e.g. Ref.\cite{Aoki:2019cca} and references therein.}. Methods based on the dispersion relation is necessary to extend the amplitude from low energy region to higher ones. Combination of $\chi$EFT and dispersion relations makes great progress in the study of the lightest scalars in the past decades.

Naturally, dispersion relation could include the contribution of FSI, for recent works on the dispersion relations referring to the $\pi\pi$ re-scattering, see e.g.~\cite{Dai:2014zta,Kang:2013jaa,Chen:2015jgl}.
For the $\sigma$ and $\kappa$, their masses are not faraway from the $\pi\pi$/$\pi K$ threshold, but their widths are rather large. This makes it difficult to extract the pole information of these resonances. One needs dispersion relation to continue the amplitude from the real axis to the deep complex-$s$ plane, including FSI.  $\chi$PT is able to constrain the amplitude in the low energy region, especially near and below the threshold. Indeed, the chiral expansion converges best inside the Mandelstam triangle region, see e.g. the discussion in Ref.\cite{Buettiker:1999ap}. Meanwhile the threshold parameters, scattering lengths and slope parameters, can be well predicted.  The l.h.c. can only be calculated directly by the $\chi$EFT and it is a key point to combine dispersion relations and $\chi$EFT. For the $\sigma$ and $\kappa$, the inelastic channel ($\bar{K} K$ or $\eta K$) is faraway and thus one only needs to deal with single channel scattering. The Roy and/or Roy-like equations, PKU factorization and other dispersive approaches are such powerful tools as they keep unitarity, crossing symmetry and analyticity.
The very existence of the $\sigma$ and $\kappa$ has been confirmed by PKU factorization in Refs.\cite{Xiao:2000kx,Zhou:2004ms,Zheng:2003rw} and by Roy equations in Refs.\cite{Colangelo:2001df,DescotesGenon:2006uk}, and their pole locations (and also the couplings) have been given in Refs.~\cite{Caprini:2005zr,Buettiker:2003pp,GarciaMartin:2011jx}.
However, the inner structure of these resonances are still unknown. Notice that the $\kappa$ has different quark flavors such as $\bar{d}$ and $s$, so it can not be a glueball. Considering that the similarity between the $\sigma$ and $\kappa$, it is very likely that the $\sigma$ is not a glueball, too.
It may be more suitable to conclude what the $\sigma$ and $\kappa$ are not rather than what they are \cite{Pelaez:2015qba}.
For the $f_0(980)$, it is around the $\bar{K}K$ threshold and thus far beyond the working region of the $\chi$PT, but the resonance is much narrower, thus other unitarization methods such as $K$-matrix and U$\chi$PT could work well. The $f_0(980)$ is narrower than the $\sigma$, but it is around the $\bar{K}K$ threshold, thus one needs coupled channel method to deal with $\pi\pi-\bar{K}K$ coupled channels. For the $a_0(980)$, there is no data of the phase shifts and inelasticity, and it is not so narrow as that of the $f_0(980)$. All these increase the difficulty to study their properties.
Efforts are needed from both the experimental and theoretical sides. So far, none of the lightest scalars behaves like a normal $q\bar{q}$ state.
There are various kinds of models
\cite{vanBeveren:1986ea,Oller:1998hw, Black:1998wt,Close:2002zu,Baru:2003qq,Maiani:2004uc,Achasov:2007fz,Hooft:2008we,Zheng:2008wb,Zhou:2010ra,Mennessier:2010xg,
Guo:2012ym,Guo:2012yt,Weinberg:2013cfa,Cohen:2014vta,Briceno:2016mjc,Badalian:2020wua} trying to reveal the nature of the scalars,
but it still remains a mystery.


\subsubsection{$\sigma$: from $\pi\pi$ scattering\label{sec:sigma}}
$\pi\pi$ scattering has been studied for more than 70 years, so why is it still so important? Pion is the lightest hadron and it is common to find
a pair of pions in the final states. What is more, for several pions in the final states, for instance three pions, the two pion scattering amplitudes are the basics to study the three pions FSI, see discussions in section \ref{sec:FSI}.  Besides, the two pseudoscalar meson scattering is one of the best way to test $\chi$PT, which is constructed by chiral symmetry, discrete symmetries, etc..
On the other side, the lightest scalars, especially the $\sigma$, appear as the intermediate states in the $\pi\pi$ scattering are rather important.
The $\sigma$ has the same quantum number as that of the vacuum and could be important for understanding how the hadrons get masses.
As early as 1955, Johnson and Teller proposed an isoscalar scalar meson exchange in the inter-nucleon interactions \cite{Johnson:1955zz}. It is now called $\sigma$. Later Gell-Mann and Levy introduced the linear $\sigma$ model \cite{GellMann:1960np} where the $\sigma$ is a scalar singlet and pions appear as the pseudoscalar Goldstone bosons due to spontaneous chiral symmetry breaking (S$\chi$SB) of $SU_L(2)\times SU_R(2)$. However, its prediction on the mass and width of the $\sigma$ has large difference from what it should be. Thus the linear sigma model is not considered as the faithful EFT of low-energy QCD \cite{Gasser:1983yg} and the $\sigma$ particle had disappeared in the PDG for many years \footnote{We are aware that, after unitarization, it is found in the linear $\sigma$ model that the $\sigma$ should persist if the meson loop diagrams are accounted for \cite{Achasov:1994iu}. This is also discussed in the three flavor linear sigma model \cite{Black:2000qq}, where unitarization is important to dress the \lq bare' mass and width to lighter and wider ones. A study of quark-level linear sigma model claims that the $\sigma$ and $\pi$ are \lq chiral partners' \cite{Scadron:2002mm}. All these works, qualitatively rather than quantitatively, suggest the existence of the $\sigma$. }. Nevertheless, there are many phenomenological analyses, such as the unitarized meson model~\cite{vanBeveren:1986ea}, unitarized quark model \cite{Tornqvist:1995kr}, and the interference model \cite{Ishida:1995xx}, Lippmann-Schwinger equation \cite{Kaminski:1993zb} etc., reveal that the low energy $\pi$-$\pi$ scattering phase shifts should be contributed by a light broad $\sigma$ state.

\begin{figure}[hpt]
\includegraphics[width=0.98\textwidth,height=0.12\textheight]{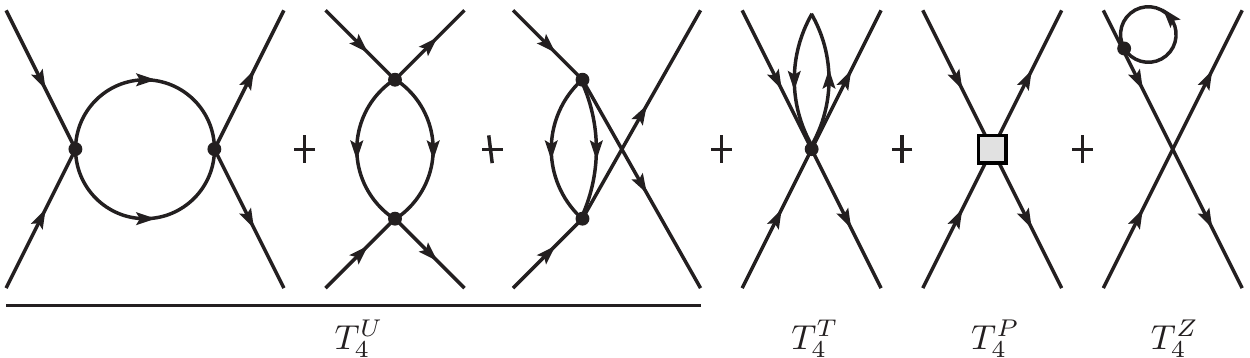}
\caption{\label{Fig:mms} The Feynmann diagrams of meson-meson scattering amplitudes at $\mathcal{O}(p^4)$. The $T_4^U$ represents the loops constructed from $\mathcal {L}_2$ vertices with four fields. It includes contributions
from s, t and u channels. $T_4^T$ is of contributions
coming from the $\mathcal {L}_2$ ChPT lagrangian with six fields and
a tadpole. $T_4^P$ is the $\mathcal{O}(p^4)$ tree diagrams from $\mathcal {L}_4$ lagrangian. $T_4^Z$ is the wave-function renormalization of leading order meson-meson scattering amplitudes.  }
\end{figure}

In 1980s the $\chi$PT is extended up to $SU(3)\times SU(3)$ \cite{Gasser:1983yg,Gasser:1984gg} and it is quite successful in describing the low energy $\pi\pi$ scattering. It is now considered as the most efficient EFT to describe the low energy strong interactions, and it supports important information about the $\sigma$. From Roy equation this is obvious. The two subtractions of the Roy's dispersion relations should be matched to the $\chi$PT in the low energy region. In some other dispersive approaches, the l.h.c. needs to be predicted/constrained by $\chi$PT, too.
The leading order lagrangian of $\chi$PT ($\mathcal{O}(p^2)$) is given as  \cite{Gasser:1984gg}
\begin{equation}\label{l2}
\mathcal
{L}_2=\frac{f_0^2}{4}\langle \partial_{\mu}U^{\dag}\partial^{\mu}U+\mathcal
{M}(U+U^{\dag})\rangle \ ,
\end{equation}
where $\langle\,\rangle$ represents the trace of the $3\times 3$ matrices, $\mathcal{M}$ is the mass matrix, and $U(\Phi)$ is the exponential parametrization
\begin{equation}
U(\Phi)=\exp(\frac{i\sqrt{2}}{F_0}\Phi)\ .
\end{equation}
Here $F_0$ is the pion decay constant and the exponential parametrization $\Phi$ is the representation of pseudo-goldstone boson fields
\begin{equation}
\Phi(x)=\left(
          \begin{array}{ccc}
            \frac{1}{\sqrt{2}}\pi^0+\frac{1}{\sqrt{6}}\eta & \pi^+ & K^+ \\
            \pi^- & -\frac{1}{\sqrt{2}}\pi^0+\frac{1}{\sqrt{6}}\eta & K^0 \\
            K^- & \bar{K}^0 & -\frac{2}{\sqrt{6}}\eta \\
          \end{array}
        \right)\ .
\end{equation}
The $\mathcal{O}(p^4)$ lagrangian of $\chi$PT can be written as \cite{Gasser:1984gg}
\begin{eqnarray}
\mathcal{L}_4&=&L_1 \langle \partial_{\mu}U^{\dag}\partial^{\mu}U \rangle^2+
L_2\langle \partial_{\mu}U^{\dag}\partial_{\nu}U\rangle~\langle\partial^{\mu}U^{\dag}\partial^{\nu}U\rangle+\nonumber\\
&&
L_3\langle \partial_{\mu}U^{\dag}\partial^{\mu}U\partial_{\nu}U^{\dag}\partial^{\nu}U\rangle+
L_4\langle \partial_{\mu}U^{\dag}\partial^{\mu}U\rangle~\langle U^{\dag}\mathcal{M}+\mathcal{M}^{\dag}U\rangle+\nonumber\\
&& L_5 \langle \partial_{\mu}U^{\dag}\partial^{\mu}U(U^{\dag}\mathcal{M}+\mathcal{M}^{\dag}U)\rangle
+L_6 \langle U^{\dag}\mathcal{M}+\mathcal{M}^{\dag}U\rangle^2+\nonumber\\
&& L_7 \langle U^{\dag}\mathcal{M}-\mathcal{M}^{\dag}U\rangle^2+
L_8 \langle \mathcal{M}^{\dag}U\mathcal{M}^{\dag}U+U^{\dag}\mathcal{M}U^{\dag}\mathcal{M}\rangle
\ .
\end{eqnarray}
Here $L_i$'s are the LECs. According to crossing symmetry and isospin decomposition, the $\pi\pi\rightarrow\pi\pi$ scattering amplitude could be expressed by one amplitude:
\begin{eqnarray}
T^{I=0}(s,t,u)&=&3T(s,t,u)+T(t,s,u)+T(u,t,s)\ ,\nonumber\\
T^{I=1}(s,t,u)&=&T(t,s,u)-T(u,t,s)\ ,\nonumber\\
T^{I=2}(s,t,u)&=&T(t,s,u)+T(u,t,s)\ , \label{eq:T;pipi}
\end{eqnarray}
where $T(s,t,u)=T(\pi^+\pi^-\rightarrow\pi^0\pi^0)$.
At leading order, one has
\begin{eqnarray}
T_{2}(s,t,u)=\frac{s-m_\pi^2}{3F_\pi^2} \,, \nonumber
\end{eqnarray}
where the subscript \lq2' represents the chiral counting. The partial wave decomposition can be obtained by applying Eq.~(\ref{eq:Roy;t}), with
\begin{eqnarray}
T_{2}^{0S}(s)=\frac{2s-m_\pi^2}{32\pi F_\pi^2} \,,\;\;\;\; T_{2}^{2S}(s)=-\frac{s-2m_\pi^2}{32\pi F_\pi^2}\,,\;\;\;\;
T_{2}^{1P}(s)=\frac{s-4m_\pi^2}{96\pi F_\pi^2} \,. \label{eq:T;pipi;Op2}
\end{eqnarray}
They are called the Weinberg's low energy theorem.
For higher order corrections, the meson-meson scattering amplitudes at $\mathcal{O}(p^4)$ are shown in Fig.\ref{Fig:mms}.
the $SU(3)$ one-loop meson-meson scattering amplitudes are calculated in Refs.~\cite{Bernard:1991zc,GomezNicola:2001as}, with $\overline{MS}-1$ scheme.
The analytical forms of the partial waves of $\pi\pi$ scattering amplitudes with $\overline{MS}$ scheme can be found in Ref.\cite{Dai:2019zao}.

Since $\chi$PT only works in the low energy region, but usually the resonances lie near the higher energy region, one thus needs different methods to extend the working region of $\chi$PT. On the other hand, to find the pole locations, one needs to continue the amplitude to the complex-$s$ plane. This is done along the unitarity cut and thus U$\chi$PT are applied to accomplish the unitarization and extracting poles. It is worth pointing out that in U$\chi$PT, the $\rho$ and $\sigma$ are not written directly in the Lagrangian, but they are found simultaneously in the RS-II when the unitarization is applied. As is well known, the $\rho$ has been recognized as a resonance for a long time. This also gives an evidence of the existence of the $\sigma$ provided that U$\chi$PT describes the S-wave as well as the P-wave.
In U$\chi$PT methods, Ref.~\cite{Dobado:1996ps} finds the $\sigma$ pole located at $440-i245$~MeV within IAM,
and Ref.~\cite{Pelaez:2003rv} gives a similar $\sigma$ pole location but the width is a bit smaller for the IAM-III case.
With Pad\'e approximation Refs.~\cite{Dai:2011bs, Dai:2012kf} both find a $\sigma$ pole in the RS-II. It is also interesting to note that the
companion shadow poles are found in these two works, with $\pi\pi-\bar{K}K$ or $\pi\pi-\bar{K}K-\eta\eta$ coupled channels. This confirms the sizable $q\bar{q}$ component. In Ref.~\cite{Guo:2011pa}, the $U(3)$ $\chi$PT is unitarized by an $N/D$ method and the $\sigma$ is found at similar position, but with the width a bit larger and closer to that of the Roy analysis~\cite{Caprini:2005zr}.
In all these papers, the pole in the RS-II has been found after unitarization, the masses of the $\sigma$ are quite close to what is obtained by dispersive approaches such as Roy equation, but the widths are smaller.
Note that these results are sensitive to the values of the LECs, and the systematic uncertainty of the LECs should be taken into account\footnote{For discussions about the relation between LECs and the resonance parameters, one is referred to Refs.\cite{Guo:2007ff,Guo:2007hm}}.
This could be one of the reason why the widths are smaller than that obtained by dispersion relation.
In short, the U$\chi$PT is qualitative rather than quantitative to study the property of the scalars. Nevertheless, it gives hints of the existence of the $\sigma$ and is also helpful to study its property.

\begin{figure}[hpt]
\begin{center}
\includegraphics[width=0.48\textwidth,height=0.3\textheight]{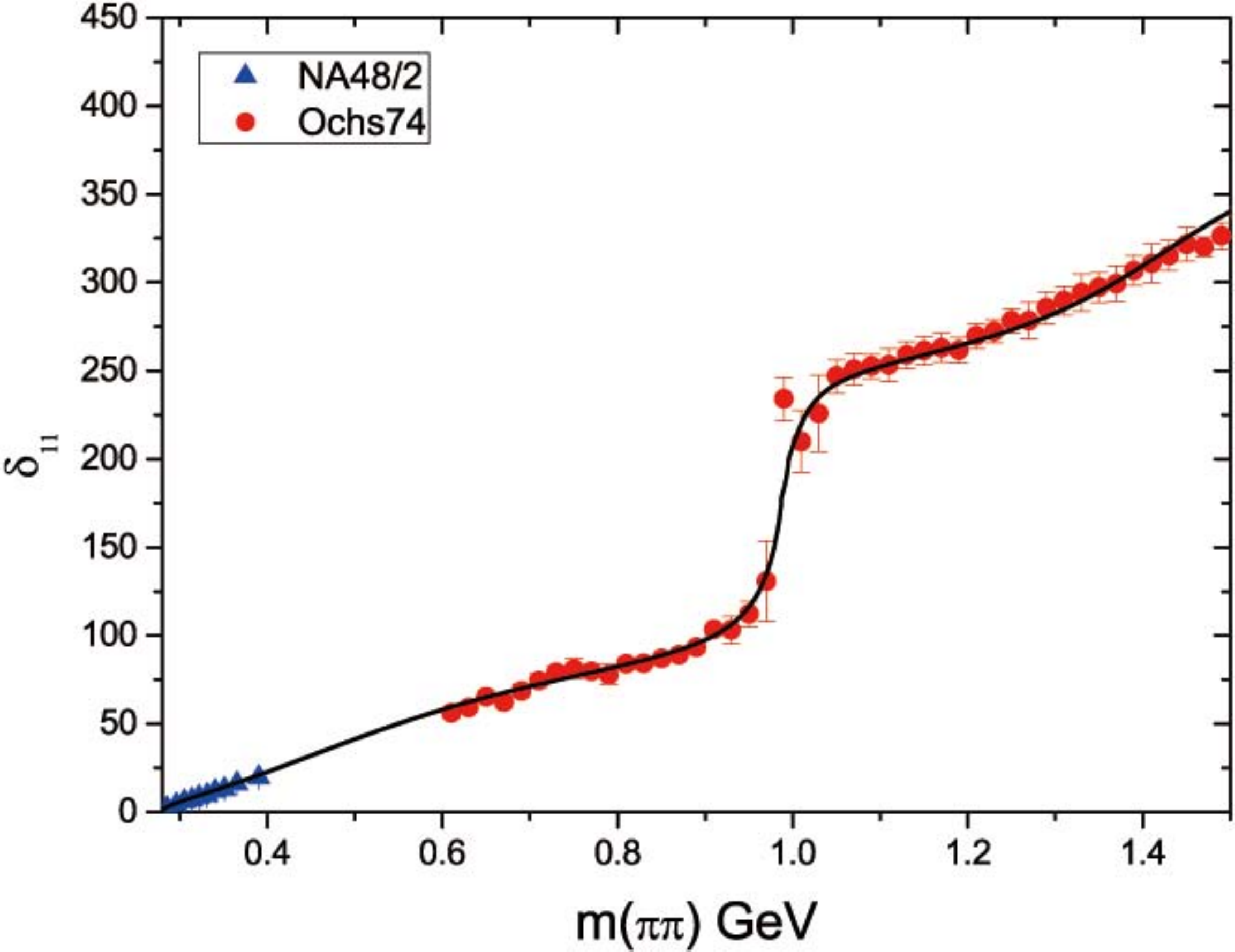}
\includegraphics[width=0.48\textwidth,height=0.3\textheight]{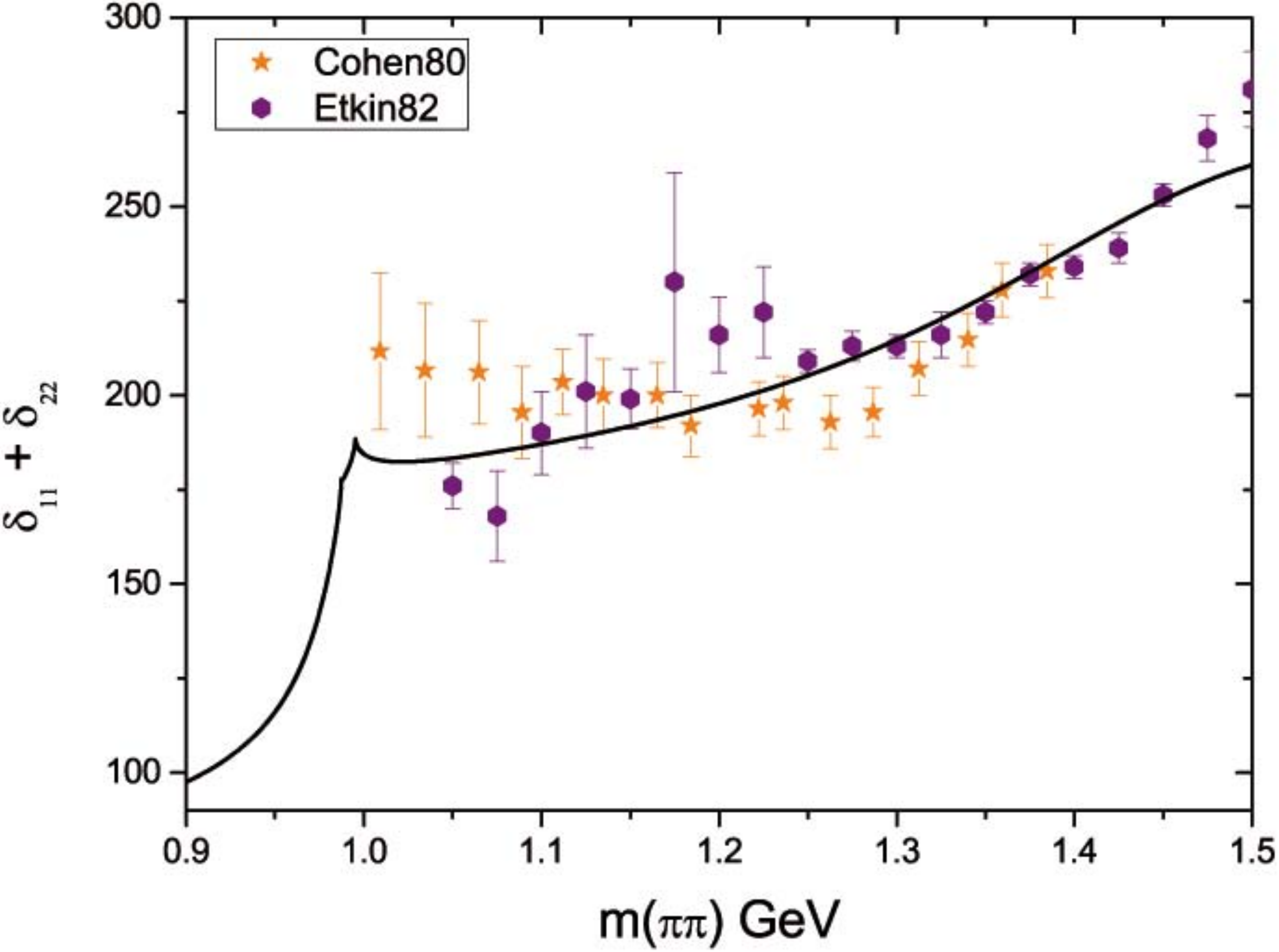}
\caption{\label{Fig:ph;pipi} The I=0 S-wave phase shifts of $\pi\pi$-$\bar{K}K$ coupled channels. The $\pi\pi$ scattering phase shifts of Ochs are from the CERN-Munich group~\cite{Hyams:1973zf,Grayer:1974cr,Hyams:1975mc},   and those of the NA48/2 data are from~\cite{Batley:2010zza}.
The $\pi\pi\to \bar{K}K$ phase shifts are from Cohen {\it et.al.} \cite{Cohen:1980cq} and Etkin {\it et.al.} \cite{Etkin:1981sg}, respectively. The black lines are taken from a $K$-matrix fit \cite{Dai:2014zta}. }
\end{center}
\end{figure}

The pole information of the $\sigma$ needs to be extracted from accurate scattering amplitude. To reach it, one needs high statistics experimental data to constrain the amplitude. In 1970s, the phase shifts of $\pi\pi$ scattering are given by Protopopescue {\it et. al.}~\cite{Protopopescu:1973sh} and CERN-Munich group \cite{Hyams:1973zf,Grayer:1974cr,Hyams:1975mc}.
These phase shifts are above 0.5~GeV and nowadays people realize that the $\sigma$ has a low mass (the real part of the pole location in the RS-II), thus data in lower energy region is necessary.  The $K_{e4}$ data given by \cite{Rosselet:1976pu} and the latest $K_{e4}$ data in the 21st century \cite{Batley:2010zza,Pislak:2003sv}, especially the latter one,  are rather precise and they are important to determine the $\sigma$ pole location.
Note that in Ref.\cite{Batley:2010zza}, the data is so accurate that one needs to consider the isospin breaking.
Some of the phase shift data has been shown in Fig.\ref{Fig:ph;pipi}.
From these data, one still needs a powerful tool to extract the resonance information in a reliable manner, that is, dispersion relation.  As the $\sigma$ may not be found by using the traditional Breit-Wigner(BW) formalism, see e.g. Ref.~\cite{Gardner:2001gc}, we also need another form to label the resonance except for the BW mass and width.
As stated in PDG \cite{Zyla:2020zbs}, the resonance is also indicated by the $T$ matrix pole location. For the $\sigma$ and $\kappa$, this is even more urgent due to their large widths, which are faraway from the real axis of the $s$-plane and the BW mass and width can be quite different in different processes.
In recent twenty years, the dispersive approaches are revived and developed.
They work rather well to accomplish the task: confirming the lightest scalars and extracting out their poles from scattering amplitudes.

In Ref.\cite{Ishida:1995xx}, the interfering amplitude method is used and the unitarity has been implemented. By fitting to the $\pi\pi$ phase shift, the existence of the $\sigma$ has been confirmed.
They give the resonance information as $M=553.3\pm0.5$~MeV, $\Gamma=242.6\pm1.2$~MeV.
In the analysis, they introduce a negative background phase shift. As will be discussed in the next sections, it is indeed  compatible with the contribution of the l.h.c. of the PKU factorization.
On the other hand, $\chi$PT could give constraints in the low energy region, including the region below the threshold where the experiment (data) can not reach.
With both $\chi$PT and experimental data, the dispersive approach combines all the knowledge we know and makes the deep study of the resonances to be possible.

\begin{figure}[hpt]
\centering
\includegraphics[width=0.6\textwidth,height=0.3\textheight]{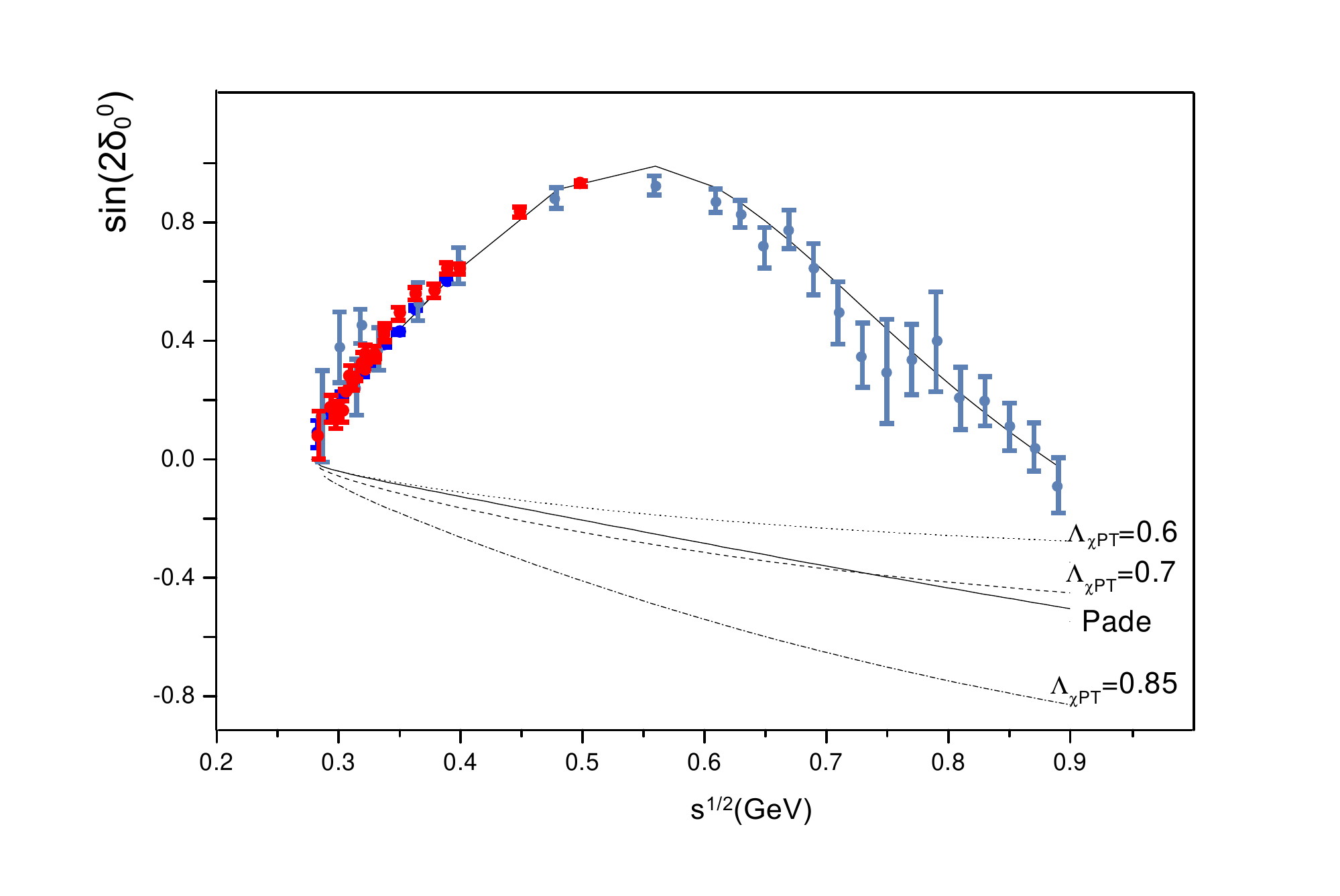}
\caption{\label{Fig:pku} Fit to the phase shifts of isoscalar S-wave of $\pi\pi$ scattering \cite{Xiao:2000kx}. The contribution of the l.h.c. with
different cut-offs and also a different method (Pad\'e approach) are plotted, too. The data is taken from \cite{Batley:2010zza,Rosselet:1976pu}. }
\end{figure}

The PKU factorization \cite{Xiao:2000kx} is a good approach to combine the experimental data and resonances together. In this way the contribution of l.h.c, r.h.c and poles are classified clearly.
The pole contributions are explicitly written in the dispersion relation and the S matrix is given as
{ \setlength{\mathindent}{0.cm}  \begin{eqnarray}
S(z)&=&\tilde{\alpha}+i\alpha\rho(z)+\sum_i \frac{\beta_i}{2(z-s_i)}
+\sum_i \frac{\rho(z)\beta_i}{2\rho(s_i)(z-s_i)}+\sum_j\frac{\rho(z_j^{II})-\rho(z)}{2\rho(z_j^{II})S'(z_j^{II})(z-z_j^{II})} \nonumber\\
 &&+\frac{1}{\pi}\int_L \frac{{\rm Im}_L F(s')d s'}{s'-s} +\frac{1}{\pi}\int_R \frac{{\rm Im}_R F(s')d s'}{s'-s}
 +\frac{1}{\pi}\int_L \frac{{\rm Im}_L \tilde{F}(s')d s'}{s'-s} +\frac{1}{\pi}\int_R \frac{{\rm Im}_R \tilde{F}(s')d s'}{s'-s} \,. \nonumber\\
\label{eq:pku,S;final}
\end{eqnarray} }
One would notice that the contribution from the pole in the RS-II has been included as $z_j^{II}$. It is implemented by the reflection property of $S(s^*)=S^*(s)$ and the analytical continuation along the unitary cut. The \lq $R$' denotes the r.h.c. starting from the first inelastic threshold of 4$\pi$, but it can be ignored below the $\bar{K}K$ threshold. The contribution of the l.h.c. is shown in Fig.\ref{Fig:pku}, from Ref.~\cite{Xiao:2000kx}.
In the analysis, the l.h.c. is calculated from $\chi$PT and the cut-offs are adjusted a bit to estimate its uncertainty.
They conclude that the l.h.c. can only supply a negative and concave contribution to the $\sin2\delta^0_0$, see the declined lines in Fig.\ref{Fig:pku}.  Since the background contribution of l.h.c. has a smooth behavior,  there should be a resonance in the RS-II to fit the convex structure of the data.
A refined analysis based on PKU factorization gives the $\sigma$ pole at $470\pm50-i285\pm25$~MeV \cite{Zhou:2004ms}.
This calculation clearly demonstrates the existence of the $\sigma$ resonance in a model independent way, according to analyticity and unitarity.

In a more sophisticated way, one needs to implement all the principles of the QFT to refine the analysis on scattering amplitude and decompose it according to the partial wave dynamics. Also it needs to be continued to the complex plane when the width of the resonance is large.  After all these steps could the pole of the resonance be extracted out exactly.
So far the Roy equation plays such a role with all the requirements discussed above. For details about Roy equation, we refer readers to Refs.~\cite{Ananthanarayan:2000ht,Pelaez:2015qba}. Here we only give a brief introduction about how Roy equation fixes the pole location of the $\sigma$ for reader's convenience.
As is well known, the Roy equation is based on fixed-$t$ dispersion relation and the functions of $t$ need to be projected into functions of $s$. With twice-subtracted dispersion relations, one has
{ \setlength{\mathindent}{1.cm} \begin{eqnarray}
\vec{T}(s,t)= \vec{f_0}(t)+\vec{f_1}(t)s+\frac{s^2}{\pi}\int_{4m_\pi^2}^{\infty}ds'\frac{{\rm Im}\vec{T}(s',t)}{s'^2(s'-s)}
+\frac{u^2}{\pi}\int_{4m_\pi^2}^{\infty}ds'\frac{C_{su}{\rm Im}\vec{T}(s',t)}{s'^2(s'-u)} \,. \label{eq:Roy;fixt}
\end{eqnarray} }
With the crossing relation $\vec{T}(s,t)=C_{su}\vec{T}(u,t)$ and the scattering lengths
  \begin{eqnarray}
T^0(4m_\pi^2,0)=a_0^0~m_\pi\,,\;\; T^1(4m_\pi^2,0)=0\,,\;\; T^2(4m_\pi^2,0)=a_0^2 ~m_\pi \,, \nonumber
\end{eqnarray}
one could get explicit expressions about the polynomial (subtraction) part, as shown in Eq.~(\ref{eq:Roy;g}).
With the partial wave projection of Eq.~(\ref{eq:Roy;t}), after a few steps, one would find the Roy equations of $\pi\pi$ scattering
\begin{eqnarray}
T^I_J(s)=k^I_J(s)+\sum_{I'=0}^{2}\sum_{J'=0}^{\infty}\int_{4m_\pi^2}^{\infty}ds' K^{II'}_{JJ'}(s){\rm Im} T^{I'}_{J'}(s') \,. \label{eq:Roy:pipi}
\end{eqnarray}
For details about the kernel functions $K^{II'}_{JJ'}(s)$, see Refs.\cite{Ananthanarayan:2000ht,Wanders:2000mn}.
Here $k^I_J(s)$ is the partial wave projection of the subtraction terms. From it one finds that the l.h.c. has disappeared and they are represented by the r.h.c. of an infinite series of partial waves. In Ref.\cite{Caprini:2005zr} it estimates the contribution of the l.h.c. to the $\sigma$ pole location, it is roughly 14\%. In Ref.\cite{Dai:2019zao} it shows that when the l.h.c. changes a bit, with unitarity kept, the variation of the $\sigma$ is reduced to about 2-3 percents by changing the l.h.c. of 100\%. We stress that this does not mean the l.h.c. has small contribution to the $\sigma$, but rather it reveals that the unitarity and analyticity give strong constraint on the l.h.c. or the crossing relations.
This is compatible with that of Ref.\cite{Fariborz:2009cq}, where the $\sigma$ is found to be $724$~MeV in a generalized linear sigma model, but it can be refined to $477-i252$~MeV according to unitarization\footnote{This paper also interestingly shows that the $\sigma$ has 40\% of $q\bar{q}$ and 60\% of tetraquark components, respectively.}.
For the integral above the matching point $s_0$ (for example $\sqrt{s_0}=0.8$~GeV), one needs other models to predict the integral and takes them as inputs. It could also be divided into two parts, one is the intermediate region of $s_0<s<s_2$, where one could choose $\sqrt{s_2}=2$~GeV. In this region, the experimental measurement is precise enough now and one could use different unitary models to fit the data and take them as input, such as analysis by $K$-matrix and Breit-Wigner formalism.
For the higher energy part $s\geq s_2$ the integration is small and Regge asymptotic behaviour can be used to estimate the contribution.
One could also expand the dispersion relations around small $s$ and match it to the results of the $\chi$PT, crosschecking with each other.

\begin{figure}[hpt]
\centering
\includegraphics[width=0.6\textwidth,height=0.3\textheight]{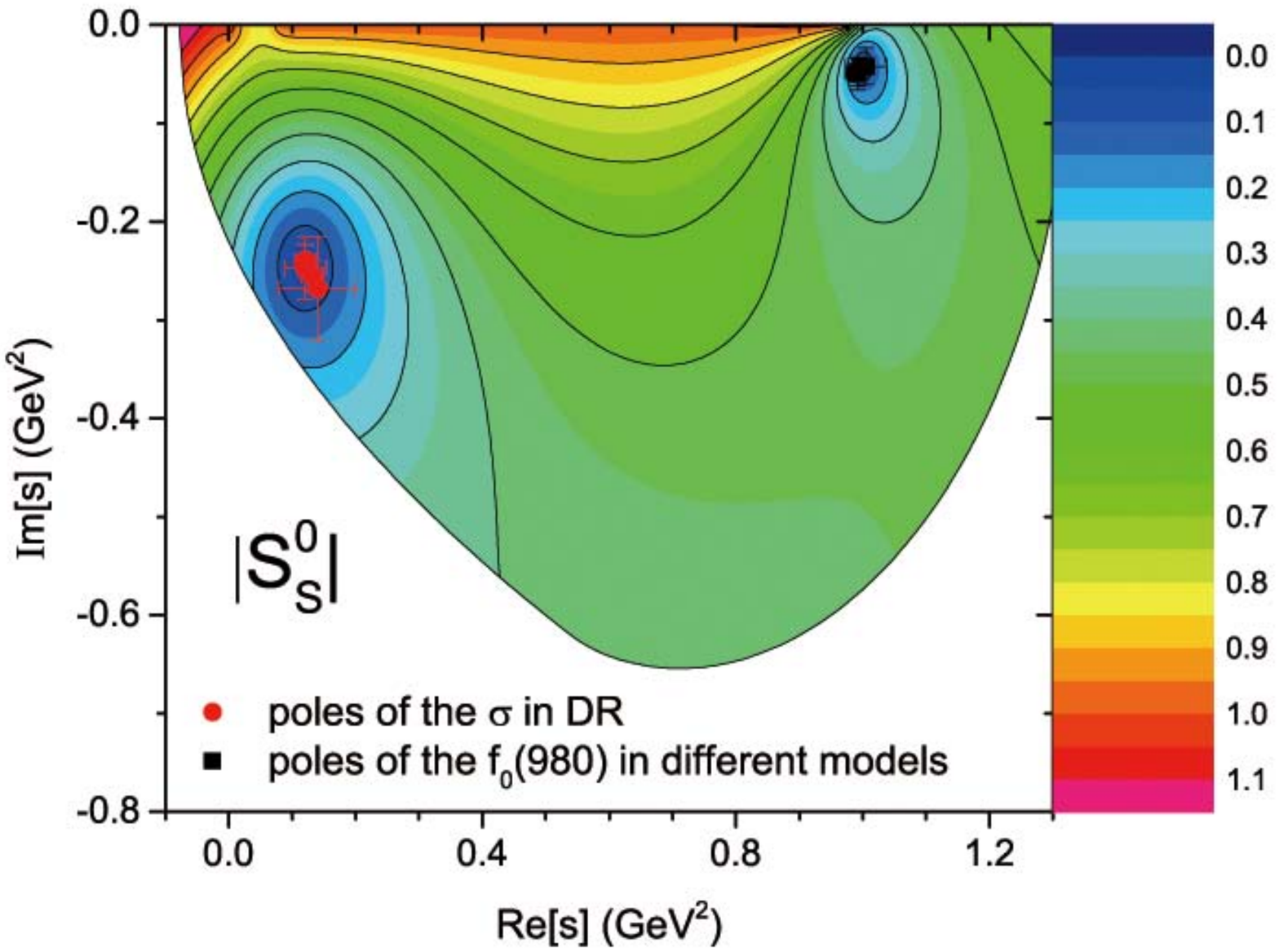}
\caption{\label{Fig:roy} The absolute values of the $S$-matrix for $IJ=00$ $\pi\pi$ scattering partial wave in the domain where Roy equation works. The $|S|$ on the upper half of s-plane are readily obtainable from the ones on the lower side according to the Schwarz reflection principle. We calculate it following \cite{Caprini:2005zr,Dai:2017uao} and add the poles given by some dispersive analysis.  The red dots represent the poles of the $\sigma$ and the black squares for the $f_0(980)$. The $\sigma$ poles given by dispersion relations are taken from Refs. \cite{Caprini:2005zr,Moussallam:2011zg,GarciaMartin:2011jx,Zhou:2004ms}, and the $f_0(980)$'s are taken from Refs.\cite{Dai:2014lza,Moussallam:2011zg,GarciaMartin:2011jx}.
 }
\end{figure}

\begin{figure}[hpt]
\includegraphics[width=0.48\textwidth,height=0.3\textheight]{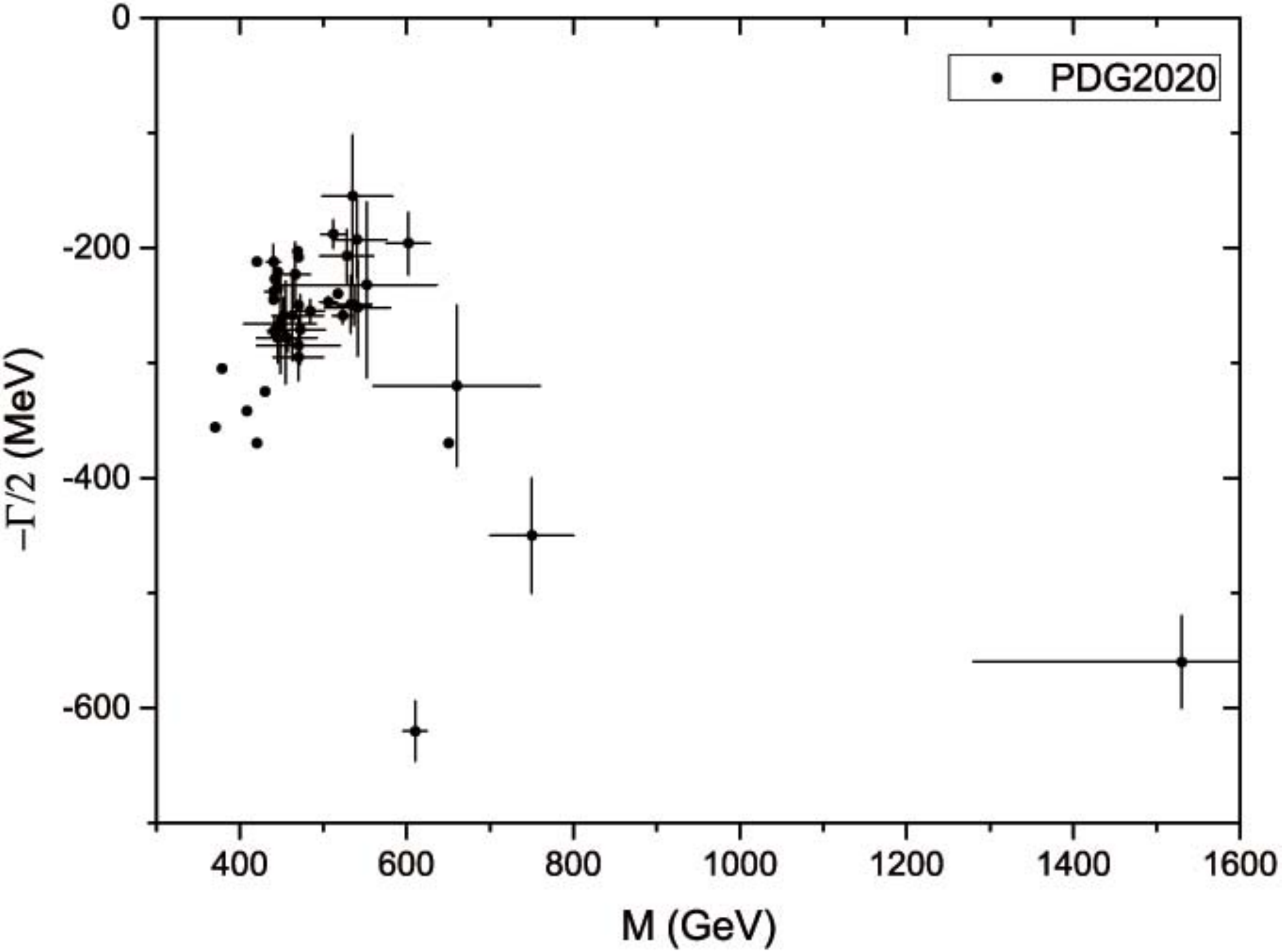}
\includegraphics[width=0.48\textwidth,height=0.3\textheight]{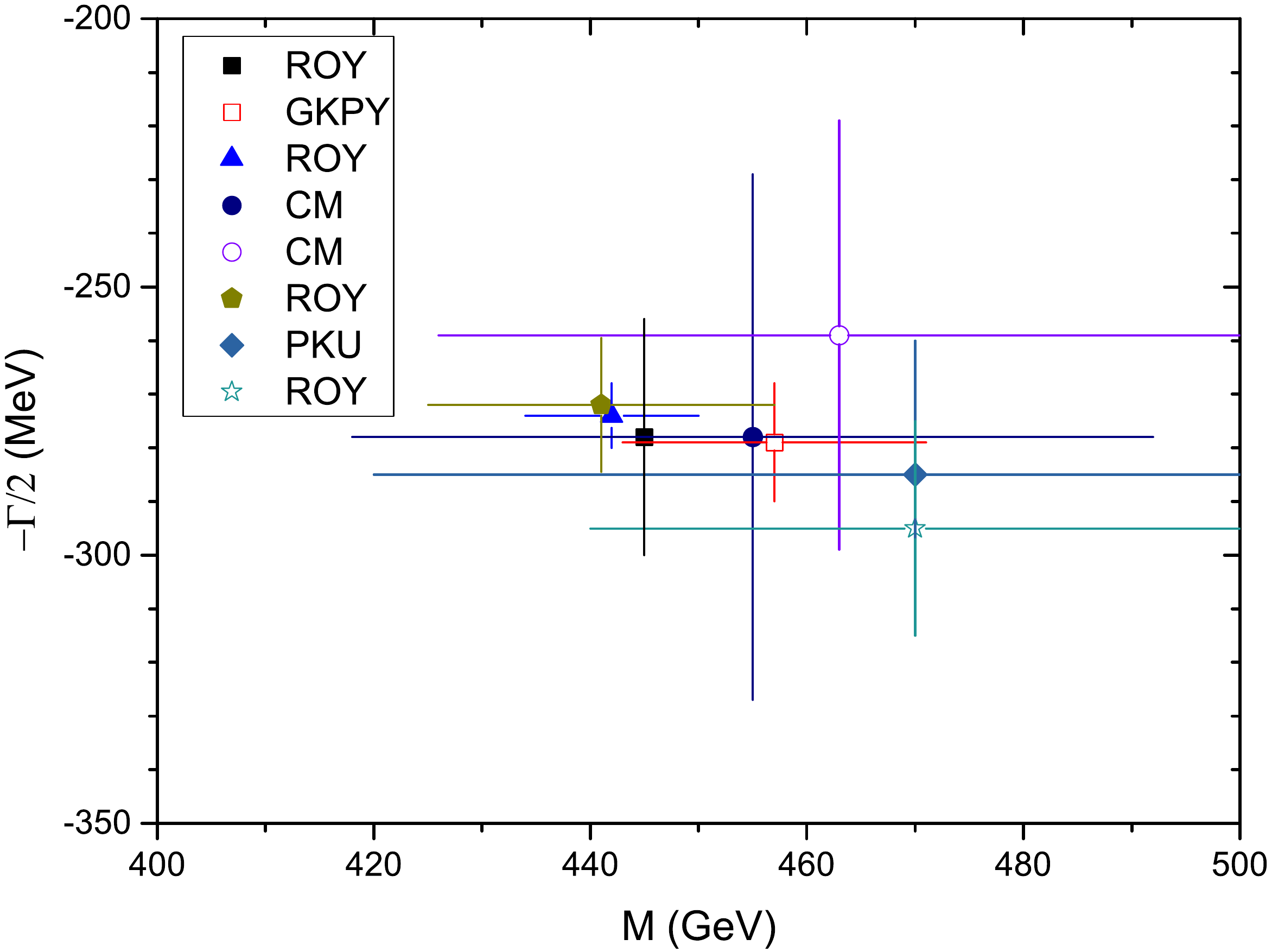}
\caption{\label{Fig:sigma} The pole locations of the $\sigma$ in the $T$ matrix. The result in the left graph is taken from PDG \cite{Zyla:2020zbs}, and the ones shown in the right graph are taken from Refs.\cite{Caprini:2005zr,Ananthanarayan:2000ht,GarciaMartin:2011jx,Zhou:2004ms,Moussallam:2011zg,Caprini:2008fc} in order. The label \lq CM' represents the conformal mapping formalism.  }
\end{figure}

The domain where Roy equation works for $\pi\pi$ scattering is restricted in the region of Fig.\ref{Fig:roy}.
It is the region where the partial wave expansion of Roy equations is convergent. The details to calculate the domain of the Roy equation can be found in Ref.\cite{Caprini:2005zr}, and earlier discussions about using the axiomatic field theory can be found in Refs.~\cite{Lehmann:1958ita,Martin:1965jj,Martin:1969ina,Mahoux:1974ej}.
It should satisfy the equation of Lehmann-Martin ellipse \cite{Lehmann:1958ita,Martin:1965jj,Martin:1969ina,Mahoux:1974ej}:
\begin{eqnarray}
&&\frac{\left[x+\frac{s'-4m_\pi^2}{2}\right]^2}{\left[\frac{s'-4m_\pi^2}{2}+\frac{16m_\pi^2 s'}{s'-4m_\pi^2}\right]^2}
+\frac{y^2}{16m_\pi^2 s'+\left(\frac{16m_\pi^2 s'}{s'-4m_\pi^2}\right)^2}=1 \,, \nonumber\\
&&\frac{\left[x+\frac{s'-4m_\pi^2}{2}\right]^2}{\left[\frac{s'-4m_\pi^2}{2}+\frac{16m_\pi^2 s'}{s'-4m_\pi^2}\right]^2}
+\frac{y^2}{\frac{4m_\pi^2 s'(s'-4m_\pi^2)}{s'-16m_\pi^2}+\left[\frac{4m_\pi^2 s'}{s'-16m_\pi^2}\right]^2}=1 \,, \nonumber
\label{eq:roy;LME}
\end{eqnarray}
with $x={\rm Re}~t$ and  $y={\rm Im}~t$. The first equation works for $4m_\pi^2\leq s' \leq 20 m_\pi^2$ and the second one works for $s'>20 m_\pi^2$.
From it one sees clearly that the $\sigma$ poles found by dispersion relations locates inside the domain.
Once the accurate partial wave is available, one could extract the precise pole location of the resonance from it.
For example, it is at $441^{+16}_{-8}-i272^{+9}_{-12.5}$~MeV \cite{Caprini:2005zr}, $445\pm25-i278^{+22}_{-18}$~MeV \cite{GarciaMartin:2011jx}, and $442^{+5}_{-8}-i274^{+6}_{-5}$~MeV \cite{Moussallam:2011zg} by Roy equations.
Prediction of the scattering length is also an important result of Roy equations, see e.g. Refs.\cite{Ananthanarayan:2000ht,Pelaez:2015qba,Basdevant:1973ru}.
And for the pioneering works about it, see references in an early compilation \cite{Nagels:1979xh}.
Another analysis with less subtractions could find the $\sigma$ at $457^{+14}_{13}-i278^{+11}_{-7}$~MeV \cite{GarciaMartin:2011jx} in the RS-II. The pole locations of the $\sigma$ by Roy equations are shown in  Table \ref{tab:sigma} and Fig.\ref{Fig:sigma}.
\begin{table}[bt]
\caption{\label{tab:sigma} Pole locations of the $\sigma$ in dispersive approaches, in units of MeV. }
\hspace{0.5cm}{\footnotesize\begin{tabular}{@{}cccc}
\br
                    PKU\cite{Zhou:2004ms}   & ROY & GKPY  \cite{GarciaMartin:2011jx}  & CM  \cite{Caprini:2008fc}   \\
\mr
$470\pm50-i285\pm25$   & $441^{+16}_{-8}-i272^{+9}_{-12.5}$\cite{Caprini:2005zr}
       & $457^{+14}_{13}-i278^{+11}_{-7}$   &  $455\pm6^{+31}_{-13}-i278\pm6^{+34}_{-43}$  \\
      &  $445\pm25-i278^{+22}_{-18}$\cite{GarciaMartin:2011jx}  &     & $463\pm6^{+31}_{-17}-i259\pm6^{+33}_{-34}$  \\
      & $442^{+5}_{-8}-i274^{+6}_{-5}$\cite{Moussallam:2011zg}  &     &  \\
\br
\end{tabular}}
\end{table}
For the graph on the left side of Fig.\ref{Fig:sigma}, the pole locations of the $\sigma$ are given by PDG \cite{Zyla:2020zbs}.
It is shown that most of models find the $\sigma$ poles roughly in the region of $M_\sigma \sim 400-550$~MeV and $\Gamma_\sigma/2\sim 200-350$~MeV, as shown in PDG \cite{Zyla:2020zbs}. For the poles found with much higher masses (for instance, 1500MeV), they are indeed caused by the puzzle $\sigma-f_0(1370)$ mixing, we refer to Refs.\cite{Jaffe:2004ph,Godfrey:1998pd,Pelaez:2016klv} and references therein for discussions about it.

 For the poles found with much larger widths, it is even outside the region where the dispersive tools work and thus it is beyond our knowledge.
Taking into account all these constraints (such as symmetry and other principles of QFT), we choose some results of dispersive approaches to give an impression where the pole locations are most likely to be\footnote{We stress that we can not take all the results and only latest dispersive approaches are included. }. From the right graph, a conservative estimate of the $\sigma$ pole location in the RS-II is in the region of $M_\sigma \sim 430-470$~MeV and $\Gamma_\sigma/2\sim 260-290$~MeV. Note that the high precision pole locations also confirm the existence of the $\sigma$ (and also $\kappa$).
The pole information of the $\sigma$ are also used to check whether an EFT works well or not. For instance, in the linear $\sigma$ model the T matrix pole of the $\sigma$ is not the same as what is obtained from Roy equation analyses.
It is also worth pointing out that the low energy constants $l_i$  of the SU(2) $\chi$PT can be well constrained by matching the $\chi$PT amplitudes and those by the Roy equations \cite{Colangelo:2001df}. Especially the $l_3$ could be important to understand the chiral quark condensate.

Of course, a view from the whole meson spectrum might shed more light on the nature of these states including the $\sigma$ resonance. There is another idea that the $\sigma$ and $f_0(1370)$ could have some connections so that they can not be treated independently~\cite{vanBeveren:1983td,Tornqvist:1995kr,Zhou:2010ra,Boglione:2002vv,Wolkanowski:2015lsa,Wolkanowski:2015jtc,Badalian:2020wua,zhou:2020,Zhou:2020pyo}, which is illustrated in a clear way in a recent progress of a relativistic Friedrichs-Lee~(FL) scheme combined with the relativistic quark pair creation~(QPC) model~\cite{zhou:2020,Zhou:2020pyo}. Although it is a model calculation, unitarity is satisfied and a dispersive relation function can be obtained by its dynamics.

In the relativistic Friedrichs-Lee-QPC scheme, it is found that, due to coupling between the scalar $(u\bar u+d\bar d)/\sqrt{2}$ state in the quark potential model~\cite{Godfrey:1985xj} and $\pi\pi$ continuum state with the same quantum numbers, $\sigma$ and $f_0(1370)$ are produced at the same time, which is called two-pole structure in~\cite{Zhou:2020pyo}. Although the pole mass of $\sigma$ is not so accurate as the values obtained from the dispersive techniques mentioned above, it has only several parameters to explain the existence of about twenty states. In this picture, $f_0(1370)$ is shifted from the bare $(u\bar u+d\bar d)/\sqrt{2}$  state, while the $\sigma$ is dynamically generated from the coupling form factor.

Besides the poles' positions, the scheme also presents a sum rule for phase shifts contributed by the two-pole structure in single-channel approximation. Usually, there is no constraint for two independent states. However, it is pointed out in this scheme, if the two pairs of poles are caused by the same bare state coupling to the continuum state, they will contribute a total phase shift of $180^\circ$.
The phase shift data of the $IJ=00$ $\pi\pi$ scattering provide
some hints
to this sum rule, because it is quite complicated for being contributed
by both the states generated from  $(u\bar u+d\bar d)/\sqrt{2}$ and
$s\bar s$ states. Suppose that the sharp-rising contribution of $f_0(980)$ be removed, which is suggested to be dynamically generated state from $s\bar s$ state, the $IJ=00$ phase shift of $\pi\pi$ scattering up to 1.6 GeV is about $180^\circ$.

In a short summary, $\chi$PT combined with dispersion relations could confirm the existence of the $\sigma$. Furthermore, the high statistics datasets referring to $\pi\pi$ final states are available and the dispersion relation works well to do the partial wave decomposition. These fix the precise pole location of the $\sigma$ in the RS-II. This is a big progress as the $\sigma$ has a wide width and its mass is not faraway from the $\pi\pi$ threshold.

\subsubsection{$\kappa$: from $\pi K$ scattering}\label{sec:kappa}
One key reason why the $\kappa$ attracted people's attention is that one needs to know the $\kappa$ to study the property the $\sigma$.
In standard nonet description of the lightest scalars ($q\bar{q}$ state), the $\kappa$ should appear as the adjoint resonance of the $\sigma$ with strange quark component.
In another aspect, if the lightest scalar nonet are $q\bar{q}$ states, the mass of the $\kappa$ should be 200~MeV heavier than the isovector $a_0$ state. However, the lightest isovector $0^{++}$ state is the $a_0(980)$, so the contradiction of the mass order is obvious. While if the lightest scalar nonet are interpreted as tetra-quark states, the mass order could be changed and the $\kappa$ is in the right position together with the other scalars \cite{Jaffe:2004ph}.
Hence, to study the property of the lightest scalars, one needs to clarify two things: one is to confirm their existence, e.g. the $\kappa$ state; the other one is to exact the pole information exactly.

It would be helpful to give a brief review of the history of the $\kappa$.
In Ref.\cite{GellMann:1962xb}, the $I=1/2$ scalar $K'$ ($\kappa$) meson is supposed.
Later it is studied by the experiments~\cite{Alexander:1962zzb,FerroLuzzi:1964jr} and by theoretical efforts~\cite{Sakurai:1963iz,Roy:1968nv}. Though these studies are not very successful to determine the width of the $\kappa$, they give evidences to suggest its existence.
In Ref.\cite{Jaffe:1976ig,Jaffe:1976ih}, the $\kappa$ is suggested to be a tetraquark state.
In Ref.\cite{Ishida:1997wn}, the interfering amplitude method is used to fit to the $\pi K$ phase shift and the unitarity is implemented. They show strong evidence of the existence of the $\kappa$ and give the resonance information as $M=905^{+65}_{-30}$~MeV, $\Gamma=470^{+185}_{-90}$~MeV.
Note that they are Breit-Wigner mass and width so that they look to be quite different from the pole information shown in PDG table~\cite{Zyla:2020zbs}.
In the analysis, they introduce a repulsive background contributing negative phase shift. Hence they need the $\kappa$ resonance to compensate the background and fit to the experimental phase shift well. Indeed this background is in compatible with the contribution of the l.h.c. discussed in PKU factorization~\cite{Xiao:2000kx,Zheng:2003rw}. In Ref.~\cite{Cherry:2000ut}, it rules out the $\kappa$ with the pole mass larger than 825~MeV with a model based on conformal mapping. It should be noted that this is better interpreted as a strong constraint on the upper limit of the mass, and it does not contradict with recent dispersive analyses where the pole mass is exactly lower than the limit.

\begin{figure}[hpt]
\includegraphics[width=0.48\textwidth,height=0.3\textheight]{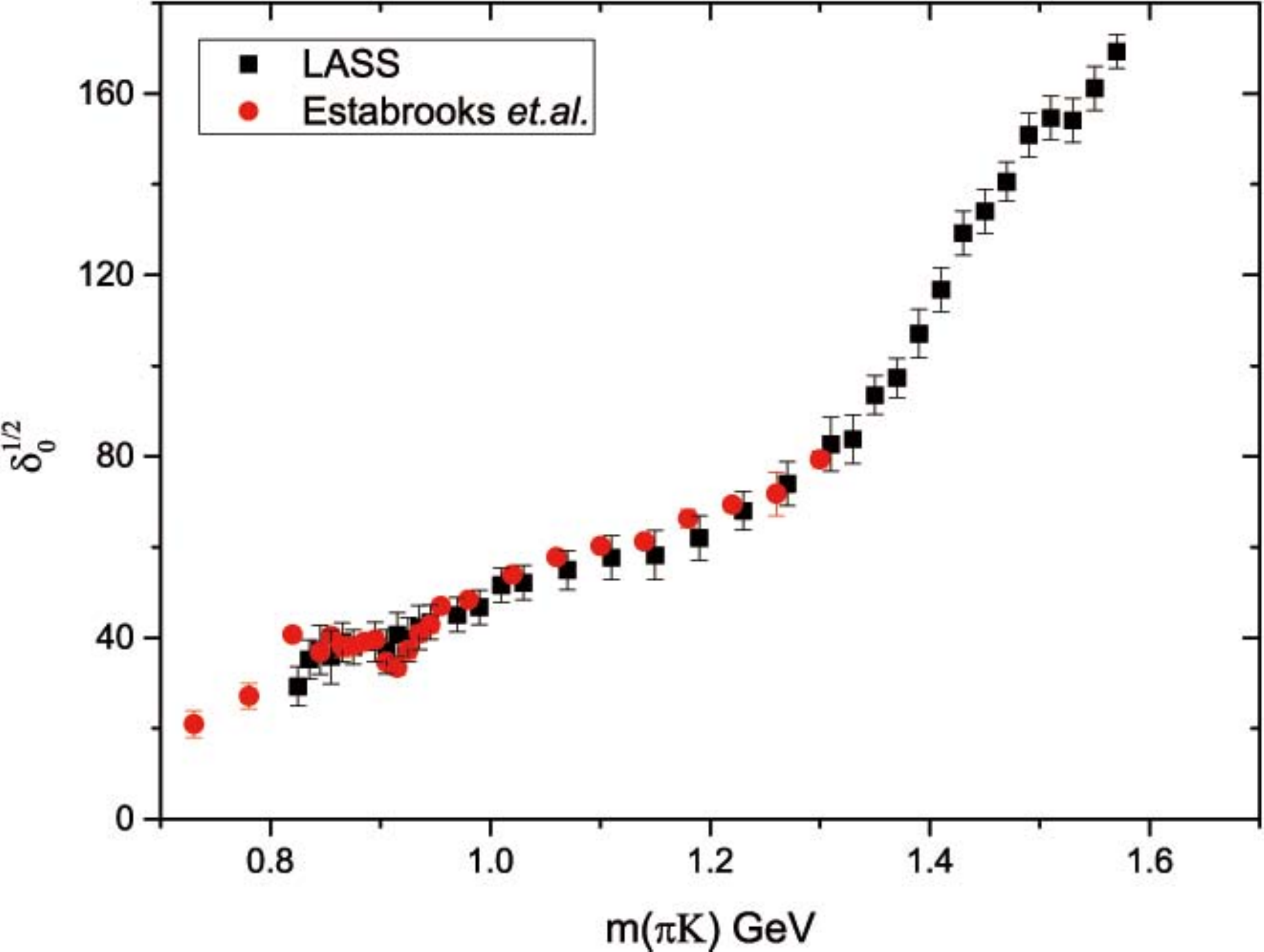}
\includegraphics[width=0.48\textwidth,height=0.3\textheight]{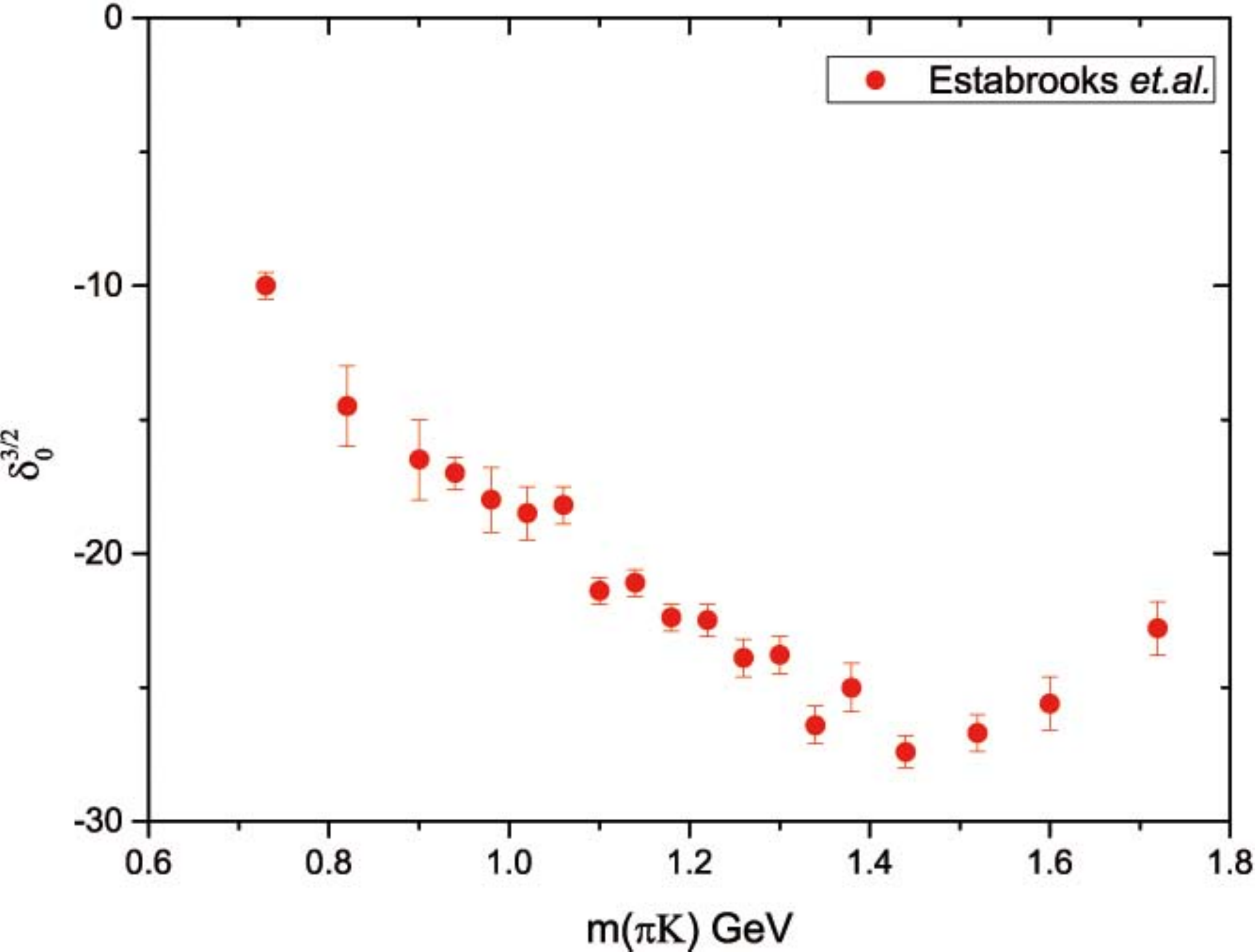}
\caption{\label{Fig:ph;piK} The $I=1/2, 3/2$ S-wave phase shifts of $\pi K$ scattering. The LASS phase shifts are from \cite{Aston:1987ir}, and that of the Estabrooks {\it et.al.} data is from~\cite{Estabrooks:1977xe}. }
\end{figure}

On the experimental side, the data from LASS~\cite{Aston:1987ir} and Ref.~\cite{Estabrooks:1977xe} gives phase shifts of $\pi K$ scattering. Most analysis of different theoretical models are based on these data. The more recent data from heavy meson decays could also be helpful to study the $\kappa$, for instance, D meson decays of $D^+\to \pi^-\pi^+\pi^+$ from E791~\cite{Aitala:2002kr} and $D^+\to K^-\pi^+\pi^+$ from CLEO~\cite{Bonvicini:2008jw},  and $J/\psi$ decays of $J/\psi \to \bar{K}^{*0}(892) K^+ \pi^-$ and  $J / \psi \to \bar{K}^{*\mp}(892) K_s \pi^\pm$ from BES/BESII~\cite{Bai:2003fv,Ablikim:2005ni,Ablikim:2010kd,Ablikim:2010ab}.
These give evidences about the existence of the $\kappa$ and they could also be used to refine the theoretical analysis due to the high statistics.
The FOCUS experiment \cite{Link:2002ev} presented that there exists a coherent $\pi K$ S-wave from the weak decay of $D$ mesons $D^+\to K^-\pi^+\mu^+\nu$. It helps to determine the $\pi K$ S-wave phase shift below 1~GeV.
Some data of the phase shifts as discussed above are shown in Fig.~\ref{Fig:ph;piK}.
According to the crossing symmetry, the $\pi K \to \pi K$ scattering amplitude is correlated with the $\pi\pi\to K\bar{K}$  scattering amplitude. One thus needs the high-statistics phase shifts and inelasticity data of $\pi\pi\to K\bar{K}$ \cite{Cohen:1980cq,Etkin:1981sg}. This has already been shown in Fig.\ref{Fig:ph;pipi}, where the $\pi\pi-K\bar{K}$ coupled channel scattering is discussed.

Having these data, one still needs a reliable tool to extract out the resonance information. The ratio of $\Gamma/M$ is still large, but it is much smaller than that of the $\sigma$. This could be the reason why the $\kappa$ can be found without dispersive approaches in some paper \cite{Ishida:1997wn,Ablikim:2005ni,Ablikim:2010kd}, but not in some other papers~\cite{Tornqvist:1995kr,Anisovich:1997qp}.
Obviously the $K$ and $\eta$ mesons containing the $s$ quark and one needs the the $SU(3)$ or even $U(3)$ chiral symmetry. The relative U$\chi$PT approach of the $\pi K$ scattering is helpful to study the the property of the $\kappa$.
The $\mathcal{O}(p^2)$ partial wave decomposition can be obtained by applying Eq.~(\ref{eq:Roy;t}), with
\begin{eqnarray}
T_{2}^{\frac{1}{2}S}(s)&=&\frac{5s^2-2s(M_K^2+M_\pi^2)-3(M_K^2-M_\pi^2)^2}{128 \pi s F_\pi^2} \,,\nonumber \\
T_{2}^{\frac{3}{2}S}(s)&=&\frac{M_\pi^2+M_K^2-s}{32\pi F_\pi^2}\,.
\end{eqnarray}
The $SU(3)$ one-loop meson-meson scattering amplitudes of $\pi K-\eta K$ coupled channels with $\overline{MS}-1$ scheme are given in Refs.\cite{Bernard:1991zc,GomezNicola:2001as}.
For example, in Ref.\cite{Pelaez:2003rv} it finds the $\kappa$ pole around $750$~MeV.
In Ref.~\cite{Dai:2011bs}, a $\kappa$ pole is found around $670$~MeV in RS-II, while in Ref.~\cite{Dai:2012kf} two poles are found both in the RS-II and RS-III. The twin pole structure might suggest a Breit-Wigner origin component.
In Ref.\cite{Guo:2011pa}, the $U(3)$ $\chi$PT is used and the $\kappa$ pole location is close to that of recent Roy-like equations~\cite{Pelaez:2020uiw}.
In all these paper, the pole in the RS-II has been found after unitarization,  the masses and widths of the $\kappa$ are close to that of the Roy-Steiner equations, with deviations up to 100~MeV.
These also give clues of the existence of the $\kappa$, since no such resonance is directly written in the chiral Lagrangian.

As discussed before, pole parameters are different from the Breit-Wigner mass and width.
The latter in different processes could vary a bit as the background could be different, but the former is unique and it should be the same in any processes.
We will focus on the dispersion relation approach in the following sections, which is almost the only model independent way to extend the amplitude to the complex energy plane and extract out the pole parameters exactly.
In Ref.\cite{Zheng:2003rw}, PKU factorization is applied and gives a strong evidence of the existence of the $\kappa$.
Like what is done for the $\sigma$, the physical $S$-matrix is expressed as $S=S^{cut}\cdot \prod_i S_i^{pole}$, and the cut contribution is parameterized as $S^{cut}=e^{2i\rho(s)f(s)}$ and $f(s)$ is written down as
{ \setlength{\mathindent}{1.5cm} \begin{eqnarray}
f(s)&=&f(s_0)+\frac{s-s_0}{2\pi i}\int_L \frac{{\rm disc}_L f(s')d s'}{(s'-s)(s'-s_0)} +\frac{s-s_0}{\pi}\int_R \frac{{\rm Im}_R f(s')d s'}{(s'-s)(s'-s_0)}\,.  \label{eq:pku;pik;f}
\end{eqnarray} }
For simplicity the subtraction point can be chosen at $s=0$, and it is proved that $f(0)\equiv 0$~\cite{Xiao:2005rg}.

The difference comparing with the case of $\pi\pi$ is that now in \lq $L$' there is an extra circular cut except for the l.h.c..
In this way, not only unitarity and analyticity, but also crossing symmetry are included through the $\chi$PT estimation on the l.h.c. as well as the circular cut.
Again it finds that the l.h.c. part can only support the negative phase shift and thus one needs the $\kappa$ to fit the phase shift well.
According to their analysis, the conclusion is model independent supposing that the scattering length of $(I,J)=(1/2,0)$ does not deviate too much from that of $\chi$PT. Later, the refined analysis \cite{Zhou:2006wm} is done, with the subtraction point is chosen at $s_0=0$.
The pole location of the $\kappa$ is found to be at  $694\pm53-i303\pm30$~MeV.

\begin{figure}[hpt]
\centering
\includegraphics[width=0.6\textwidth,height=0.3\textheight,angle=0]{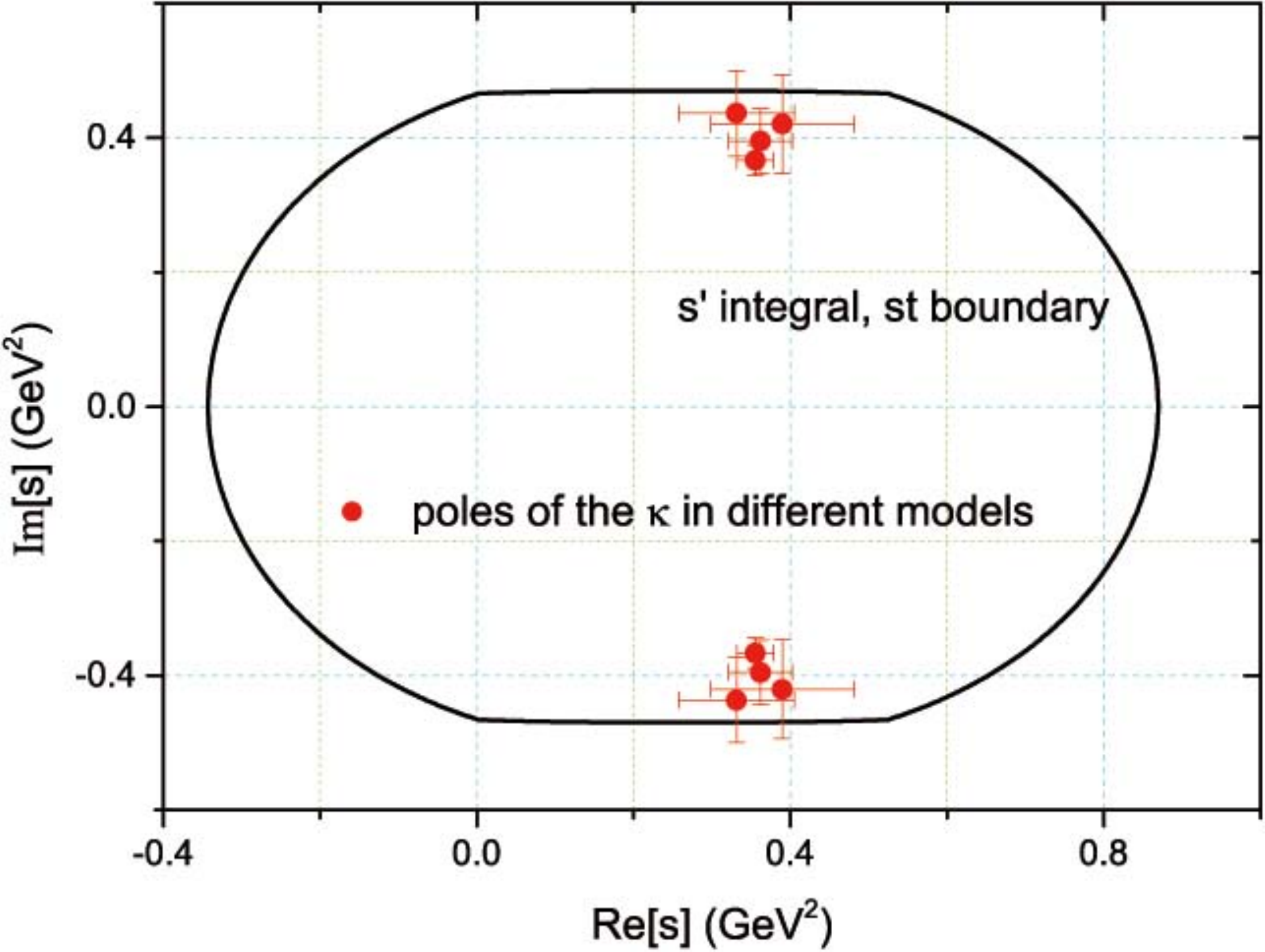}
\caption{\label{Fig:royS} The domain where Roy-Steiner equation works.  It is calculated by the Lehmann-Martin ellipse.   The unit is $m_\pi^2$. The domain is calculated following Ref.\cite{DescotesGenon:2006uk}, adding the $\kappa$ pole locations given by different models. The earlier discussion about the domain can be also found in Refs.~\cite{Lehmann:1958ita,Martin:1965jj,Martin:1969ina,Mahoux:1974ej}. The poles are taken from Refs.~\cite{DescotesGenon:2006uk,Pelaez:2016klv,Zhou:2006wm,Bugg:2009uk}.
 }
\end{figure}

To date, a rather precise work to determine the $\kappa$ pole is done by the Roy-Steiner equation \cite{DescotesGenon:2006uk,Buettiker:2003pp}.
In their analysis, the fixed-$us$ approach ($su=b$) is used to write down the once subtracted dispersion relation for the $\pi K \to \pi K$ amplitude $F^+(s,t,u)$ as\cite{DescotesGenon:2006uk}
{ \setlength{\mathindent}{1.cm} \begin{eqnarray}
F^+(s,t_b(s))&=& f^+(b) + t_b(s) h^+(b) \nonumber\\
&+& {1\over\pi} \int_{m_+^2}^\infty ds'\left[
 {2s'-2\Sigma+t_b(s)\over (s'-s)(s'-b/s)}-{2s'-2\Sigma-t_b(s)\over s^{'2}-2\Sigma s'+b} \right]{\rm Im}_s F^+(s',t_b(s'))
\nonumber\\
&+& {t_b(s)^2\over \pi}\int_{4m_\pi^2}^\infty{dt'\over t^{'2}(t'-t_b(s))}{\rm Im}_t F^+(s'_b(t'),t')\,, \label{eq:RS;su}
\end{eqnarray} }
with
\begin{eqnarray}
t_b(s)= 2\Sigma-s -{b\over s},\quad  s'_b(t')= {1\over2} \left(2\Sigma-t' +\sqrt{ (2\Sigma-t')^2-4b}\right)\ . \nonumber
\end{eqnarray}
And it is similar for the  $F^-(s,t,u)$.
The discontinuity functions ${\rm Im}_s F^+(s',t_b(s'))$ and ${\rm Im}_t F^+(s'_b(t'),t')$ can be decomposed according to the partial wave projection. They relate to $\pi K\to\pi K$ and $\pi\pi\to \bar{K}K$ scattering, respectively. After decomposition, the partial waves are obtained
{ \setlength{\mathindent}{2.cm} \begin{eqnarray}
f_0^{1\over2}(s) &=& {1\over2} m_+ a_0^{1\over2} + {1\over12} m_+ (a_0^{1\over2}
-a_0^{3\over2}) { (s-m_+^2)(5s +3m_-^2)\over (m_+^2-m_-^2)\, s } \nonumber\\
&+&{1\over\pi}\int_{m_+^2}^\infty ds' \sum_{l=0}^\infty \left\{ K_{0l}^{1\over2} (s,s') {\rm Im} f^{1\over2}_l(s') + K_{0l}^{3\over2} (s,s') {\rm Im} f^{3\over2}_l(s') \right\} \nonumber\\
&+&{1\over\pi}\int_{4m_\pi^2}^\infty dt' \sum_{l=0}^\infty \left\{   K_{0 2l}^0(s,t'){\rm Im} g^0_{2l}(t')
+  K_{0 2l+1}^1(s,t'){\rm Im} g^1_{2l+1}(t') \right\}\ , \label{eq:RS;PW}
\end{eqnarray} }
where $f_l^I(s)$ is the partial wave of $\pi K\to \pi  K$ and $g_l^I$ is that of the $\pi\pi\to\bar{K}K$. $a_0^I$ is the scattering length and $K^I_{ij}(s,s')$ is the kernel function can be found in Ref.\cite{DescotesGenon:2006uk}. The $\pi K$ scattering amplitudes are also rather important to constrain the LECs in SU(3) $\chi$PT, see e.g. Ref.\cite{Buettiker:2003pp}. Indeed the LECs of $L_{1,2,3,4}$ are predicted by matching the subthreshold expansion parameters of Roy-Steiner equations with those of $\chi$PT. The amplitudes can be calculated directly in the complex-$s$ plane as the analyticity is kept by dispersion relation.
The working domain of the Roy-Steiner equation (fixed-$us$) is shown in Fig.\ref{Fig:royS}.
It is worth noting that in the fixed-$us$ dispersion relation, the working domain is significantly enlarged comparing with the fixed-$t$ dispersion relation along the ${\rm Im}s$ direction.
As shown in Ref.~\cite{DescotesGenon:2006uk}, the $\kappa$ pole, for instance with the pole location at $0.356\pm0.024-i0.366\pm0.023$~GeV$^2$ (or $658\pm13-i279\pm12$~MeV)~\footnote{We simply use the error propagation formula to estimate the uncertainty by changing the number of $M-i \Gamma/2$ in Ref.\cite{DescotesGenon:2006uk} into the form of ${\rm Re}s-i{\rm Im}s$. }, is outside the working domain of the fixed-$t$ dispersion relation. This problem has been solved by using the fixed-$us$ one.
For ${\rm Re}s=0.356$GeV~$^2$, the boundary of the working domain is at ${\rm Im}s\thickapprox-0.48$GeV~$^2$, fairly below the imaginary part of the pole.
Indeed in most of the dispersive researches it is found that the $\kappa$ pole locates around the $\pi K$ threshold (roughly 20.73$M_\pi^2$ or $0.404$~GeV$^2$), and the boundary of the working domain is roughly at $|{\rm Im}s|\sim 0.40- 0.48$GeV$^2$, being able to keep the pole location of the $\kappa$ inside.
It is worth pointing out that the S-wave was not well described below the matching point for the lack of experimental data in the low energy region, see Fig.\ref{Fig:ph;piK}. Nevertheless, unitarity combined with Roy-Steiner equations compensate for it, hence an accurate pole location extraction of the $\kappa$ is available.

\begin{figure}[hpt]
\includegraphics[width=0.48\textwidth,height=0.3\textheight]{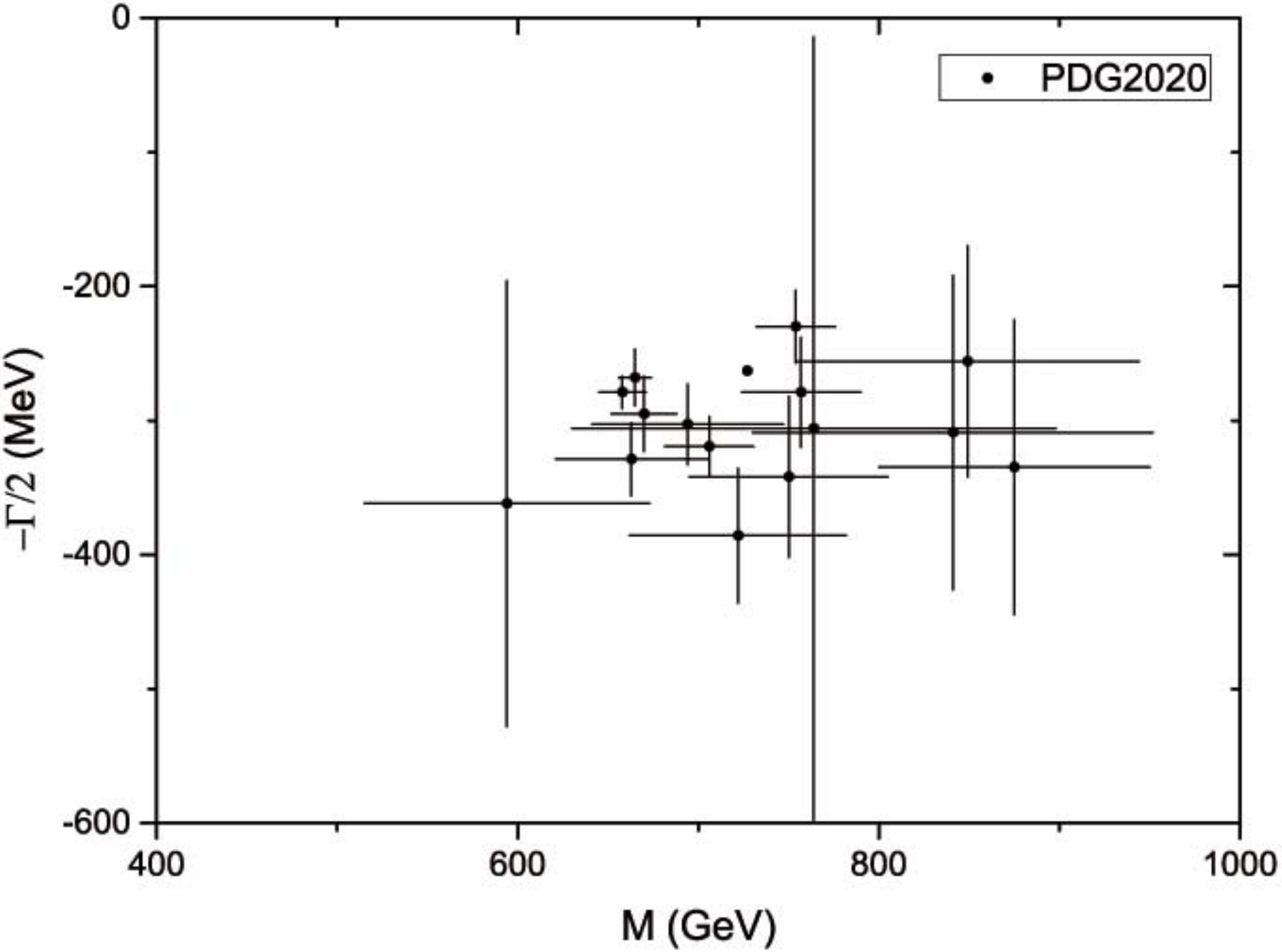}
\includegraphics[width=0.48\textwidth,height=0.3\textheight]{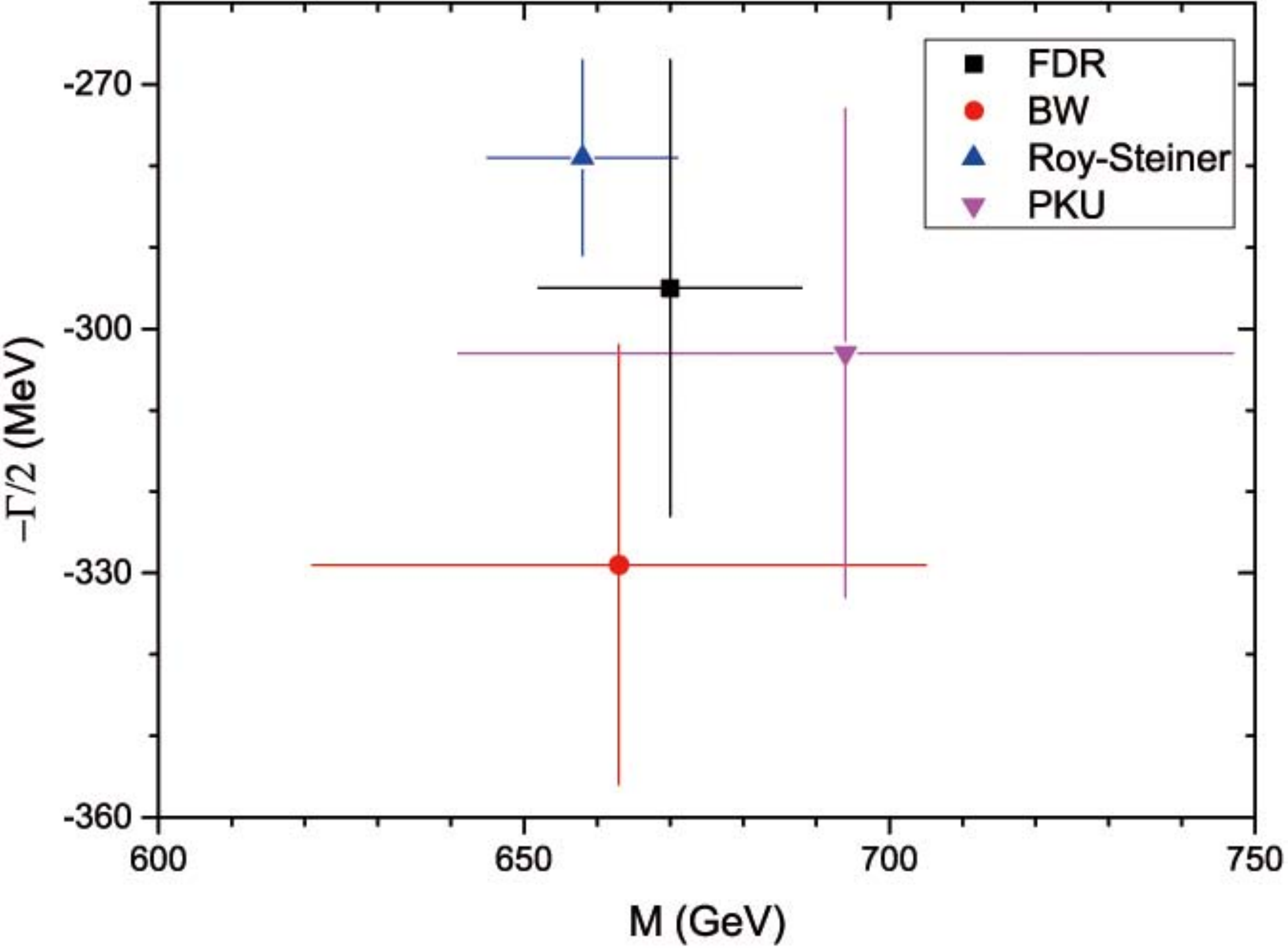}
\caption{\label{Fig:kappa} The pole locations of the $\sigma$ in the $T$ matrix. The result in the left graph is taken from PDG \cite{Zyla:2020zbs}, and the ones shown in the right graph are taken from Refs.\cite{DescotesGenon:2006uk,Pelaez:2016klv,Zhou:2006wm,Bugg:2009uk} in order.   }
\end{figure}

The pole locations of the $\kappa$ given by PDG \cite{Zyla:2020zbs} are shown in Fig.\ref{Fig:kappa}.
We also choose some dispersive approach to give an impression where the pole locations are. They are shown in Table \ref{tab:kappa} and also in the right side graph of Fig.\ref{Fig:kappa}.
\begin{table}[bt]
\caption{\label{tab:kappa} Pole locations of the $\kappa$ in dispersive approaches, in units of MeV. }
\hspace{0.8cm}{\footnotesize\begin{tabular}{@{}cccc}
\br
PKU\cite{Zhou:2006wm}   & Roy-Steiner\cite{DescotesGenon:2006uk} & Roy-like \cite{Pelaez:2020uiw}  & FDR  \cite{Pelaez:2016klv}   \\
\mr
$694\pm53-i303\pm30$   & $658\pm13-i279\pm12$  & $650\pm7-i279\pm16$   & $670\pm18-i295\pm28$  \\
\br
\end{tabular}}
\end{table}
It is crude to ignore some of the works, but we can only include some latest ones with dispersive approach.
For example, the pole of the $\kappa$ is at $658\pm13-i279\pm12$~MeV by Roy-Steiner equation \cite{DescotesGenon:2006uk}.
Very recently, the unsubtracted and subtracted  Roy-like equations with hyperbolic (and also fixed-$t$) dispersion relation are applied to the $\pi K$ scattering and they find it to be $650\pm7-i279\pm16$~MeV  \cite{Pelaez:2020uiw}.
A conservative estimation of the pole location of the $\kappa$ in the RS-II is in the region of $M\sim630-710$~MeV and $\Gamma/2\sim260-320$~MeV.
The pole locations of the $\kappa$ are taken from Refs.\cite{DescotesGenon:2006uk,Pelaez:2016klv,Zhou:2006wm,Bugg:2009uk}.

There are also other interesting studies referring to the $\kappa$. In Ref.\cite{vanBeveren:1983td,Badalian:2020wua,Zhou:2020pyo}, $\kappa$ and the $K_0^*(1430)$ both come from the $\bar d s$ state coupling to the $IJ=1/2\ 0$ $\pi K$ continuum. $\kappa$ appears as a dynamically generated pole and the bare $\bar d s$ is shifted to become the $K_0^*(1430)$, and both pole's positions are in coincidence with the values obtained from the dispersive techniques or experiments. More interesting observation is that the phase shift of $IJ=1/2\ 0$ $\pi K$ scattering exhibit a total phase shift of about $180^\circ$ up to about 1.7 GeV, as shown in Fig.~\ref{Fig:ph;piK}, which is consistent with the sum rule of two-pole structure in Ref.~\cite{Zhou:2020pyo}. In Ref.~\cite{Giacosa:2018vbw}, whether the $\kappa$ should be included into thermal hadronic gas is discussed. It is found that its thermodynamical property is canceled by a repulsion in the $I=3/2$ channel. Thus one could just ignore it at nonzero temperature.
For the latest lattice study, we refer readers to~Refs.\cite{Fu:2011xb,Wilson:2019wfr,Rendon:2020rtw} and references therein. A combination of dispersive approach with lattice study is necessary to continue the $\pi K$ S-wave amplitude to the complex-$s$ plane.

\subsubsection{$f_0(980)$: from $\pi\pi-\bar{K}K$ coupled channels } \label{sec:f980}
The $f_0(980)$ is much more confident comparing with the $\sigma$. One reason is that its width is much smaller and also there is a distinct rapid growth of the phase shift around the $\bar{K}K$ threshold, which is a normal characteristic property of a narrow resonance. Of course, one needs to study the $\pi\pi-\bar{K}K$ coupled channel and the coupled channel unitarity should to be implemented.
The $K$-matrix is an explicit and easy way to implement the coupled channel unitarity.  Ref.~\cite{Morgan:1993td} use $K$-matrix and AMP method to implement the coupled channel unitarity and the theorem of FSI. It fits to the $\pi\pi$ phase shifts and inelasticities as well as various production processes. The width of the $f_0(980)$ is found to be narrow as $988\pm10-i24\pm6$~MeV in RS-II. They also found one pole in Fit 1 and twin poles in more favor Fit 2.
The situation is quite the same as done in Ref.~\cite{Zou:1993az}, the $T$ amplitude is represented similar to Dalitz-Tuan~\cite{Dalitz:1960du} formalism and the AMP method is used to construct the central production amplitude $pp\to pp(\pi\pi,K\bar{K})$. It separates the pole term for $f_0(980)$ from the background term, and finds a RS-II pole very close to that of Ref.~\cite{Morgan:1993td}. It also finds a faraway pole in the RS-III.
In the Roy equations, the coupled channel unitarity is not implemented. But luckily the $f_0(980)$ is just around the $\bar{K}K$ threshold and it is inside the working domain of the Roy equations, see Fig.~\ref{Fig:roy}. In Ref.~\cite{GarciaMartin:2011jx}, the pole is found at
$1003^{+5}_{-27}-i~21^{+10}_{-8}$~MeV by Roy equations and $996^{+7}_{-7}-i~25^{+10}_{-6}$~MeV with one less subtraction. In Ref.\cite{Moussallam:2011zg} the pole is found at $996^{+4}_{-14}-i~24^{+11}_{-3}$~MeV by Roy equations. All of the poles are in the RS-II as only $\pi\pi$ single channel is included.
We listed all these pole locations as well as those from other models in Table \ref{tab:f980}.
\begin{table}[bt]
\caption{\label{tab:f980} Pole locations of the $f_0(980)$ in Riemann sheet II from different models, in units of MeV. }
{\footnotesize\begin{tabular}{c@{}c@{}c@{}c}
\br
 U$\chi$PT        & K-matrix & Roy/Roy-like & other models      \\
\mr
$994-i14$\cite{Oller:1998hw}   & $988\pm10-i24\pm6$\cite{Morgan:1993td}   &  $996^{+4}_{-14}-i~24^{+11}_{-3}$\cite{Moussallam:2011zg}  &    $993.2\pm6.5\pm6.9-i50$\cite{Ishida:1995xx} \\
$973^{+39}_{-127}-i11^{+189}_{-11}$\cite{Pelaez:2003rv}  & $988-i23$\cite{Zou:1993az}
 &  $996^{+7}_{-7}-i~25^{+10}_{-6}$\cite{GarciaMartin:2011jx}   & $1015-i15$\cite{Janssen:1994wn} \\
  $974-i25$\cite{Dai:2011bs}   &   $998\pm3-i21\pm3$ \cite{Dai:2014lza}
 &  $1003^{+5}_{-27}-i~21^{+10}_{-8}$\cite{GarciaMartin:2011jx}   &  $982.13-i21.67$\cite{Ahmed:2020kmp}   \\
   $981^{+9}_{-7}-i22^{+5}_{-7}$\cite{Guo:2011pa}   &
 &     &  $997.7\pm1.1-i21.7\pm1.9$\cite{Dai:2017tew}  \\
  &  &     &   $997.6\pm0.1-i20.3\pm1.3$\cite{Dai:2019zao}     \\
\br
\end{tabular}}
\end{table}

The pole of the $f_0(980)$ is quite close to the real axis, and thus the requirement on the analyticity could be properly \lq reduced'.  Following it the unitarized $\chi$PT is a good way to analyze the amplitude and extract out the information of the resonances, with unitarity implemented. Taking Pad\'e approximation as an example, the coupled-channel unitarity condition is given as
\begin{eqnarray}\label{unitarity}
\mathrm{Im} T_{11}&=&T_{11}\rho_1 T_{11}^{*}\theta(s-4m_{\pi}^2)+T_{12}\rho_2 T_{21}^{*}\theta(s-4m_{K}^2)\ ,\nonumber\\
\mathrm{Im} T_{12}&=&T_{11}\rho_1 T_{12}^{*}\theta(s-4m_{\pi}^2)+T_{12}\rho_2 T_{22}^{*}\theta(s-4m_{K}^2)\ ,\nonumber\\
\mathrm{Im} T_{22}&=&T_{21}\rho_1 T_{12}^{*}\theta(s-4m_{\pi}^2)+T_{22}\rho_2 T_{22}^{*}\theta(s-4m_{K}^2)\ . \label{eq;f980;uni}
\end{eqnarray}
And the [1,1] matrix Pad\'e approximation is constructed as
\begin{equation}
T=T^{(2)}\cdot[T^{(2)}-T^{(4)}]^{-1}\cdot T^{(2)}\ . \label{eq;pade11}
\end{equation}
Here the $T^{(2)}$ is the $\pi\pi-\bar{K}K$ coupled channel amplitudes of $\mathcal{O}(p^2)$ and $T^{(4)}$ is that of $\mathcal{O}(p^4)$.
The isoscalar $\mathcal{O}(p^2)$ partial waves of $\pi\pi$ scattering are given in Eq.~(\ref{eq:T;pipi;Op2}), and those of the $\pi\pi\to\bar{K}K$ and $\bar{K}K\to\bar{K}K$ are as follows:
\begin{eqnarray}
T_{2,\pi\pi\to\bar{K}K}^{0S}(s)=\frac{\sqrt{3}s}{64 \pi F_\pi^2} \,,\;\;\;\;
T_{2,\bar{K}K\to\bar{K}K}^{0S}(s)=\frac{3s}{64 \pi F_\pi^2}\,. \label{eq:T;KK}
\end{eqnarray}
If $T^{(4)}/T^{(2)}$ is small and do the expansion, one would find that the first two terms of the $T$ amplitude is $T^{(2)}+T^{(4)}$, restoring the perturbation calculation of $\chi$PT in the low energy region up to $\mathcal{O}(p^4)$. Notice that the ${\rm Im}T^{(4)}=T^{(2)}\rho T^{(2)}$ and taking it into Eq.~(\ref{eq;pade11}) one would find that the unitarity is restored.
The $SU(3)$ one-loop meson-meson scattering amplitudes are calculated in Refs.\cite{Bernard:1991zc,GomezNicola:2001as} and the analytical forms of the partial waves of $\pi\pi$ scattering amplitudes can be found in Ref.\cite{Dai:2019zao}.
The complete $SU(3)$ one-loop meson-meson scattering amplitudes of $\pi \pi-\bar{K} K$ coupled channels with $\overline{MS}-1$ scheme are given in Ref.\cite{GomezNicola:2001as}.

In Ref.~\cite{Pelaez:2003rv} and references therein, there is a list about the pole locations of the $f_0(980)$ in their Table IV. For instance the $f_0(980)$ locates at $994-i14$~MeV in the IAM approximation~\cite{Oller:1998hw}, with smaller masses and widths  in the IAM I and IAM III. The differences are caused by the way to deal with the pion (kaon and $\eta$) decay constants and the LEC $L_4$.
In Ref.~\cite{Dai:2011bs}, the Pad\'e approximation is used and the $f_0(980)$ is found in the RS-II with pole location of $974-i25$~MeV, there is no shadow poles in other RSs. As stated there, a couple channel Breit--Wigner resonance should exhibit two poles on different Riemann sheets and
meet each other on the real axis when $N_c=\infty$. Hence it supports the $\bar{K}K$ molecule structure.
The single pole structure is the same in a triple channel ($\pi\pi-\bar{K}K-\eta\eta$) case~\cite{Dai:2012kf},  but with a much smaller width and a bit larger mass. All these find poles of the $f_0(980)$ in the RS-II, but the position is not stable.
It should be pointed out that in Pad\'e approximation or IAM,
the amplitude of the $\chi$PT is analytical but the unitarization takes the left hand cut in
$(-\infty, 4m_K^2-4m_{\pi}^2]$ of $T_{22}$ into the other scattering amplitudes.
Besides, it introduces spurious poles on the complex $s$-plane, and it violates crossing symmetry as the resummation is done in the $s$-channel series.
Therefore one should be careful to make conclusions on the output of the unitarized amplitudes. As discussed before, the U$\chi$PT methods are more qualitative rather than quantitative to study the $f_0(980)$.
Nevertheless, by fitting to the experimental data, the LECs are fixed and the amplitudes can be used to extract the poles. Note that in the very beginning only the pseudoscalar mesons ($\pi$, $\eta$ ,$K$) are written down in the chiral Lagrangian and no scalar or vector resonances are included, thus the poles we extracted also give clues of the existence of the $f_0(980)$. However, as pointed out by Ref.\cite{Su:2007au}, discussing the physical meaning of a dynamical generated pole from unitarization amplitude should be rather cautious, as the property of the unitarized amplitude could be highly model-dependent.
Nevertheless, finding poles only in the RS-II is a clue about the $\bar{K}K$ molecule structure of the $f_0(980)$ and indeed exclude a dominant $\bar{q}q$ bound state component due to the pole counting rule \cite{Morgan:1992ge}.

In Ref.~\cite{Oller:1998zr}, a $N/D$ chiral unitary approach based on $SU(3)$ $\chi$PT is used and they find the $f_0(980)$ at $987-i14$~MeV. Note that the l.h.c. has been ignored there.
In Ref.~\cite{Guo:2011pa}, this $N/D$ method is \lq refined' to unitarize the meson-meson scattering amplitudes within $U(3)$ $\chi$PT, where the l.h.c. by perturbative calculation is kept. It reads
\begin{eqnarray}
{T_J^{I}}(s)^{-1} = N_J^I(s)^{-1}+ g(s)~,\label{eq:NOD;U3T}
\end{eqnarray}
where the $g(s)$ relates to the unitary cut and is given by the dispersion relation
\begin{eqnarray}
 g(s) = g(s_0) - \frac{s-s_0}{\pi}\int_{s_{\rm th}}^{\infty} \frac{\rho(s')}{(s'-s)(s'-s_0)} d s'\,,\label{eq:NOD;U3R}
\end{eqnarray}
with ${\rm Im} g(s) = -\rho(s)$.
The $N_J^I(s)$ includes the l.h.c., and could be estimated by the $\chi$PT and/or R$\chi$T
\begin{eqnarray}
N_J^I(s) = {T_J^I}(s)^{\rm (2)+Res+Loop}+ T_J^I(s)^{(2)}  \,\, g(s) \, \, T_J^I(s)^{(2)} \,. \label{eq:NOD;U3L}
\end{eqnarray}
In their analysis, the $f_0(980)$ is found to be located at $981^{+9}_{-7}-i22^{+5}_{-7}$~MeV.
It relates to $\pi\pi-\bar{K}K-\eta\eta-\eta\eta'-\eta'\eta'$ coupled channels and there lacks experimental data on the channels scattering into $\eta\eta$, $\eta\eta'$ and $\eta'\eta'$ channels. Hence it is complicated to determine all the amplitudes.
Nevertheless, this work confirms the existence of the $f_0(980)$ and are helpful to determine the pole location and study the $f_0(980)$ coupling to the $\pi\pi$, $\bar{K}K$ and $\eta\eta$ channels, where the latter two are found to be almost equally large.

In Ref.~\cite{Ishida:1995xx}, they give the $f_0(980)$ as $M=993.2\pm6.5\pm6.9$~MeV, $\Gamma=100$~MeV with the interfering amplitude method where unitarity has been implemented.
In Ref.~\cite{Janssen:1994wn}, the Lippmann-Schwinger equation is used to solve the scattering amplitude.
They use a modified J$\ddot{u}$lich model to describe the potential of the $\pi\pi$ scattering. They found a single pole of the $f_0(980)$ at $1015-i15$~MeV in the RS-II. Though the mass is a bit higher than the $\bar{K}K$ threshold, it will go back to $985-i0$~MeV in the zero $\pi\pi/\bar{K}K$ coupling limit. This supports the $\bar{K}K$ molecule picture.
Very recently Ref.~\cite{Ahmed:2020kmp} use the on-shell Bethe-Salpeter equation to solve the $\pi\pi-\bar{K}K$ coupled channel amplitudes.
With these they find the poles are somehow stable by varying the cut-off. For instance, the
$f_0(980)$ is at $982.13-i21.67$~MeV, with $q_{max}=1080$~MeV.
They have similar conclusion as that of Ref.~\cite{Janssen:1994wn} about the nature of the $f_0(980)$: $K\bar{K}$ molecule origin. In Ref.~\cite{Zhou:2020pyo}, in the single-channel approximation, $f_0(980)$ appears to be a bound state just below the $K\bar K$ threshold when the bare $s\bar s$ state is considered to be coupled with the $K\bar K$ continuum. The bare $s\bar s$ state is regarded to be shifted to $f_0(1710)$. But a coupled channel analysis is required to obtain a more accurate description of $IJ=00$ $\pi\pi$ scattering phase shift. Until then, their argumentations could be reinforced.

As discussed above, the $f_0(980)$ is close to the $\bar{K}K$ threshold and also close to the real axis. On the other hand, there are a lot of data to refine the analysis, both the phase shifts and inelasticity datasets from Refs.~\cite{Batley:2010zza,Hyams:1973zf,Grayer:1974cr,Hyams:1975mc,Cohen:1980cq,Etkin:1981sg} and the data of scattering/decaying processes relating to the $\pi\pi-\bar{K}K$ FSI, such as the BESIII's invariant mass distribution of $J/\Psi\to\phi K^+ K^-, \pi^+\pi^-$~\cite{Ablikim:2004wn}  as well as the BaBar's Dalitz plot analysis of $D_s^+\to\pi^+\pi^-\pi^+$~\cite{Aubert:2008ao} and $D_s^+\to K^+K^-\pi^+$~\cite{delAmoSanchez:2010yp}. According to Eq.~(\ref{eq:FSI;AMP}), the $\phi$ or $\pi$ can be treated as a spectator and thus the $\pi\pi$ and $\bar{K}K$ FSI can be used to constrain the $S$-wave amplitudes. Especially the $D_s^+\to K^+K^-\pi^+$~\cite{delAmoSanchez:2010yp} have high statistics and give more information about the amplitude around the $\bar{K}K$  threshold.
These makes the $K$-matrix, especially the one with Chew-Mandelstam functions, would be a good way to describe the physics and extract the pole of the $f_0(980)$. This is done in Ref.~\cite{Dai:2014zta,Dai:2014lza}, where the conformal mapping parametrization of the $\pi\pi\to\pi\pi$ $S$-wave ~\cite{GarciaMartin:2011cn} as well as the $\pi\pi\to \bar{K}K$  $S$-wave scattering amplitude analyzed by the Roy-Steiner equation~\cite{Buettiker:2003pp} are included as constraints, too. The effects of isospin breaking, the mass difference between the neutral and charged kaon, is included by differing the $K^+K^-$ and $\bar{K}^0K^0$ thresholds  rather than treating the kaons as isospin invariant with a common mass. After all these steps the outcome is a rather precise knowledge about the $f_0(980)$ pole location. It is found at $998\pm3-i21\pm3$~MeV in the RS-II~\cite{Dai:2014lza}. According to the pole counting rule, this supports the molecule structure.

Since the $\pi\pi$-$\bar{K}K$ coupled channel amplitudes are well determined in Ref.~\cite{Dai:2014zta}, at least near the real axis, one can extend it to the complex-$s$ plane with dispersion relation. In Ref.~\cite{Dai:2017uao}, the Omn$\grave{e}$s function is calculated by the phase of the $\pi\pi$ scattering instead of the phase shift, and then the contribution of all the unknown l.h.c. and distant r.h.c. are absorbed into the polynomials.
By fitting to the amplitudes determined well in the real axis, coupled channel unitarity could be restored.
The data of the phase shift and inelasticity~\cite{Batley:2010zza,Hyams:1973zf,Grayer:1974cr,Hyams:1975mc,Cohen:1980cq,Etkin:1981sg} are fitted to constrain the parameters, too. Besides, the amplitude in the complex-$s$ plane, are constrained by fitting to that of the Roy equation analysis.
All these make the analysis reliable and the working domain is extended deeper to the complex-$s$ plane comparing with that of the the $K$-matrix formalism\footnote{This method is also used to analyze the other partial waves of $\pi\pi$ scattering in Refs.\cite{Dai:2019zao,Dai:2017tew,Dai:2018fmx}.}. The $f_0(980)$ is found to be at $997.7\pm1.1-i21.7\pm1.9$~MeV in the RS-II.
Later, the Omn$\grave{e}$s function of phase is combined with the partial wave dispersion relation on the $\ln T(s)$ and one could solve the two body scattering amplitudes in a simple way. See in section~\ref{sec:Omnes} and the details in Ref.~\cite{Dai:2019zao}. From that analysis the $f_0(980)$ is found to be at $997.6\pm0.1-i20.3\pm1.3$~MeV in the RS-II. It is shown that the unitarity cut is much more important than the l.h.c. to determine the pole locations of the $f_0(980)$ and also the $\sigma$. There are some other interesting ways to study the light scalars. For instance in Ref.~\cite{Fariborz:2009cq}, the $f_0(980)$ is found to be $1085$~MeV having 95\% $\bar{q}\bar{q}qq$ component with a generalized linear sigma model.

In a short summary, there are various kinds of models to conclude that the $f_0(980)$ is to be around the $\bar{K}K$ threshold ($992$~MeV) and the width is smaller than $50$ MeV. It could make an even more constructive conclusion that the $f_0(980)$ is close to the $\bar{K}^0K^0$ threshold ($995$~MeV) and the width is near $40$~MeV. There is always only one pole in the RS-II (in $\pi\pi-\bar{K}K$ coupled channels) and it is most likely to be a $\bar{K}K$ molecule.

\subsubsection{$a_0(980)$: from $\pi\eta-\bar{K}K$ coupled channels }\label{sec:a980}
The $a_0(980)$ appears as the intermediate states of $\pi\eta-\bar{K}K$ coupled channel scattering, with $IJ=10$. However, unlike the $f_0(980)$, it does not have data on the phase shifts and inelasticity, which makes the analysis more complicated and the conclusion not easy to obtain. There is a $\pi^+\eta$ effective mass distribution data from the
$pp\rightarrow p(\eta\pi^+\pi^-)p$ reaction by WA76~\cite{Armstrong:1991rg}, which can be used to constrain the $\pi\eta$ scattering amplitude.
In the past decades, the data referring to the processes that decay/scatter  into the $\pi\eta X$ has increased dramatically. For instance the processes of $\gamma\gamma\to\pi^0\eta$ \cite{Uehara:2009cf}, $\phi\to\pi^0\eta\gamma$ \cite{Ambrosino:2009py}, $J/\psi\to \gamma\eta\pi^{0}$ \cite{Ablikim:2016exh}, $J/\psi\to \phi\eta\pi^{0}$ \cite{Ablikim:2018pik}, and $\chi_{c1} \rightarrow \eta \pi^+ \pi^-$ \cite{Adams:2011sq,Kornicer:2016axs}.
These are rather helpful to study the resonances. There is another good point to make things easier, that is, the total isospin of $\pi\eta$ is $I=1$, thus the partial waves are roughly one half less than other processes such as the $\pi\pi$ and $K\bar{K}$ channels. However, there are still some difficulties to be conquered. To do a comprehensive amplitude analysis, lots of the processes need not only the S-wave but also the higher partial waves, but they are ignored in lots of the analysis except for a few groups such as in Ref.~\cite{Danilkin:2017lyn}.

In the chiral unitary approach, the pseudoscalar meson-meson scattering are calculated within $SU(3)$ $\chi$PT.
The isovector $\mathcal{O}(p^2)$ partial waves of the $\pi\eta-\bar{K}K$ coupled channels are as follows:
{ \setlength{\mathindent}{-0.1cm} \begin{eqnarray}
T_{2,\pi\eta\to\pi\eta}^{1S}(s)=\frac{M_\pi^2}{48 \pi F_\pi^2} \,,\;
T_{2,\pi\eta\to\bar{K}K}^{1S}(s)=\frac{8M_K^2+M_\pi^2+3M_\eta^2-9s}{96\sqrt{6}\pi F_\pi^2} \,,\;
T_{2,\bar{K}K\to\bar{K}K}^{1S}(s)=\frac{s}{64 \pi F_\pi^2}\,. \nonumber\\ \label{eq:T;KK}
\end{eqnarray} }
The analysis of $\pi\eta$ scattering up to $\mathcal{O}(p^4)$ was first given in Ref.\cite{Bernard:1991xb}.  With decomposition of isospin and spin, the partial waves of allowed quantum number are obtained. The chiral unitary approach automatically decomposes partial waves and also takes into account the chiral dynamics and unitarity. In Ref.\cite{Pelaez:2003rv} and references therein, the $a_0(980)$ locates at $1055-i21$~MeV in the IAM approximation \cite{Oller:1998hw}, while in IAM I there is no such a pole and the structure is possible to be caused by cusp effects. A pole of $1117^{+24}_{-320}-i12^{+43}_{-12}$~MeV is found in the IAM II and $1091^{+19}_{-45}-i52^{+21}_{-40}$~MeV in the IAM III. They are in the RS-II. In the Pad\'e approximation \cite{Dai:2011bs}, it finds a pole at $1131-i79$~MeV in the RS-IV and a very faraway companion shadow pole in the the RS-III, these two poles will meet each other at the real axis in the limit of large $N_C$. This reveals the Breit-Wigner origin of the $a_0(980)$\footnote{It is worth noting that the pole counting first proposed by Ref.\cite{Morgan:1992ge} has somehow been generalized by Refs.\cite{Dai:2011bs,Dai:2012kf} as companion shadow poles do not need to be very close to the threshold and they are related to the pole (being closest to the physical region) in the large $N_C$ limit. The companion shadow poles are always an evidence of the Breit-Wigner component. }. Following it, Ref.\cite{Dai:2012kf} confirms the twin pole structure of the $a_0(980)$, but with the width of the $a_0(980)$ pole even larger and the shadow pole a bit closer. In Ref.\cite{Oller:1998zr}, an $N/D$ chiral unitary approach based on $SU(3)$ $\chi$PT is used and they find a pole at $1053.13-i24$~MeV, while in Ref.\cite{Guo:2011pa}, the same approach based on $U(3)$ $\chi$PT finds a pole at $1012^{+25}_{-7}-i16^{+50}_{-13}$~MeV. We listed all these pole locations as well as those discussed below in Table~\ref{tab:a980}.
\begin{table}[bt]
\caption{\label{tab:a980} Pole locations of the $a_0(980)$ from different models, in units of MeV. }
{\footnotesize\begin{tabular}{cccc}
\br
           &  SU(3) U$\chi$PT        & U(3) U$\chi$PT & other models    \\
\mr
\multirow{3}{*}{RS-II\rule[-0.3cm]{0cm}{8mm}}  & $1055-i21$\cite{Oller:1998hw}    &  $1012^{+25}_{-7}-i16^{+50}_{-13}$\cite{Guo:2011pa}    &  $991-i 101$\cite{Janssen:1994wn} \\
\multirow{3}{*}{\rule[-0.3cm]{0cm}{8mm}}       & $1117^{+24}_{-320}-i12^{+43}_{-12}$\cite{Pelaez:2003rv}  &  $1037^{+17}_{-14}-i44^{+6}_{-9}$(LO)\cite{Guo:2016zep} &  $974.50-i57.31$\cite{Ahmed:2020kmp}  \\
\multirow{3}{*}{\rule[-0.3cm]{0cm}{8mm}}       & $1091^{+19}_{-45}-i52^{+21}_{-40}$\cite{Pelaez:2003rv}  &  & $994\pm2-i25.4\pm5.0$\cite{Albaladejo:2015aca}   \\ \hline
RS-III     &    &  &   $958\pm13-i60.8\pm11.5$\cite{Albaladejo:2015aca}  \\ \hline
RS-IV    &  $1131-i79$\cite{Dai:2011bs}  & $1019^{+22}_{-8}-i24^{+57}_{-17}$(NLO)\cite{Guo:2016zep}  &  $1120^{+20}_{-70}-i140^{+40}_{-65}$\cite{Danilkin:2017lyn}   \\
\br
\end{tabular}}
\end{table}

In Ref.~\cite{Guo:2016zep}, the unitarized $U(3)$ $\chi$PT  has been used to extrapolate the lattice simulation of $\pi\eta$ scattering (with pion mass 391~MeV) to the physical region. Meanwhile the experimental data of $\gamma\gamma\to\pi^0\eta$ has been fitted to constrain the amplitudes. When the pion mass is fixed to be physical, they find a pole at $1037^{+17}_{-14}-i44^{+6}_{-9}$~MeV in the RS-II for the leading order case, and a pole at $1019^{+22}_{-8}-i24^{+57}_{-17}$~MeV on the RS-IV for the next-to leading order case.
In Ref.~\cite{Albaladejo:2015aca}, a $K$-matrix type chiral unitary approach is used, where the amplitude in the low energy region is enforced to be matched with $SU(3)$ $\chi$PT up to NLO in the low energy region. Comparing with the normal $K$-matrix formalism, it reduces parameters and in fact it only has 6 unknown parameters. With chiral symmetry in the low energy region and coupled channel unitarity, it finds two poles in their best fit: one at $994\pm2-i25.4\pm5.0$~MeV in RS-II and the other one at $958\pm13-i60.8\pm11.5$~MeV in RS-III. Indeed this also supports a Breit-Wiger origin of the $a_0(980)$ according to the pole counting rule~\cite{Morgan:1992ge}. From U$\chi$PT, it tends to support the existence of the $a_0(980)$, but the pole location is not stable and we need effort both from the theoretical side to reduce the model dependence and from the experimental side to have more precise data.

There are some other ways to study the $a_0(980)$ in different models.
In Ref.~\cite{Danilkin:2017lyn} the $\gamma\gamma\to\pi^0\eta$ is studied within the Muskhelishvili-Omn$\grave{e}$s dispersive approach. The D-wave is described by a Breit-Wigner formalism where the $a_2(1270)$ is included. The crossing symmetry is implemented by analyzing the $\eta\to\pi^0\gamma\gamma$ process together. They finally find a pole at $1120^{+20}_{-70}-i140^{+40}_{-65}$~MeV in RS-IV.
In Ref.~\cite{Janssen:1994wn}, they find the $a_0(980)$ to be at $991-i 101$~MeV in RS-II by solving the Lippmann-Schwinger equation.
The direct $\bar{K}K$ interaction is not essential for generating such a pole and the $\pi\eta-\bar{K}K$  transition potential is more likely to be the origin of the $a_0(980)$. Thus the $a_0(980)$ is suggested to be a dynamically generated threshold effect.
Ref.~\cite{Ahmed:2020kmp} solves the $\pi\eta-\bar{K}K$ coupled channel amplitudes using the on-shell Bethe-Salpeter equation. They find the $a_0(980)$ to be at $974.50-i57.31$~MeV, with $q_{max}=1080$~MeV. The pole is stable by varying the cut-off.
They have a similar conclusion as that of Ref.~\cite{Janssen:1994wn} about the nature of the $a_0(980)$ state. Ref.~\cite{Liang:2016hmr} also finds a clear cusp form of the $a_0(980)$ in the $\pi\eta$ invariant mass distribution for the $\chi_{c1}\to\eta\pi^+\pi^-$, with a chiral unitary approach.

In a short word, the pole location of the $a_0(980)$ is unstable, and there is no final conclusion on its nature, i.e., cusp, tetraquark, normal Breit-Wigner resonance, $K\bar{K}$ molecule, etc.. Different from the $f_0(980)$, which is extracted as a resonance pole on RS-II below the $K\bar K$ threshold in many different analyses, the $a_0(980)$ is usually regarded as a resonance pole on RS-II above the $K\bar K$ threshold with a larger pole width. It is the threshold effect which leads to a narrow peak below the $K\bar K$ threshold in the $\pi\eta$ events number in Ref.~\cite{Armstrong:1991rg}, as illustrated in the Flatt\'e effect~\cite{Flatte:1976xu}.
To get further understanding about the resonance, one needs more accurate data on the experimental side where the information of the $\pi\eta$ scattering could be extracted out reliably. Also one needs better understanding about the strong interaction in the non-perturbation QCD region, too. Urgently one needs more comprehensive partial wave analysis on the data already measured such as the $\gamma\gamma\to\pi\eta$, $\phi\to\pi\eta\gamma$ and so on. The process of $\Lambda_c^+\to \pi^+ \eta\Lambda$ can be an ideal process to study the $a_0(980)$ \cite{Xie:2016evi}, too.
Further more, the $f_0(980)-a_0(980)$ mixing \cite{Wu:2010za} reveals new connection between the two resonances. The high statistics measurements \cite{Ablikim:2018pik} confirms the mixing and would improve the understanding of the nature of them.

\subsection{Light baryon resonances\label{sec:3.2}}
\subsubsection{A subthreshold resonance: from $\pi N$ scattering}
Recently, in Refs.~\cite{Wang:2017agd,Wang:2018gul,Wang:2018nwi}, a novel subthreshold resonance named $N^\ast(890)$ was found in the $S_{11}$ wave of the $\pi N$ interaction when applying the PKU representation~\cite{Xiao:2000kx,He:2002ut,Zheng:2003rw,Zheng:2003rv,Zhou:2004ms,Zhou:2006wm} to analyze the covariant chiral amplitude of  elastic \(\pi N\) scattering~\cite{Alarcon:2012kn,Chen:2012nx,Yao:2016vbz,Siemens:2017opr} at low energies.

\begin{figure}[hpt]
\centering{\includegraphics[width=0.65\textwidth]{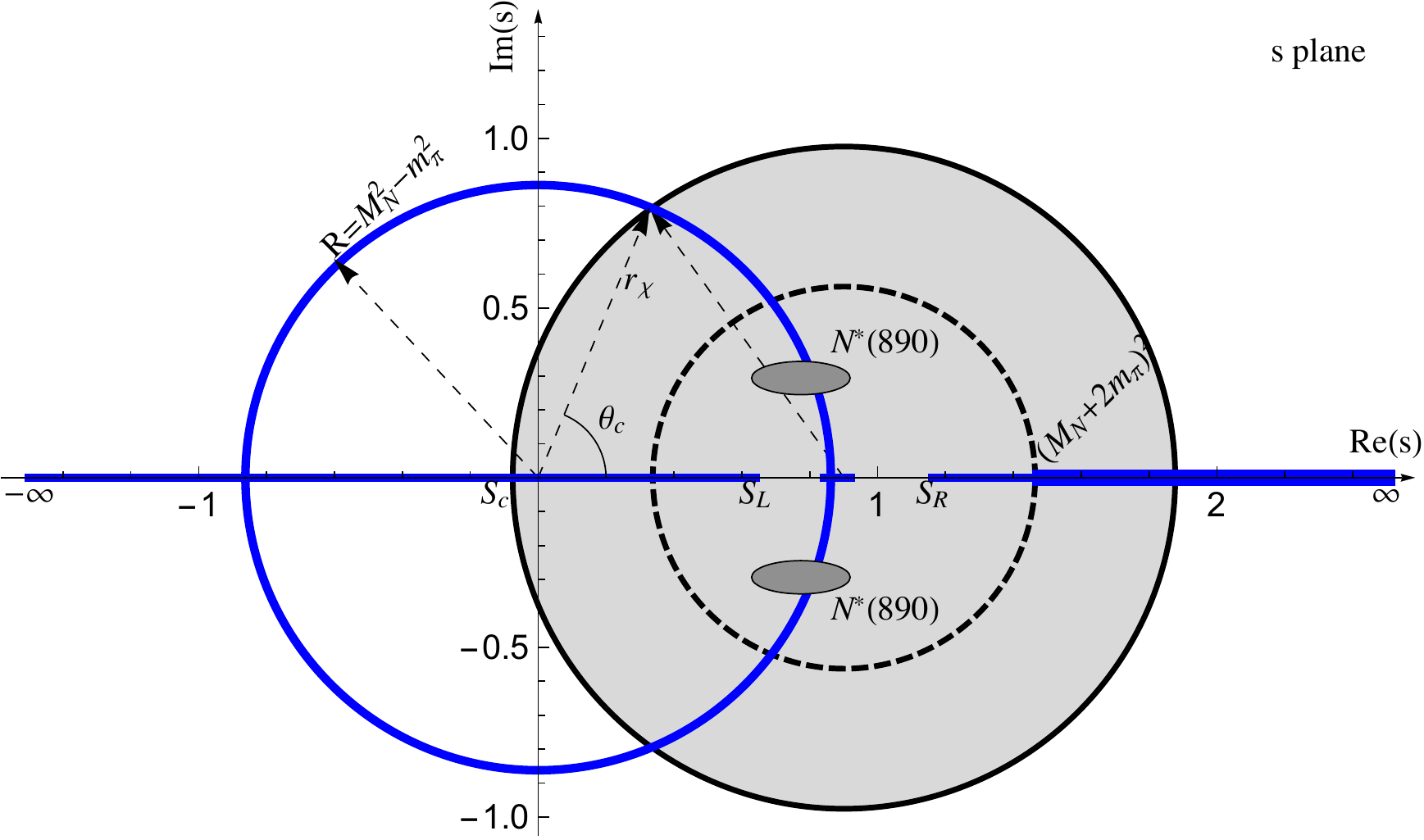}}
\caption{\label{fig:piNcuts} Schematic drawing of the branch cuts of the partial wave $\pi N$ elastic scattering amplitude on the $s$ plane. The cuts are represented by the blue thick lines and the blue circle. The right hand cut corresponds to the line $[s_R,+\infty)$, while the others are left hand cuts. The gray disk indicates the B$\chi$PT valid region constrained by the convergence radius $r_\chi\sim m_R^2-s_\chi$, where $s_\chi=M_N^2+m_\pi^2$ and $m_R$ is pole mass of the Roper resonance. In addition, the dashed circle is drawn with $r_\chi$ estimated by the $\Delta$ resonance.   The symbol $s_c$ denotes the intersection of the on-axis left-hand cuts and the B$\chi$PT convergence circle. The pair of heavy gray ellipses represents the pole position of the $N^*(890)$ resonance. }
\end{figure}

In the PKU formalism,  partial-wave $\pi N$ elastic scattering $S$ matrix can be cast as
\begin{equation}\label{eq:Spku}
S(s)=\prod_bS_b(s)\cdot\prod_vS_v(s)\cdot\prod_rS_r(s)\cdot e^{2\text{i}\rho(s)f(s)}\ \mbox{, }
\end{equation}
where the kinematic factor $\rho(s)={\sqrt{s-s_L}\sqrt{s-s_R}}/{s}$ with $s_R\equiv (M_N+m_\pi)^2$ and $s_L\equiv(M_N-m_\pi)^2$. Here the indices $b$, $v$ and $r$ stand for bound states, virtual states and resonances, respectively. The explicit expressions of their contributions to the $S$ matrix are specified in Eqs.~\eqref{eq:PKUbvr} and~\eqref{z0depen}. The exponential term is a background that encodes the information of l.h.c. as well as r.h.c above inelastic thresholds. The branch cuts of the partial-wave pion-nucleon scattering amplitudes, derived following the method given in Refs.~\cite{MacDowell:1959zza,Kennedy:1961}, are schematically drawn in Fig.~\ref{fig:piNcuts}.

\begin{figure}[hpt]
\centering{\includegraphics[width=0.45\textwidth]{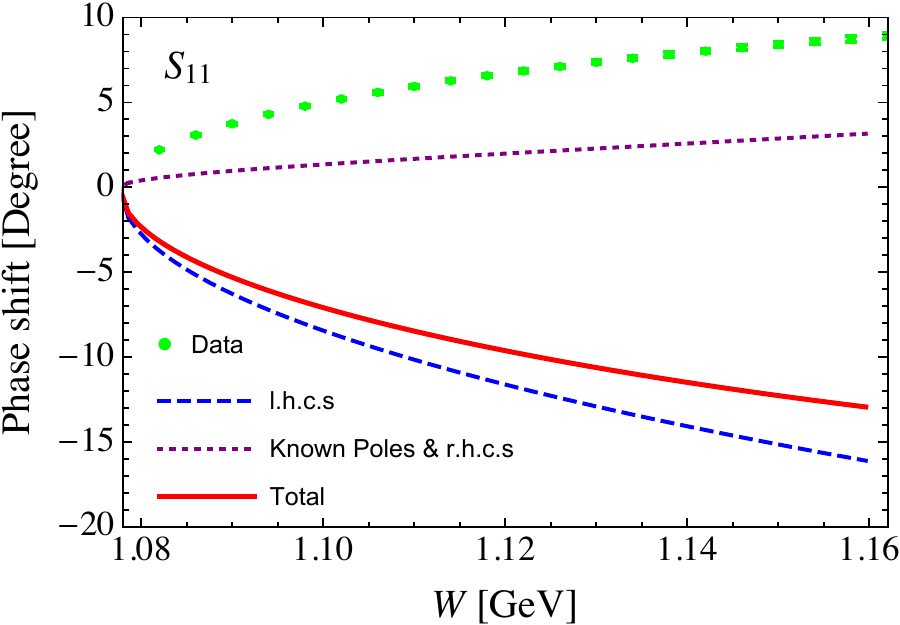}}~~~
\centering{\includegraphics[width=0.435\textwidth]{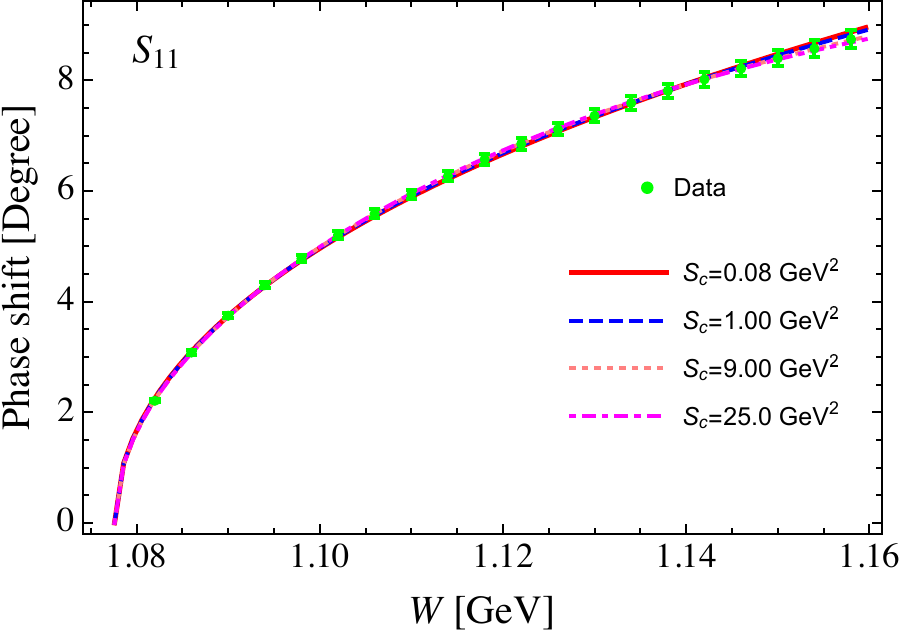}}
\caption{\label{fig:S11} Left panel: known contributions vs data~\cite{Hoferichter:2015hva} ($s_c=-0.08$ GeV$^2$). Right panel: fits to the data with  different cut-off parameter $s_c$.}
\end{figure}

The function $f(s)$ satisfies the following dispersion relation,
\bea
f(s)&=&-\frac{s}{\pi}\int_{s_{c}}^{s_L} \frac{\ln|S(s')|ds'}{2\rho(s')s'(s'-s)}
\nonumber\\
&&+\frac{s}{\pi}\int_0^{\theta_c}\frac{\ln[S_{in}(\theta)/S_{out}(\theta)]}{2i\rho(s')(s'-s)\mid_{s'=(M_N^2-m_\pi^2)e^{i\theta}}}d\theta\nonumber\\
&&+\frac{s}{\pi}\int_{(M_N^2-m_\pi^2)^2/M_N^2}^{M_N^2+2m_\pi^2} \frac{\text{Arg}[S(s')]ds'}{2is'\rho(s')(s'-s)}\nonumber\\
&&+\frac{s}{\pi}\int_{(2m_\pi+M_N)^2}^{\Lambda_R^2} \frac{\ln[1/\eta(s')]ds'}{2\rho(s')s'(s'-s)}\ \mbox{, }\label{fdisperpiNp3}
\eea
where $S_{out}$ and $S_{in}$ are the $S$-matrix along the circular cut  calculated from outside and inside, respectively. Here $0<\eta<1$ is the inelasticity along the inelastic r.h.c. ranging from $(2m_\pi+M_N)^2$ to infinity.  Furthermore, $s_c$, $\theta_c$  and $\Lambda_R$  are cut-off parameters for the kinematical cut $(-\infty,s_L]$, the circular cut and the inelastic r.h.c., in order. In practice, the first three terms, i.e. the left-hand cut contributions, can usually be estimated by B$\chi$PT, while the last term is calculated using experimental data on the inelasticity~\cite{Arndt:2006bf}. For $S_{11}$ wave, contributions from known poles ($N^*(1535)$, $N^*(1650)$ and $N^*(1895)$), left-hand cuts and right-hand inelastic cuts were computed numerically as shown in Fig.~\ref{fig:S11}. Note that the l.h.c contribution is estimated using the $\mathcal{O}(p^3)$ chiral amplitude obtained from convariant B$\chi$PT.\footnote{In fact, there is a kinematical singularity at $s=0$ in the partial wave amplitudes, causing partly by integrating out heavy degrees of freedom. }  It can be seen that their sum of known contributions differs from the data which is unable to resolved by fine-tuning parameters in the numerical analysis. This issue can be addressed by adding a hidden pole contribution in the production representation.  In addition, the parameters of the hidden pole should be determined by fitting data. Such a procedure leads to the novel subthreshold resonance $N^*(890)$ with the pole position being~\cite{Wang:2018nwi}
\bea\label{eq:Nst890}
\sqrt{z_r}=(895\pm 81)-{i}\,(164\pm23)\ {\rm MeV}.
\eea
It should be stressed that the pole position is quite stable against the cutoff parameter $s_c$. The fitting results change slightly for $s_c\in[0.08,25.0]$~GeV$^2$.

The possible existence of the $N^\ast(890)$ resonance is owing to a proper incorporation of l.h.c. and inelastic r.h.c cut contribution. An improved implementation of analyticity is made compared to other unitarization methods. This is verified in Ref.~\cite{Ma:2020sym} that the $N^\ast(890)$ resonance still keeps its existence even in a $K$-matrix parametrization if a better treatment of analyticity is executed. Though the existence of $N^\ast(890)$ is guaranteed by the requirements of analyticity and data, it still deserves further investigation on its nature. In Ref.~\cite{Ma:2020hpe}, the couplings of $N^\ast(890$) to the $\gamma N$ and $\pi N$ systems were extracted using a dispersive representation of the process $\gamma N\to\pi N$. It is found that, compared to $N^*(1535)$, $N^\ast(890$) couples more strongly to the $\pi N$ system. It is also worth noting that the $N^\ast$ resonance might be a candidate for members of S-wave baryon octet (including hyperon $\Sigma(1480)$ and $\Xi(1620)$), proposed by Azimov and denoted by $N^\prime$ therein~\cite{Azimov:1970ei,Azimov:2003bb}, and searched for in despair for 50 years since then\footnote{We would like to thank Igor Strakovsky for pointing out the related information on the $N^\prime$ state.}.

\subsubsection{$\Lambda(1405)$: from $\bar{K}N$ scattering}\label{sec:1405}
The $\Lambda(1405)$ was interpreted as a dynamically generated resonance, emerging in the coupled meson-baryon channels with strangeness $S=1$ and isospin $I=0$, by Refs.~\cite{Dalitz:1959dn,Kim:1965zzd,Jones:1977yk},  into which continuous efforts have been devoted in order to gain more insights in the past years, see e.g. Ref.~\cite{Hyodo:2011ur}. Utilizing chiral effective theory and non-perturbative resummation techniques,  modern partial wave analyses of the SU(3) meson-baryon interaction has obtained impressive interpretation on the nature of the $\Lambda(1405)$ state and the related experimental data~\cite{Kaiser:1996js,Oset:1997it,Oller:2000fj,Meissner:2000re,Ikeda:2011pi,Mai:2012dt,Ikeda:2012au,Guo:2012vv,Mai:2014xna}.

A first study of the pole structure of the $\Lambda(1405)$ within the context of chiral dynamics at lowest order was done in Ref.~\cite{Oller:2000fj}. The leading order~(LO) chiral $\phi_i B_a\to\phi_j B_b$ amplitudes reads
\bea\label{eq:LOKNV}
V_{\rm LO}({ij,ab})=V_s(ij,ab)+V_d(ij,ab)+V_c(ij,ab)\ ,
\eea
where the subscripts correspond to seagull, direct and crossed diagrams. The amplitudes $V_{s,d,c}$ can be obtained straightforwardly from the LO chiral effective Lagrangians and their explicit expressions read
{\bea
T_s(ij,ab)&=&\sum_{k=1}^8f_{abk}f_{ijk}\frac{NN^\prime}{F_0^2}\nonumber\\
&\times&\chi_b\bigg[W-m_0+(\vec{p}\cdot\vec{p}^\prime+i(\vec{p}^\prime\times\vec{p})\cdot\vec{\sigma})\frac{W+m_0}{(NN^\prime)^2}\bigg]\chi_a\ ,\\
T_d(ij,ab)&=&-\sum_{k=1}^8(D\,d_{jkb}+iF\,f_{jkb})(D\,d_{ika}-iFf_{ika})\frac{NN^\prime}{F_0^2}\nonumber\\
&\times&\chi_b\bigg[-2m_0+
\frac{s+3m_0^2}{W+m_0}\,\frac{\vec{p}\cdot\vec{p}^\prime+i(\vec{p}^\prime\times\vec{p})\cdot\vec{\sigma}}{(NN^\prime)^2}\nonumber\\
&&\hspace{0.6cm}\times
\bigg(2m_0+\frac{s+3m_0^2}{W-m_0}\bigg)
\bigg]\chi_a\ ,\\
T_c(ij,ab)&=&-\sum_{k=1}^8(D\,d_{ikb}+iF\,f_{ikb})(D\,d_{jka}-iFf_{jka})
\frac{NN^\prime}{F_0^2}\nonumber
\\
&\times&\chi_b\bigg[-2m_0-
\frac{(W-m_0)(u+3m_0^2)}{u-m_0^2}+\frac{\vec{p}\cdot\vec{p}^\prime+i(\vec{p}^\prime\times\vec{p})\cdot\vec{\sigma}}{(NN^\prime)^2}\nonumber\\
&&\hspace{0.6cm}\times
\bigg(2m_0-\frac{W+m_0}{u-m_0^2}(u+3m_0^2)\bigg)
\bigg]\chi_a\ ,
\eea
with $N=\sqrt{E_0+m_0}$ and $N^\prime=\sqrt{E_0^\prime+m_0}$. For more definitions of the notations involved in the above amplitudes, see Ref.~\cite{Oller:2000fj}.
}

The $S$-wave partial wave potential is simply given by
\bea
\mathcal{V}_{\rm LO}({s})=\frac{1}{8\pi}\sum_{\sigma=1}^2\int{\rm d}\Omega\, V_{\rm LO}({s},\Omega;
\sigma,\sigma)\ ,\label{eq:ollerV}
\eea
which is just a perturbative quantity and usually violates unitarity.  To restore unitarity,  an approach was deduced by the authors of Ref.~\cite{Oller:2000fj} as demonstrated below. Partial wave unitarity ensures that the inverse of the realistic $S$-wave amplitude, denoted by $T(s)$, should satisfy
\bea
{\rm Im}T^{-1}(s)_{ij}=-\rho(s)_i\delta_{ij}\ ,
\eea
with $\rho_i\equiv q_i/(8\pi\sqrt{s})$ and $q_i$ is the modulus of the c.m. momentum.
Hence one can write a once-subtracted dispersion relation for the ${\rm Im}T^{-1}(s)$, which is
\bea
T^{-1}(s)_{ij}
&=&\mathcal{T}^{-1}(s)_{ij}-\delta_{ij}\left\{\tilde{a}_i(s_0)+\frac{s-s_0}{\pi}\int_{s_i}^{\infty}{\rm d}s^\prime\frac{\rho(s^\prime)_{i}}{(s^\prime-s)(s^\prime-s_0)}\right\}\nonumber\\
&\equiv&\mathcal{T}^{-1}(s)_{ij}+g(s)_i\,\delta_{ij}\label{ollerUnitarization}\ .
\eea
Here $\mathcal{T}^{-1}(s)_{ij}$ indicates other contributions coming from local, pole terms and crossed channel dynamics but without right-hand cut, and $g(s)_i$ can actually be identified to the familiar scalar loop integral:
\bea
g(s)_i&=&i\int\frac{{\rm d}^4q}{(2\pi)^4}\frac{1}{(q^2-M_i^2+i\epsilon)((P-q)^2-m_i^2+i\epsilon)}\nonumber\\
&=&\frac{1}{16\pi^2}\bigg\{a_i(\mu)+\log\frac{m_i^2}{\mu^2}+\frac{M_i^2-m_i^2+s}{2s}\log\frac{M_i^2}{m_i^2}\nonumber\\
&&+\frac{q_i}{\sqrt{s}}\log\frac{m_i^2+M_i^2-s-2\sqrt{s}\,q_i}{m_i^2+M_i^2-s+2\sqrt{s}\,q_i}\bigg\}
\eea
where $M_i$ and $m_i$ are the meson and baryon masses in the $i_{\rm th}$ channel. Eq.~\eqref{ollerUnitarization} can be rewritten in the following matrix form
\bea
{ T}(s)=\left[1+\mathcal{T}(s)\cdot g(s)\right]^{-1}\cdot\mathcal{T}(s)\ ,\label{OMuni}
\eea
with $T(s)=(T(s)_{ij})$, $\mathcal{T}(s)=(\mathcal{T}(s)_{ij})$ and $g(s)={\rm diag}(g(s)_1,\cdots,g(s)_i,\cdots)$. Since the LO chiral amplitude $\mathcal{V}({s})$ in Eq.~\eqref{eq:ollerV} is considered as input in Ref.~\cite{Oller:2000fj}, the kernel $\mathcal{T}(s)$ is just set as
\bea
\mathcal{T}(s)=\mathcal{V}_{\rm LO}(s)\ .
\eea
If further adding higher-order chiral perturbative terms, one should be reminded that the right-hand cut contribution should be subtracted.
The above proposed unitarization approach can be either regarded as an improved version of the $K$ matrix or a BS equation method under on-shell approximation as shown in section~\ref{sec:2.2}.

Based on the above formalism, it was found that in the same Riemann sheet there are two poles of the unitarized $S$-wave amplitude close to the $\Lambda(1405)$ resonance, as it was the case within the bag model~\cite{Fink:1989uk}. The lower pole is located at $z_{R}=1390-i66$~MeV in the complex plane, while the higher pole is situated at $z_{R}=1426-i16$~MeV~\cite{Jido:2003cb},  both contributing to the $\pi \Sigma$ distribution. The two poles attribute their existence to two attractive forces in the coupled channels~\cite{Jido:2003cb,Hyodo:2007jq} that relies much on the sign and the strength of the used leading order interaction given in Eq.~\ref{eq:LOKNV}. Nevertheless, the robustness of the two-pole structure of the $\Lambda(1405)$ state was verified by going beyond LO~\cite{Ikeda:2011pi,Mai:2012dt,Ikeda:2012au,Guo:2012vv,Mai:2014xna}, by taking into account off-shell effects with the use of BS equation~\cite{Mai:2012dt,Mai:2014xna}, and by incorporating new experimental data~\cite{Bazzi:2011zj,Bazzi:2012eq}. The upgraded results on the $\Lambda(1405)$ are compiled in Table~\ref{tab:1405}. For comparison, poles obtained with LO chiral potential~\cite{Oller:2000fj,Jido:2003cb} and the measured value summarized from production experiments in PDG~\cite{Zyla:2020zbs} are also listed. A comparison of different unitarization results for the $\Lambda$(1405) is also given in Ref.~\cite{Cieply:2016jby}.
\begin{table}[hbt]
\caption{\label{tab:1405} Pole positions of the $\Lambda(1405)$ resonance from partial-wave chiral dynamics, in units of MeV. }
\begin{indented}
\item[]\begin{tabular}{@{}cll}
\br
&\multicolumn{2}{c}{pole position}\\
\cline{2-3}
 Approach used& Lower pole & Higher pole\\
\mr
NLO+BSE (on-shell appr.)~\cite{Ikeda:2011pi,Ikeda:2012au}&$1381^{+18}_{-6}-i\,81^{+19}_{-8}$& $1424^{+7}_{-23}-i\,26^{+3}_{-14}$
\\
NLO+BSE (on-shell appr.),~Fit II of~\cite{Guo:2012vv}& $1388^{+9}_{-9}-i\,114^{+24}_{-25}$ & $1421^{+3}_{-2}-i\,19^{+8}_{-5}$
\\
NLO+BSE,  solution \#2 of~\cite{Mai:2014xna} &$1330^{+4}_{-5}-i\,56^{+17}_{-11}$ & $1434^{+2}_{-2}-i\,10^{+2}_{-1}$
\\
NLO+BSE, solution \#4 of~\cite{Mai:2014xna} &$1325^{+15}_{-15}-i\,90^{+12}_{-18}$ & $1429^{+8}_{-7}-i\,12^{+2}_{-3}$\\
\mr
LO +BSE (on-shell appr.)~\cite{Oller:2000fj,Jido:2003cb} &$1390-i\,66$ & $1426-i\,16$\\
\mr
 Production experiments in PDG~\cite{Zyla:2020zbs}& \multicolumn{2}{c}{$1405.1^{+1.3}_{-1.0}-i\, 25.3^{+1.0}_{-1.0}$}
\\
\br
\end{tabular}
\end{indented}
\end{table}

\subsection{Exotic states\label{sec:3.3}}
\subsubsection{$D^*_{s0}(2317)$ and $D^*_0(2400)$: from $\mathcal{P}\phi$ scattering}\label{sec:D2317}
Since 2003, many new excited charmed states have been observed experimentally~\cite{Aubert:2003fg,Krokovny:2003zq,Besson:2003cp,Bevan:2014iga}, and further experiments are still going on with the purpose of  precisely investigating their properties or searching for more new states. Some of those observed charmed meson are at odd with expectations from the conventional quark-potential models, of which the most fascinating one is $D^*_{s0}(2317)$ with quantum numbers $J^P=0^+$. This charm-strange scalar state was first observed by the BABAR
Collaboration by analysing the inclusive $D_s^+\pi^0$ invariant mass distribution and
later it was confirmed by CLEO and Belle
Collaborations~\cite{Aubert:2003fg,Besson:2003cp,Krokovny:2003zq}. The $D^*_{s0}(2317)$ couples to
the $DK$ channel and decays mainly into the isospin breaking channel $D_s\pi$
being below the $DK$ threshold.  Many investigations
were triggered in consequence, aiming at revealing the nature of the $D_{s0}^*(2317)$ meson
as well as other newly observed charmed states with $J^P=0^+$.  For instance, the $D_{s0}^*(2317)$ has been
suggested to be interpreted as a $DK$ molecule~\cite{Barnes:2003dj}. Were this
interpretation true, one can learn much about the interaction between the charmed mesons
$\mathcal{P}\in\{D/D_s\}$
 and  Goldstone bosons $\phi\in\{\pi/K/\eta\}$ from studying the $D_{s0}^*(2317)$  and related states.
This path has been followed in Refs.~\cite{Kolomeitsev:2003ac,Hofmann:2003je,
Guo:2006fu,Gamermann:2006nm,Guo:2008gp,Guo:2009ct,Cleven:2010aw,
Cleven:2014oka} where the $S$-wave interaction between $P$ mesons and
Goldstone bosons $\phi$ was investigated
systematically up to the next-to-leading order (NLO) using chiral effective theory for heavy mesons~\cite{Burdman:1992gh,Wise:1992hn,Yan:1992gz} in
combination with a unitarization approach such as the one in
Ref.~\cite{Oller:2000fj}.

In the meantime, lattice QCD~\cite{Moir:2013ub,Cichy:2015tma} has made significant progress in the study of $\mathcal{P}\phi$ interaction.
The L{\"u}scher formalism and its extension to coupled channels (see e.g. Refs.~\cite{Liu:2005kr,Lage:2009zv} for early works) are adopted to calculate
scattering lengths and recently phase shifts for the $\mathcal{P}\phi$ interaction at unphysical quark
masses~\cite{Liu:2008rza,Liu:2012zya,Mohler:2012na,Mohler:2013rwa,
Lang:2014yfa,Moir:2016srx}. The first calculation only concerns the isospin-3/2 $D\pi$, $D_s\pi$, $D_s$K, isospin-0 and isospin-1 $D\bar{K}$ channels~\cite{Liu:2008rza,Liu:2012zya}, which are free of disconnected Wick contractions.
Those channels with disconnected Wick contractions
such as isospin-1/2 $D\pi$ and isospin-0 $DK$ channel were calculated later by other lattice groups~\cite{Mohler:2012na,Mohler:2013rwa,Lang:2014yfa}.
The lattice results  are indispensable for us to explore the underlying charm physics. They can be used to determine the unknown LECs in the chiral
Lagrangian~\cite{Wang:2012bu,Liu:2012zya,Altenbuchinger:2013vwa,Yao:2015qia,
Guo:2015dha}.

At present, the widely-used NLO potentials for the processes of $\mathcal{P}_1(p_1)\phi_1(p_2)\to \mathcal{P}_2(p_3)\phi_2(p_4)$ with strangeness $S$ and isospin $I$ read
\bea
\mathcal{V}^{(S,I)}(s,t) _{\mathcal{P}_1\phi_1\to\mathcal{P}_2\phi_2}&=& \frac1{F^2} \bigg\{\frac{\mathcal{C}_{\rm LO}}{4}(s-u) - 4 \mathcal{C}_0 h_0 +
2 \mathcal{C}_1 h_1 - 2\mathcal{C}_{24} \mathcal{H}_{24} (s,t)\nonumber\\
&&\hspace{1cm}+ 2\mathcal{C}_{35} \mathcal{H}_{35}(s,t)\bigg\}\label{eq:pot}
\eea
where $s\equiv(p_1+p_2)^2$, $t\equiv(p_1-p_3)^2$ are standard Mandelstam variables. We refer the readers to Ref.~\cite{Liu:2012zya} or Ref.~\cite{Altenbuchinger:2013vwa} for the coefficients, which are derived from the relevant chiral effective Lagrangians. The functions $\mathcal{H}_{24}(s,t)$ and $\mathcal{H}_{35}(s,t)$ are given by
\bea
\mathcal{H}_{24}(s,t) &=& 2 h_2 p_2\cdot p_4 + h_4 (p_1\cdot p_2 p_3\cdot p_4 +
p_1\cdot p_4 p_2\cdot p_3)\,, \\
\mathcal{H}_{35}(s,t) &=& h_3 p_2\cdot p_4 + h_5
(p_1\cdot p_2 p_3\cdot p_4 + p_1\cdot p_4 p_2\cdot p_3)\,.
\eea
Note that the most recent unitarized amplitudes based on one-loop (i.e. NNLO) potentials can be found in
Refs.~\cite{Yao:2015qia,Du:2017ttu}. However, as pointed out in
Ref.~\cite{Du:2016tgp}, the LECs involved in the NNLO analyses can not be well determined
due to the lack of precise data.

For a given angular momentum $L$, the partial wave projection of the potentials in Eq.~\eqref{eq:pot} is given by
\bea\label{pwv}
\mathcal{V}_{L,\,\mathcal{P}_1\phi_1\to \mathcal{P}_2\phi_2}^{(S,I)}(s)
= \frac{1}{2}\int_{-1}^{+1}{\rm
d}\cos\theta\,P_J(\cos\theta)\, \mathcal{V}^{(S,I)}_{\mathcal{P}_1\phi_1\to
\mathcal{P}_2\phi_2}(s,t(s,\cos\theta))\,,
\label{eq:pwp}
\eea
with $\theta$ the scattering angle between the initial and final particles
in the center-of-mass frame.  The Mandelstam variable $t$ can be expressed as a function of $s$ and $\cos\theta$, i.e.
\bea
t(s,\cos\theta)&=& m_{\mathcal{P}_1}^2+m_{\mathcal{P}_2}^2-
\frac{1}{2s}\left(s+m_{\mathcal{P}_1}^2-m_{\phi_1}^2\right)
\left(s+m_{\mathcal{P}_2}^2-m_{\phi_2}^2\right) \nonumber\\
\al\al
-\frac{\cos\theta}{2s}\sqrt{\lambda(s,m_{\mathcal{P}_1}^2,M_{\phi_1}^2)
\lambda(s,m_{\mathcal{P}_2}^2,m_{\phi_2}^2)}\ .
\label{eq:t}
\eea
where $\lambda(a,b,c)=a^2+b^2+c^2-2ab-2bc-2ac$ is the K\"all\'{e}n function.

According to Eq.~\eqref{eq:BS;OS} displayed in section~\ref{sec:2}, the unitarized two-body scattering amplitude has the following form~\cite{Oller:2000fj}
\begin{eqnarray} \label{defut}
 \mathcal{T}(s) = \big[ 1 - \mathcal{V}(s)\cdot g(s) \big]^{-1}\cdot \mathcal{V}(s)\,,
\end{eqnarray}
where $\mathcal{V}(s)$ represents the partial wave potentials in Eq.~\eqref{pwv}. For brevity, the super- and subscripts have been suppressed. The function $g(s)$ encodes information on the unitarity cuts generated by the intermediate two-particle states under consideration.

 The $S$-wave scattering lengths can be obtained from the
scattering amplitudes $\mathcal{T}(s)$ through
 \bea
 a^{(S,I)}_{\mathcal{P}\phi\to \mathcal{P}\phi}=-\frac{1}{8\pi(m_\mathcal{P}+m_{\phi})}\mathcal{T}^{(S,I)}_{L=0}(s_{\rm thr})_{\mathcal{P}\phi\to \mathcal{P}\phi}\ ,
 \eea
with $s_{\rm thr}=(m_\mathcal{P}+m_\phi)^2$. This formula is quite suitable to describe the existing lattice data just by replacing all the quantities in the formula by their corresponding pion mass dependent ones, see e.g. Ref.~\cite{Yao:2015qia} for details. With the above theoretical equations, fits can be performed to the lattice data and good agreement was achieved as shown in Fig.~4 of Ref.~\cite{Liu:2012zya}. Most importantly, all the unknown parameters in the unitarized scattering amplitudes were pinned down, enabling one to analyze the underlying dynamics.  Below, we will only review the dynamically generated poles in the channels with quantum numbers $(S,I)=(1,0)$ and $(S,I)=(0,1/2)$.

The most interesting channel of $\mathcal{P}\phi$ interaction is the one  $(S,I)=(1,0)$, where the $D^*_{s0}(2317)$ resides. In this channel, a dynamically generated pole was found  with the pole position $2315^{+18}_{-28}$ MeV, which is very close to the reported mass of the $D^*_{s0}(2317)$, $(2317.8\pm0.5)$ MeV~\cite{Zyla:2020zbs}. This finding strongly supports the molecule nature of the $D^*_{s0}(2317)$ state from the viewpoint of partial wave dynamics with the help of chiral effective theory and untarization approach. In the coupled $D\pi$-$D\eta$-$D_s\bar{K}$ channel with $(S,I)=(0,1/2)$, it was demonstrated in Ref.~\cite{Albaladejo:2016lbb} that two poles can be found in the $D^*_0(2400)$ region using the parameters fixed in Ref.~\cite{Liu:2012zya}. In addition, energy levels in this channel were computed in Ref.~\cite{Albaladejo:2016lbb}, and a remarkable agreement with the existing lattice results given in Ref.~\cite{Moir:2016srx} was established. This agreement can be regarded as a strong evidence that the particle denoted as $D_0^*(2400)$ in PDG~\cite{Olive:2016xmw} actually corresponds to two states with poles located at $(2105^{+6}_{-8}-i102^{+10}_{-12})$ MeV and
$(2451^{+36}_{-26}-i134^{+7}_{-8})$ MeV, respectively~\cite{Albaladejo:2016lbb}, similar to the well-known two-pole structure of the $\Lambda(1405)$~\cite{Oller:2000fj}. Very recently, a brief review on the two-pole structure in QCD is presented in Ref.~\cite{Meissner:2020khl}. It should be pointed out that the $\mathcal{P}\phi$ interaction was also studied with NLO potentials obtained in U(3) sector and $N_c$ trajectories of the poles in the $(S,I)=(1,0)$ and $(0,1/2)$ channels are obtained. It is found that, with $N_C$ increasing, neither the $D^*_{s0}(2317)$ pole nor the poles in the $(0,1/2)$ channel tend to fall down to the real axis, which indicate that each of the states do not behave like a quark-antiquark meson at large $N_c$.

 Based on the U$\chi$PT results, a way towards a new paradigm for heavy-light meson spectroscopy is discussed in Ref.~\cite{Du:2017zvv}. It was pointed out that various puzzles in the charmed-meson spectrum, e.g. the abnormal mass ordering of the charm-strange $D^*_{s0}(2317)$ and the nonstrange $D_{0}^*(2400)$ states, can be resolved naturally if the SU(3) multiplets for the lightest $0^+$ states owe their existence to the non-perturbative partial-wave dynamics of the $\mathcal{P}\phi$ interaction.  The resolution for the $0^+$ meson also holds true for the lightest $1^-$ states given that heavy quark spin symmetry is enforced. If further implementing heavy quark flavour symmetry,  same conclusions can be made in the bottom as the charm sector. In Table~\ref{tab:hsmeson} and Table~\ref{tab:nsmeson}, U$\chi$PT results of the heavy-quark-symmetry partners of the $D^*_{s0}(2317)$ and $D^*_{0}(2400)$ are shown. For comparison, the latest LQCD and experimental results are also compiled in the two tables.

\begin{table}[bt]
\caption{\label{tab:hsmeson} Masses of lowest positive-parity heavy-strange mesons, in units of MeV. For comparison, U$\chi$PT results, the measured values and latest LQCD results are shown.}
\begin{indented}
\item[]\begin{tabular}{@{}cccc}
\br
& U$\chi$PT~\cite{Du:2017zvv} & \text{RPP}~\cite{Patrignani:2016xqp} & \text{LQCD} \\
\mr
$D_{s0}^*$ & $2315^{+18}_{-28}$
& $2317.7\pm 0.6$ & $2348^{+7}_{-4}$~{\cite{Bali:2017pdv}} \\
$D_{s1}$ &  $2456^{+15}_{-21}$  & $2459.5\pm0.6$ &  $2451\pm4$~{\cite{Bali:2017pdv}}
\\
$B_{s0}^*$   & $5720^{+16}_{-23}$ & - & $5711\pm23$~{\cite{Lang:2015hza}} \\
$B_{s1}$ &  $5772^{+15}_{-21}$  & - & $5750\pm25$~\text{\cite{Lang:2015hza}} \\
\br
\end{tabular}
\end{indented}
\end{table}

\begin{table}[bt]
\caption{\label{tab:nsmeson} Pole positions of lowest positive-parity non-strange mesons, in units of MeV. For comparison, U$\chi$PT results, the measured values and latest LQCD results are shown. The poles are given as ($M,\Gamma/2$), with $M$ the mass and $\Gamma$ the total decay width. }
\begin{indented}
\item[]\begin{tabular}{@{}cccc}
\br
  & \text{lower pole}~\cite{Du:2017zvv} & \text{higher pole}~\cite{Du:2017zvv} & \text{RPP}~\cite{Patrignani:2016xqp} \\ \hline
$D_0^*$ & $\left(2105^{+6}_{-8}, 102^{+10}_{-11}\right)$
& $\left(2451^{+35}_{-26},134^{+7}_{-8}\right)$ & $(2318\pm29,134\pm20) $\\
$D_1$ &  $\left(2247^{+5}_{-6}, 107^{+11}_{-10} \right)$  & $\left(2555^{+47}_{-30},
203^{+8}_{-9}\right)$ & $(2427\pm40,192^{+65}_{-55})$
\\
$B_0^*$   & $\left(5535^{+9}_{-11},113^{+15}_{-17} \right) $
&  $\left(5852^{+16}_{-19},36\pm5\right)$ & - \\
$B_1$ &   $\left( 5584^{+9}_{-11}, 119^{+14}_{-17} \right) $
      &  $\left(5912^{+15}_{-18}, 42^{+5}_{-4}\right)$ & -\\
\br
\end{tabular}
\end{indented}
\end{table}

\subsubsection{Doubly charmed baryons}
 In quark model with $(u, d,s,c)$ quarks,  three doubly charmed ground-state baryons with $J^P={\frac{1}{2}}^+$ show up in the $\mathbf{20}_M$-plet representation of flavour SU(4) group~\cite{Zyla:2020zbs}. The three baryons are named as $\Xi_{cc}^{++}$, $\Xi_{cc}^+$ and $\Omega_{cc}^{+}$. Their masses should be close to each other, since their corresponding quark constituents are $[ccu]$, $[ccd]$ and $[ccs]$ in order. In 2002, it was reported by SELEX Collaboration~\cite{Mattson:2002vu} that the $\Xi_{cc}^+$ was first observed  and its measured mass is $3519\pm2$~MeV~\cite{Ocherashvili:2004hi}. Unfortunately, for a very long time, this state was not confirmed by any other experimental collaborations such as
FOCUS~\cite{Ratti:2003ez}, Belle~\cite{Chistov:2006zj}, Babar~\cite{Aubert:2006qw} and LHCb~\cite{Aaij:2013voa}. Furthermore, the experimental value deviates from theoretical results determined by effective potential method~\cite{Karliner:2014gca}, relativistic quark model~\cite{Ebert:2002ig}, lattice QCD~\cite{Liu:2009jc,Brown:2014ena}, heavy quark effective theory~\cite{Korner:1994nh}, etc. Therefore, the realistic existence of doubly charmed baryon, particularly the $\Xi_{cc}^+$ state, was called into question. Actually, the SELEX results were also put in question using heavy quark-diquark symmetry, see Ref.~\cite{Brodsky:2011zs}.

The question was solved in 2017 that the doubly charged state $\Xi_{cc}^{++}$ with a mass of $3621.4\pm0.78$~MeV, consistent with previous theoretical determinations within 1-$\sigma$ uncertainty~\cite{Karliner:2014gca,Ebert:2002ig,Brown:2014ena}, was observed by LHCb collaboration~\cite{Aaij:2017ueg}. Renewed interest has been triggered in studying doubly charmed baryons. Relevant theoretical works have been successively accumulated, e.g., in the investigations of weak decays~\cite{Wang:2017azm,Wang:2017mqp}, strong and radiative decays~\cite{Li:2017pxa,Xiao:2017udy}, magnetic moments~\cite{Li:2017cfz} and so on. Within the context of relativistic baryon $\chi$PT, the masses and electromagnetic form factors of the doubly charmed baryons were studied at loop level in Refs.~\cite{Sun:2016wzh,Yao:2018ifh,Blin:2018pmj}. In particular, partial wave dynamics of the interaction between the doubly charmed baryons and Goldstone bosons was first analyzed in Ref.~\cite{Guo:2017vcf}.
{For the process denoted by $\psi_{cc,A}(p)\phi_i(q)\to \psi_{cc,B}(p^\prime)\phi_j(q^\prime)$, the explicit expression of the LO potential reads
\setlength{\mathindent}{0.cm}
\bea
T^{Ai\to Bj} = -\frac{1}{F_0^2}\bar{u}_B\bigg[\mathcal{F}_{wt}(\slashed{q}+\slashed{q}^\prime)+g_A^2\left(\mathcal{F}_s\slashed{q}^\prime\gamma_5\frac{1}{\slashed{p}-m}\slashed{q}\gamma_5+(\slashed{q}\leftrightarrow\slashed{q}^\prime,s\leftrightarrow u)\right)
\bigg]u_A
\eea
where $\mathcal{F}_{wt}$, $\mathcal{F}_{s}$ and $\mathcal{F}_{u}$ are symmetry coefficients~\cite{Guo:2017vcf}. S-wave exotic doubly charmed baryon excitations were predicted by applying the on-shell-approximation BS approach introduced in section~\ref{sec:2.3} to the above chiral potentials.
}
Furthermore, by using NLO chiral potentials and taking into account the P-wave excitations between the two charm quarks, a refined analysis was done in Ref.~\cite{Yan:2018zdt}, leading to two possible quasistable states in the spectrum of negative-parity doubly charmed baryons.

\subsubsection{$P_c$ states\label{sec:Pc}}
The LHCb Collaboration first observed two hidden charm pentaquark states,
$P_c(4380)^+$ and $P_c(4450)^+$, in the $J/\psi p$ invariant mass distribution
in 2015~\cite{Aaij:2015tga}. Obviously, the $P_c$ states contain at least five quark component $\bar{c}cuud$ and thus the discovery trigged lots of theoretical
discussion, but their internal structure is still under debate.
Four years later, LHCb just updated more precisely search for the puzzling structures~\cite{Aaij:2019vzc}.
The decay events collected now by Run 1 and Run 2 are about ten times as that of Run 1.
As a result, the bin size has been decreased from 15 to 2~MeV.
With the high statistics one new narrow state
$P_c(4312)^+$ was found. What makes it more interesting is that the
$P_c(4450)^+$ split into two overlapping peaks, $P_c(4440)^+$ and $P_c(4457)^+$,
with the analysis of more data set.

In Refs.~\cite{Guo:2015umn,Liu:2015fea,Guo:2016bkl,Bayar:2016ftu}, it was shown that the triangle diagram can contribute to the
old $P_c(4450)^+$ peak, while similar
contribution also exists for the newly observed $P_c(4457)^+$ peak~\cite{Aaij:2019vzc}.
Besides, hadronic molecular interpretation on the
$P_c(4450)^+$~\cite{Roca:2015dva,He:2015cea,Burns:2015dwa,Xiao:2015fia,Lu:2016nnt, Lin:2017mtz,Geng:2017hxc,Liu:2018zzu}
and the new $P^+_c$
structure~\cite{Chen:2019bip,Chen:2019asm,Liu:2019tjn,He:2019ify,Huang:2019jlf,Xiao:2019mvs,Shimizu:2019ptd,Guo:2019kdc,Du:2019pij}  was claimed,
due to the close proximity of the $P^+_c$ structures to the $\bar{D}^{(*)}\Sigma_c$ thresholds.
Note that methods of partial wave dynamics lead to the molecular interpretations obtained in most of the above-mentioned works and the driven amplitudes are very simple, which are merely constants. For a recent review on hadronic molecules, see Ref.~\cite{Guo:2017jvc}. It was pointed out by Ref.~\cite{Guo:2019fdo} to check the molecule structure of these states in the isospin breaking decays.
In contrast, Ref.\cite{Fernandez-Ramirez:2019koa} prefers the $P_c(4312)$ to be a virtual state, due to the fact that the attractive effect of the $\Sigma_c^+ \bar{D}^0$ channel is not strong enough to form a bound state. Ref.\cite{Kuang:2020bnk} performed an amplitude analysis of the $J/\Psi p$ ~-~ $\bar{D}^0 \Sigma_c^+$~-~$\bar{D}^{*0} \Sigma_c^+$ triple channels based on $K$-matrix and AMP method when constructing the decay amplitudes. High quality of fit was obtained and pole locations were extracted, e.g. the $P_c(4440)$ at $4444.09^{+2.53}_{-1.48}-i3.10^{+0.53}_{-1.33}$~MeV and
the $P_c(4312)$ at $4314.31^{+2.06}_{-1.10}-i1.43^{+1.50}_{-0.57}$~MeV.
Furthermore, one pole for the $P_c(4312)$, four poles for the $P_c(4440)$ and no pole for the $P_c(4457)$ are found in this triple-channel study. With the pole counting rule it suggests that the $P_c(4312)$, $P_c(4440)$ and $P_c(4457)$ are molecule, compact pentaquark, and threshold behaviour, in order.
It should be noticed that in the compact pentaquark picture~\cite{Ali:2019npk}, the spin-parities of the $P^+_c$ states
structure would be $3/2^-$, $3/2^+$ and $5/2^+$ for the $P_c(4312)^+$,  $P_c(4312)^+$ and  $P_c(4312)^+$, respectively. They are different from the molecular interpretation. Even in molecular hypothesis, different models give somehow different description of quantum numbers for the $P^+_c$ structure. For example, Ref.~\cite{Chen:2019asm} used the one-boson-exchange model and got that the $P_c(4440)^+$ and $P_c(4457)^+$ correspond to $\bar{D}^*\Sigma_c$ with $J^P = 1/2^-$ and $\bar{D}^*\Sigma_c$ with $J^P = 3/2^-$, respectively. While Ref.~\cite{Huang:2019jlf} got opposite quantum numbers for the two by using quark delocalization color screening model.

Actually, some literature predicted the existence of the narrow $\bar{D}^0\Sigma_c^+$ and $\bar{D}^{*0}\Sigma_c^+$ states long before this discovery~\cite{Wang:2011rga,Yang:2011wz,Wu:2012md}. Thereinto, by using various unitarization methods with EFT results as inputs, partial wave analysis is one of the most primary approaches to properly yield theoretical predictions on exotic states including the $P_c$ states, as illustrated in Ref.~\cite{Wu:2012md}. And there are plenty of follow-up and improved investigations alone this line, e.g.~Refs.~\cite{Roca:2015dva,Xiao:2015fia,Liu:2019tjn,Xiao:2019mvs}. For more discussions on pentaquarks or tetraquarks, see e.g. Refs.~\cite{Chen:2016qju,Liu:2019zoy}.


\section{Summary and outlook\label{sec:sum}}
In studying the non-perturbative problems in the low energy region where perturbative QCD does not work, there emerge long-history controversies about whether or not the resonances as $\sigma$ or $\kappa$ exist, which blurs the categorization of the hadron states with their quark configurations. Until recently, great efforts have been made using $\chi$EFTs and dispersive relations, respecting fundamental principles of $S$-matrix theory and QCD. It leads to a general consensus that these problems finally settle down and precise pole locations are determined.  Nevertheless, a trustworthy
explanation of why these poles including the $a_0(980)$ and $f_0(980)$ appear is still desirable. Furthermore, modern developments of unitarization of $\chi$EFTs with heavy hadron states could also successfully expand the application region of $\chi$EFTs. In the next few years, measurements of large statistics of experimental data of LHCb, Belle, and BESIII, etc, will be promising to establish more states with exotic quantum numbers, which may bring new challenges to our knowledge of how the quarks and gluons combine to form hadron states. As we experienced when studying light scalars, the approaches respecting unitarity, analyticity, crossing symmetry, and chiral symmetry will still be expected to play an important role to reach a solid conclusion when facing these new challenges.

\ack
We would like to thank our collaborators for sharing their insights into the topics discussed in this review. We are grateful to Prof. Ulf-G.~Mei{\ss}ner for instructive discussions and careful reading of the manuscript. Discussions with Profs. S.~Descotes-Genon and G.~Colangelo about Roy and Roy-Steiner equations are also appreciated. This work is supported by National Nature Science Foundations of China (NSFC) under Contract Nos. 11905258, 11975028, 11805059, 11675051,  11975075, and 10925522, by Joint Large Scale Scientific Facility Funds of the NSFC and Chinese Academy of Sciences (CAS) under Contract No.~U1932110, and by the Fundamental Research Funds for the Central Universities.
\noindent


\section*{References}
\bibliographystyle{unsrt}
\bibliography{PWDbib}
\end{document}